\documentclass[fleqn,usenatbib]{mnras}

\usepackage{mathptmx}

\usepackage[T1]{fontenc}

\DeclareRobustCommand{\VAN}[3]{#2}
\let\VANthebibliography\thebibliography
\def\thebibliography{\DeclareRobustCommand{\VAN}[3]{##3}\VANthebibliography}

\usepackage{graphicx}	
\usepackage{amsmath}	
\usepackage{amssymb}	
\usepackage{amsfonts}
\usepackage[lining,semibold,scaled=1.0]{ebgaramond-maths}
\usepackage{garamondlibre}
\usepackage[colorinlistoftodos,prependcaption,textsize=tiny]{todonotes}
\usepackage{marginnote}
\usepackage{rotating}
\usepackage{adjustbox}
\usepackage{multirow}
\usepackage{array}
\usepackage{float}
\usepackage{mathtools}
\usepackage{textcomp}
\usepackage{stmaryrd}
\usepackage{subcaption}
\usepackage{longtable}
\usepackage{tabularx}
\usepackage{afterpage}
\usepackage{pdflscape}
\usepackage{rotating}
\usepackage{orcidlink}
\usepackage{hyperref}
\usepackage{amsmath}
\usepackage{tikz}
\usepackage[dvipsnames]{xcolor}
\usepackage[skins,theorems,most,listings]{tcolorbox}
\usepackage{lineno}
\usepackage{xpatch}
\usepackage{mathrsfs}
\usepackage{xspace}
\usepackage[noabbrev, nameinlink]{cleveref}
\usepackage{xparse}

\usetikzlibrary{arrows.meta, shapes.misc, patterns, shapes.geometric, decorations.pathmorphing,decorations.pathreplacing,fadings,backgrounds}
\tcbset{highlight math style={enhanced,colframe=red,colback=white,arc=0pt,boxrule=1pt}}
\renewcommand{\star}{\bigstar}
\newcommand{\dd}{~\mathrm{d}\xspace}

\newcommand{\Moyr}{\text{$\mathrm{M_{\odot}~yr^{-1}}$}\xspace}
\newcommand{\Moyrk}{\text{\xspace$\mathrm{M_{\odot}~yr^{-1}~kpc^{-2}}$\xspace}}

\newcommand{\Lsun}{\mathrm{L_{\odot}}\xspace}

\newcommand{\code}[1]{%
    \texttt{#1}
}
\newcommand{\parenthesis}[1]{\left( #1 \right)}

\newcommand{\HII}{\mathrm{H\,{II}}}
\defcitealias{Lucatelli_2024}{LP1}
\makeatletter
\newcommand{\citealias}[2]{%
    \hyperlink{cite.\the\c@refsection @#1}{#2}%
}
\makeatother

\definecolor{uncertain}{RGB}{255, 140, 0}      
\definecolor{amber}{RGB}{255, 191, 0}       
\definecolor{warning}{RGB}{255, 140, 26}    
\definecolor{review}{RGB}{138, 43, 226}

\crefname{chapter}{Chapter}{Chapters}
\crefname{section}{Section}{Sections}
\crefname{subsection}{Section}{Sections}
\crefname{subsubsection}{Section}{Sections}
\crefname{appendix}{Appendix}{Appendices}
\crefname{figure}{Figure}{Figures}
\crefname{table}{Table}{Tables}
\crefname{equation}{Equation}{Equations}

\Crefname{chapter}{Chapter}{Chapters}
\Crefname{section}{Section}{Sections}
\Crefname{subsection}{Section}{Sections}
\Crefname{subsubsection}{Section}{Sections}
\Crefname{appendix}{Appendix}{Appendices}
\Crefname{figure}{Figure}{Figures}
\Crefname{table}{Table}{Tables}
\Crefname{equation}{Equation}{Equations}

\creflabelformat{equation}{#2#1#3}
\crefrangelabelformat{equation}{#3#1#4 to #5#2#6}



\NewDocumentCommand{\parencite}{o o m}{%
	\IfNoValueTF{#1}{%
		\citep{#3}%
	}{%
		\IfNoValueTF{#2}{%
			\citep[#1]{#3}%
		}{%
			\citep[#1][#2]{#3}%
		}%
	}%
}

\NewDocumentCommand{\textcite}{o o m}{%
	\IfNoValueTF{#1}{%
		\citet{#3}%
	}{%
		\IfNoValueTF{#2}{%
			\citet[#1]{#3}%
		}{%
			\citet[#1][#2]{#3}%
		}%
	}%
}

\usepackage{soulutf8}
\usepackage{framed}

\definecolor{revcolor}{RGB}{255,255,200}  
\sethlcolor{revcolor}
\newif\ifdraft
\draftfalse

\ifdraft
    \newcommand{\rev}[1]{{\color{orange}{#1}}}
    
    \newenvironment{revised}{%
        \MakeFramed{\advance\hsize-\width\FrameRestore}%
    }{%
        \endMakeFramed%
    }
    
    \newenvironment{revblock}{%
        \setlength{\fboxsep}{3pt}%
        \MakeFramed{\advance\hsize-\width\FrameRestore}%
    }{%
        \endMakeFramed%
    }
\else
    \newcommand{\rev}[1]{#1}

\fi

\title[Nuclear and Diffuse emission in U/LIRGS]{The PARADIGM Project II: Characterising Nuclear and Diffuse Radio Components in Local U/LIRGs}

\author[G. Lucatelli et al.]{Geferson Lucatelli$^{\orcidlink{0000-0002-2410-1776}1,2}$\thanks{E-mail: gefersonlucatelli@gmail.com | lucatelli@iaa.csic.es (GL)},
Rob Beswick$^{1}$,
Javier Moldon$^{\orcidlink{0000-0002-8079-7608}1,2}$,
Antxon Alberdi$^{2}$,
Miguel \'A. P\'erez-Torres$^{\orcidlink{0000-0001-5654-0266} 2,3,4}$,
\newauthor
Santiago del Palacio$^{\orcidlink{0000-0002-5761-2417}5}$,
Kelvin Wandia$^{\orcidlink{0000-0003-4338-2611}1}$,
Susanne Aalto$^{5}$,
L. Barcos-Mu\~noz$^{7,8}$,
D. Williams-Baldwin$^{\orcidlink{0000-0001-7361-0246}1}$,
\newauthor
J. E. Conway$^{6}$,
Cristina Romero-Ca\~nizales$^{\orcidlink{0000-0001-6301-9073}\ 9}$,
Eskil Varenius$^{\orcidlink{0000-0002-3248-9467}\ 5}$,
Hans-Rainer Klöckner$^{\orcidlink{0000-0002-0648-2704}\ 10}$,
\newauthor
Simon T. Garrington$^{1}$,
Willem A. Baan$^{12,13}$ and
Ylva M. Pihlstrom$^{\orcidlink{0000-0003-0615-1785}\ 14}$\thanks{YMP is also an Adjunct Astronomer at the National Radio Astronomy Observatory.}
\ \\
$^{1}$Jodrell Bank Centre for Astrophysics, School of Physics and Astronomy, The University of Manchester, Manchester M13 9PL, UK\\
$^{2}$Instituto de Astrof\'isica de Andaluc\'ia (IAA-CSIC), Glorieta de la Astronom\'ia s/n, E-18008 Granada, Spain\\
$^{3}$Facultad de Ciencias, Universidad de Zaragoza, Pedro Cerbuna 12, E-50009 Zaragoza, Spain \\
$^{4}$School of Sciences, European University Cyprus, Diogenes street, Engomi, 1516 Nicosia, Cyprus\\
$^{5}$Department of Space, Earth and Environment, Chalmers University of Technology SE-412 96 Gothenburg, Sweden\\
$^{6}$Department of Space, Earth and Environment, Chalmers University of Technology, Onsala Space Observatory, SE-439 92 Onsala, Sweden\\
$^{7}$National Radio Astronomy Observatory, 520 Edgemont Road, Charlottesville, VA  22903, USA \\
$^{8}$Department of Astronomy, University of Virginia, 530 McCormick Road, Charlottesville, VA 22903, USA \\
$^{9}$Institute of Astronomy and Astrophysics, Academia Sinica, 11F of Astronomy-Mathematics Building, Taiwan, R.O.C.\\
$^{10}$Max-Planck-Institut für Radioastronomie, Auf dem Hügel 69, 53121 Bonn, Germany\\
$^{12}$Netherlands Institute for Radio Astronomy (ASTRON), NL-7991 PD Dwingeloo, the Netherlands\\
$^{13}$Xinjiang Astronomical Observatory, Chinese Academy of Sciences, 150 Science 1-Street, 830011 Urumqi, China\\
$^{14}$Department of Physics and Astronomy, University of New Mexico, Albuquerque, NM 87131, USA\\
}

\date{ 
    Accepted 2026 May 5. Received 2026 May 13; in original form 2026 March 27
    }

\pubyear{2026}

\begin{document}
\label{firstpage}
\pagerange{\pageref{firstpage}--\pageref{lastpage}}
\maketitle
\begin{abstract}
\rev{Disentangling SF and AGN emission is essential for understanding galaxy evolution, yet remains challenging in merging systems where both processes are enhanced and spatially intertwined. Galaxy mergers drive gas inflows that simultaneously fuel nuclear starbursts and black hole accretion, shaping morphology from nuclear ($\lesssim 250$~pc) to large-scale ($\gtrsim 500$~pc) regions. Radio interferometry provides an unobscured view, but separating compact nuclear starbursts, AGN, and diffuse star formation requires multiscale, multi-frequency observations.}
We present a systematic method to characterise multiscale radio properties in 15 local ($z\lesssim 0.1$) Luminous and Ultra-Luminous Infrared Galaxies (U/LIRGs) ($L_{\mathrm{IR}} > 10^{11}\mathrm{L}_{\odot}$). Using \emph{e}-MERLIN and VLA at 1.4, 6.0 and 33.0~GHz, we probe physical scales from $\sim 10$--$250$~pc to $\sim 0.5$--$3.0$~kpc. We decompose radio emission into nuclear (compact cores and nuclear extended) and large-scale (total and \rev{diffuse}) components, comparing morphological properties (emission fractions, sizes, luminosities, surface densities) and investigating correlations with source classes, merger stages, and infrared luminosities.
We find: i) nuclear emission contributes $\sim$50\% of total radio emission on average; ii) total multiscale \rev{diffuse} emission (SF-related) contributes $\sim$80\% to total power; iii) nuclear emission components act together to correlate with total radio and infrared luminosities, which increase with merger stage, whilst diffuse emission at larger scales shows no clear dependence on nuclear processes; \rev{iv) sources with radio excess (lower $q_{\mathrm{IR}}$) show lower nuclear luminosity ratios $L_{\mathrm{R,33}}^{\mathrm{N}}/L_{\mathrm{R,6}}^{\mathrm{N}}$, indicating a deficit of high-frequency radio emission; since 33.0~GHz traces recent star formation, this suggests the radio excess is dominated by non-thermal emission at lower frequencies, likely AGN-related, rather than enhanced star formation.}
\end{abstract}



\section{Introduction}
\label{sec:paper_2_introduction}
Understanding the co-evolution of star formation (SF) and active galactic nuclei (AGN) is fundamental to our knowledge of galaxy evolution \parencite{Kormendy_2013,Hopkins_2008,Fabian_2012,Heckman_2014}. Galaxy mergers provide unique environments where this SF-AGN connection becomes evident, as gravitational interactions drive gas inflows that can simultaneously fuel nuclear starbursts and black hole accretion \parencite{Sanders1996,Hopkins_2006,Kormendy_2013,Hopkins_2006,DiMatteo_2005}. However, disentangling the relative contributions of SF and AGN requires observations across multiple spatial scales, from compact nuclear regions where AGN and/or nuclear starbursts dominate, to extended regions where large-scale star formation is evident \parencite{Herrero2012,loreto2015,Santaella_2015,Ramirez_Olivencia_2022}.

Luminous and Ultra-Luminous Infrared Galaxies (U/LIRGs), having infrared luminosities of $L_{\mathrm{IR}}[8$--$1000~\mu$m] $\gtrsim 10^{11}~\mathrm{L_{\odot}}$ \parencite{Sanders1988,Sanders1996} represent ideal laboratories for such studies, as they host both intense star formation and frequent AGN activity \parencite{Sanders_2003,Veilleux_2009}. However, these systems are heavily obscured at optical and infrared wavelengths due to high dust content \parencite{Genzel_1998,Armus2009}. Radio interferometry offers an attractive solution, providing unobscured views of both SF and AGN emission across the required spatial scales \parencite{Condon1991,Condon1992,Murphy2012}. Previous radio studies have revealed the complex nature of U/LIRGs, showing compact nuclear sources embedded within extended emission \parencite{Smith_1998,Clemens_2008,Vardoulaki2015,Herrero_Illana_2017,loreto2017}, yet systematic multiscale analyses remain limited \parencite{loreto2017,Song2021,Song_2022}. 

Such analyses are crucial because different physical mechanisms coexist across multiple spatial scales of a system, from nuclear to large-scale diffuse regions. 
Different physical processes influence gas transport across these scales,
resulting in characteristic features and environmental conditions, particularly evident in merging scenarios, where major mergers of gas-rich galaxies efficiently transport gas from kpc scales to nuclear regions \parencite[e.g.,][]{Hopkins_2010b,Alexander_2012,Hopkins_2013}.

When sufficient matter accretes onto central super-massive black holes (SMBHs) it can trigger intense nuclear activity, typically confined to small spatial scales \parencite[$\lesssim $ 1 pc;][]{Hopkins2010}. This often produces bright core emission characteristic of an AGN \parencite[e.g.,][]{Padovani_2017,Hickox_2018}. 
At scales $\gtrsim 1$~pc, the feedback between SF and AGN activity becomes noticeable \parencite{Fabian_2012,Harrison_2024}. The AGN returns energy and material to the interstellar medium (ISM) through outflows \parencite{Harrison_2018}, which can fuel SF processes at multiple scales \parencite{Cicone_2014,Harrison_2017}, although it can also suppress SF activity \parencite[e.g.,][]{Sturm_2011,Best_2014}.

In these merging systems, both the SF and AGN activity is enhanced as the interaction evolves \parencite{Treister_2012,Matteo_2007,Ellison_2013}. In the local Universe, a common evolutionary phase in mergers produces systems with strong radio and infrared  such as U/LIRGs. The SF rates in these systems are significantly higher ($\gtrsim$ 10 $\Moyr$ up to $\sim 500~\Moyr$) than in normal spiral star-forming galaxies ($\lesssim$ 1 $\Moyr$), hence they are a laboratory of extreme physical processes, in particular SF. Similarly, AGN activity increases with infrared luminosity \parencite{Medling2014,Veilleux_2009,Nardini_2010,Iwasawa2011}, resulting in more compact and dust-obscured cores \parencite{Stierwalt2013,Falstad_2021}.

Local U/LIRGs ($\lesssim$ 250 Mpc) are ideal candidates to study relevant physical processes in radio-emitting sources. Their intense activity is typically confined to regions smaller than 1~kpc and often optically obscured by dust \parencite{Ramos_Almeida_2017,Hickox_2018}. Radio observations are unaffected by dust obscuration, and high-angular resolution interferometric observations of local sources can resolve structures down to physical scales of $\sim$ 1~pc. For a comprehensive analysis, multiple telescopes, interferometric arrays and frequencies can be combined \parencite[e.g.,][]{Muxlow2005,Muxlow2020,Lucatelli_2024} to quantify emission properties on continuous spatial scales, probing various morphological structures and emission mechanisms \parencite[e.g.,][]{Herrero_Illana_2017}.

In particular, radio emission in the $1.0$ to $33.0~\mathrm{GHz}$ range arises primarily from two physical processes: non-thermal synchrotron emission and thermal free-free emission \parencite{Murphy_2009,Condon_2016}. Both components trace star formation processes, though on different timescales and under different physical conditions. Synchrotron emission traces past star formation through supernova remnants whilst free-free emission traces current star formation from ionised gas around young massive stars \parencite[e.g.,][]{Osterbrock_2006,Murphy2012,Tabatabaei2017}. Multi-wavelength observations are essential to accurately distinguish between these components and enable proper physical interpretation of the observed emission. In some U/LIRGs, nuclear SF can contribute more than half of the total SF budget \parencite{Lucatelli_2024}, with surface densities reaching extreme values of $\sim 10^3~\Moyrk$ \parencite{loreto2015,loreto2017,Perrotta2021,Crocker2018,Lucatelli_2024}.

Radio emission is closely linked to infrared radiation through the infrared-radio correlation---a linear relation that connects these luminosities across a wide range of magnitudes \parencite{Bell_2003}. Deviations in the correlation indicate that either infrared luminosities are not proportional to star SF or radio luminosities have an excess due to bright AGN cores and/or radio jets \parencite{Yun_2001}. Therefore, understanding this connection requires decomposing the total emission into individual processes at different spatial scales and wavelengths \parencite[e.g.,][]{Murphy2012,Marvil_2015,Tabatabaei2017,Linden2019,Song2021,Song_2022}. This decomposition is crucial for establishing accurate SF rates and AGN fraction estimates and characterising the evolutionary phases of these systems.

Morphological image analysis and spectral energy distribution (SED) modelling reveal the structural content of radio sources and distinguish between emission mechanisms, particularly separating SF from AGN emission \parencite{Moric_2010}. These analyses can unravel the diverse physical properties of galaxies,
allowing the study of their properties as a function of the merger stage \parencite{Murphy_2013}, and characterise, for example, the evolution of the relative contribution of SF and AGN activity.

\rev{
Multi-frequency radio observations have been extensively used to characterise different emission mechanisms across multiple radio sources and spatial scales. \textcite{Murphy_2018} used 33~GHz VLA observations to study star-forming regions at $30$--$300$~pc resolution in sources at distances $\lesssim 30$~Mpc, finding predominantly thermal emission ($\gtrsim$~90\%) that provides extinction-free measures of SF. \textcite{Linden2019} employed multi-wavelength data to characterise SF in extranuclear regions of Star Forming Galaxies (SFGs) and U/LIRGs (located at $\lesssim 140$~Mpc), while \textcite{Song2021} examined nuclear star-forming rings (of SFGs and LIRGs at $\lesssim 70$~Mpc), finding significant variations in SF efficiency. \textcite{Song_2022} successfully differentiated between AGN and SF mechanisms in compact U/LIRGs (68 systems, at distances $\sim 17$--$400$~Mpc), highlighting the importance of proper emission decomposition.

In a statistical approach, \textcite{Vega_2008} analysed 30 U/LIRGs using SED modelling from near-infrared to radio wavelengths. Their work revealed that U/LIRGs typically contain compact, dust-enshrouded SBs with high extinction and approximately 60\% containing detectable AGN components, though SF remained dominant in most systems.

Several detailed case studies have provided valuable insights. \textcite{Murphy2011} investigated NGC~6946 using thermal free-free emission (26--40~GHz) to establish calibration relations for measuring SF. \textcite{Ramirez_Olivencia_2022} investigated Arp~299 from 150~MHz to 8.4~GHz, demonstrating the effectiveness of wide-frequency coverage to map emission on multiple spatial scales. More recently, \textcite{Chen_2024} conducted multi-component SED analysis of NGC~1365, showing how free-free and synchrotron fractions correlate with nuclear outflow activity. 

\textcite{Herrero-Illana_2014} conducted a comprehensive investigation of NGC~1614, revealing a circumnuclear ring of SF with no AGN activity, while \textcite{Herrero_Illana_2017} characterised 11 LIRGs, confirming that most LIRG luminosity derives from SF. \textcite{Tabatabaei2017} decomposed thermal and non-thermal radio emission in nearby galaxies, establishing improved SF calibrations with reduced uncertainty.

Few studies have combined high-resolution observations across multiple spatial scales. Exceptions include the multiscale work by \textcite{romero_canizales_2012}, which used \emph{e}-MERLIN and VLBI observations at multiple frequencies to unveil the nature of the nuclear region of IC\,883, pointing out a a mixed influence of a nuclear SB and an AGN, further charactering the AGN properties with multi-instrument and multi-frequency observations in \textcite{canizales2017}. In \textcite{Varenius2016}, the authors studied Arp~220 from 150~MHz to 33~GHz, mapping structures from $\lesssim$~100~pc to $\gtrsim$~2~kpc, and \textcite{loreto2015} conducted high-resolution measurements of the two nuclear disks of Arp\,220, identifying remarkably high SF surface densities approaching theoretical Eddington limits. \textcite{loreto2017} extended this analysis to 22 local major mergers, showing how energetically dominant regions become more compact and centrally concentrated as mergers progress. Recently, \textcite{Gajovic_2024} performed a spatially resolved spectral analysis of M~51 from 1.4--15~GHz, revealing significant spatial variations in spectral index and thermal fractions across the galaxy.

}

\rev{A} key challenge to study these systems is to distinguish between the potential confusion between the radio emission of a compact nuclear starburst (SB) and the AGN core \parencite{Haan2013a,Dullo2023,Morabito_2022,Herrero_Illana_2017}. This confusion can lead to misinterpretations of the physical processes at play, particularly in systems with extreme nuclear activity. Furthermore, diffuse emission from SF and AGN jets can exhibit similar radio properties, making them difficult to distinguish \parencite{Condon1992}. These complexities often make difficult to separate different emission mechanisms in radio sources \parencite[e.g.,][]{Moric_2010,Panessa2019,Wang2023,Baldi_2021}.

In this study we aim to characterise the radio emission in local U/LIRGs in terms of compact and extended features across observable physical scales, based on the multiscale analysis conducted in \textcite{Lucatelli_2024} (hereafter \citetalias{Lucatelli_2024}). Our primary objective is to distinguish emission components across different spatial scales. At large scales, we aim to separate diffuse emission from SF processes (diffuse SF, extranuclear \rev{rings}/disks, etc, on scales of $\gtrsim 1.0$~kpc) from extended AGN components such as jets. At nuclear scales ($\lesssim 200$~pc), we seek to distinguish compact starburst (SB) emission, associated with circumnuclear disks (CNDs) (which may contain also  supernovae and supernova remnants) from AGN core emission and compact AGN components such as mini jets (if visible). We introduce a multiscale tracer for the total extended emission. Through this approach, we seek to provide insights into the characterisation of radio sources, presenting some preliminary understanding of the physical mechanisms in systems with extreme nuclear activity while providing a framework for comprehensive physical investigations in future studies.

We follow a methodical approach to disentangle the complex multiscale radio emission properties in U/LIRGs. 
Our observational framework is presented in \cref{sec:paper_2_data_and_calibration}, where we describe our sample selection criteria and the radio observations. The analytical process focuses on the spatial decomposition of radio structures using image processing techniques (\cref{sec:paper_2_image_processing}). In \cref{sec:paper_2_results} we discuss the relationships between emission properties at different spatial scales, identifying connections between nuclear activity and large-scale diffuse emission. This systematic decomposition allows us to disentangle the relative contributions of AGN activity and emission related to SF processes across multiple spatial scales. Finally, in \cref{sec:paper_2_conclusion} we present the conclusions of our work. In \cref{app:paper_2_data_projects} we summarise the observational VLA code projects. In \cref{app:paper_2_image_processing} we discuss the self-calibration strategy which was critical to improve imaging quality. We also add some details on source characterisation. In \cref{app:paper_2_images_and_tabular_data} we provide all the high-angular resolution radio maps made with \emph{e}-MERLIN at $\sim$ 6.0~GHz, showcasing detailed structures and features of each source. For distance calculation, we adopt the $\Lambda$CMD model with the following constants: $\Omega_\Lambda$ = 0.692, $\Omega_{m_0}$ = 0.308 and $H_0$ = 67.8 km s$^{-1}$ Mpc$^{-1}$.

\section{Observational Data \emph{\&} Calibration}
\label{sec:paper_2_data_and_calibration}
In this work, we use multi-frequency and multiscale observations from VLA and \emph{e}-MERLIN to study the radio emission of local U/LIRGs. Details of the observational data and a summary of the sources studied are provided below. For reference, \cref{tab:data_vla} lists the project codes for all VLA data. The present analysis focuses specifically on a multiscale study at three key frequencies: 1.4, 6.0, and 33~GHz. Our primary goal is to comprehensively decompose the radio emission across various spatial scales at these frequencies, allowing us to distinguish between compact and extended emission components. 

\rev{
\subsection{Sample Selection}
}
A subset of 15 U/LIRG systems was selected from the 42 systems in the LIRGI Sample \parencite{Conway2008}\footnote{\url{https://www.e-merlin.ac.uk/legacy/proposals/e-MERLIN_Legacy_LIRGI.pdf}.}. This selection includes the four systems previously discussed in \citetalias{Lucatelli_2024}, along with 11 additional systems presented in \cref{tab:source_classes}. The distances span a range from $\sim 72$~Mpc (MCG+12-02-001) to $\sim 238.9$~Mpc (IRASF17132+5313), corresponding to linear scales of $\sim 34.6$~pc and $\sim 112.7$~pc per $0.1$", respectively. This selection aims to accommodate: i) systems at various merger stages; ii) the availability of data across different scales and frequencies, particularly \emph{e}-MERLIN-L observations; iii) diverse morphologies with respect to their nuclear activity; and iv) a focus on some U/LIRGs that have not yet been extensively studied. Although the number of systems is limited, they offer valuable insights and motivations for an in-depth investigation of U/LIRGs.

\begin{table}
	\centering
	\caption{Basic source information of our sub-sample.}
	\label{tab:source_classes}\label{tab:source_info}
	\begin{subtable}[h]{1.0\linewidth}
		\centering
        \resizebox{\textwidth}{!}{%
			\begin{tabular}{l|cccccc}
	\hline
	Source Name          & RM                      & SC                        & MS         & $\log L_{\mathrm{IR}}$&  $D_{L}$       \\[0.3ex]        \rule{0pt}{1.0em}
	&                         &                           &            & $\mathrm{[L_{\odot}]}$&  [Mpc]         \\[0.3ex] \hline \rule{0pt}{1.0em}
	UGC\,5101            &  AGN/SB                 & L, Sy1.5                  & 4          & 11.95                 & 182.3          \\[0.3ex] \hline \rule{0pt}{1.0em}
	MCG+12-02-001        & ---                     & $\HII$                    & 3          & 11.44                 & 72.0           \\[0.3ex]        \rule{0pt}{1.0em}
	(S-CE)               & UN (AGN/SB?)            &  						   & ---        &                       &                \\[0.3ex]        \rule{0pt}{1.0em}
	(S-WE)               & SB                      &						   & ---        &                       &                \\[0.3ex] \hline \rule{0pt}{1.0em}
	Mrk\,331             & AGN/SB                  & $\HII$, Sy2               & 1          & 11.41                 & 81.8           \\[0.3ex] \hline \rule{0pt}{1.0em}
	NGC\,5256            & --                      & Sy2                       & 2          & 11.49                 & 133.1          \\[0.3ex]        \rule{0pt}{1.0em}
	(CE)                 & SB                      & ---                       & ---        &                       &                \\[0.3ex]        \rule{0pt}{1.0em}
	(NE)                 & AGN/SB                  & L                         & ---        &                       &                \\[0.3ex]        \rule{0pt}{1.0em}
	(SWE)                & AGN/SB                  & Sy2                       & ---        &                       &                \\[0.3ex] \hline \rule{0pt}{1.0em}
	UGC\,8696            & ---			           & L, Sy2                    & 4      	& 12.14                 & 178.0          \\[0.3ex] 	   \rule{0pt}{1.0em}
	(NWE)                & AGN/SB                  &	  	                   & ---      	&                       &                \\[0.3ex] 	   \rule{0pt}{1.0em}
	(SE)                 & AGN/SB                  &			               & ---      	&                       &                \\[0.3ex] \hline \rule{0pt}{1.0em}
	NGC\,7674            & AGN                     & $\HII$, Sy2               & 1  		& 11.50                 & 129.1          \\[0.3ex] \hline \rule{0pt}{1.0em}
	UGC\,04881           & ---		               & L, $\HII$                 & 3   	    & 11.69                 & 182.2          \\[0.3ex]        \rule{0pt}{1.0em}
	(NE)                 & AGN/SB                  & L   		               & ---    	&                       &                \\[0.3ex]        \rule{0pt}{1.0em}
	(SWE)                & SB                      & $\HII$                    & --- 		&                       &                \\[0.3ex] \hline \rule{0pt}{1.0em}
	IRAS\,23436+5257     & ---                     & ?                         & 3			& 11.51                 & 154.0          \\[0.3ex]        \rule{0pt}{1.0em}
	(N)                  & UN (SB?)                &                           & ---		&                       &                \\[0.3ex]        \rule{0pt}{1.0em}
	(SE)                 & SB                      &                           & ---		&                       &                \\[0.3ex] \hline \rule{0pt}{1.0em}
	NGC\,6670            & ---                     & ($\HII$ ?)                & 2			& 11.60                 & 134.7          \\[0.3ex]        \rule{0pt}{1.0em}
	(E)                  & UN (AGN/SB?)            &                           & ---		&                       &                \\[0.3ex]        \rule{0pt}{1.0em}
	(WE)                 & UN (SB?)                &                           & ---		&                       &                \\[0.3ex] \hline \rule{0pt}{1.0em}
	VV\,250              & ---                     & $\HII$                    & 2			& 11.74                 & 146.2          \\[0.3ex] 	      \rule{0pt}{1.0em}
	(E)                  & SB                      &                           & ---		&                       &                \\[0.3ex] 	      \rule{0pt}{1.0em}
	(WE)                 & UN (SB?)                &                           & ---		&                       &                \\[0.3ex] \hline \rule{0pt}{1.0em}
	VV\,705              & ---                     & $\HII$                    & 2			& 11.89                 & 190.6          \\[0.3ex] 	      \rule{0pt}{1.0em}
	(N)                  & SB                      &                           & ---	    &                       &                \\[0.3ex] 	      \rule{0pt}{1.0em}
	(S)                  & AGN/SB                  &                           & ---        &                       &                \\[0.3ex] \hline \rule{0pt}{1.0em}
	IIIZw035             & AGN/SB                  & $\HII$, L                 & 1	        & 11.56                 & 120.3          \\[0.3ex] \hline \rule{0pt}{1.0em}
	IRAS\,F17132+5313    & ---                     & $\HII$                    & 2	        & 11.89                 & 238.9          \\[0.3ex] 	      \rule{0pt}{1.0em}
	(NE)                 & SB                      & $\HII$                    & ---	    &                       &                \\[0.3ex] 	      \rule{0pt}{1.0em}
	(SWE)                & AGN/SB                  & ?                         & ---	    &                       &                \\[0.3ex] \hline \rule{0pt}{1.0em}
	IRAS\,20351+2521     & UN (AGN/SB?)            & ?                         & 0 		    & 11.54                 & 155.3          \\[0.3ex] \hline \rule{0pt}{1.0em}
	IIZw096              & ---                     & $\HII$                    & 3			& 11.87                 & 165.9          \\[0.3ex]        \rule{0pt}{1.0em}
	(NE)                 & AGN/SB                  &                           & ---	    &                       &                \\[0.3ex] 	      \rule{0pt}{1.0em}
	(SWE)                & SB                      &                           & ---        &                       &                \\[0.3ex] \hline 
\end{tabular}
}
\caption*{\emph{Notes}: 
Column 1 -- Source name;
Column 2 -- Radio morphology (RM): AGN: active galactic nucleus; SB: starburst; AGN/SB: composite; UN: ``unknown'', without clear information from the literature.
Column 3 -- Spectral classes (SC): L - LINER; Sy1.5 - Seyfert 1.5; Sy2 - Seyfert 2 - $\HII$ - source with $\HII$ characteristics; UN - ``unknown'', without information from the literature. For more information, see \textcite{Ramos_Padilla_2023,Yamada_2013,Gonzalez_Martin_2015,Gonzalez_Martin_2017,Dietrich2018} and references therein.
Column 4 -- Merger stage (MS), taken from \textcite{Haan2013a}: 0 -- isolated; 1 -- early stage; 2 -- mid stage; 3 -- late stage; 4 -- post-merger;
Column 5 -- Infrared luminosity ($L_{\mathrm{IR}}$) in units of solar luminosities ($\mathrm{L_{\odot}}$);
Column 6 -- Luminosity distance ($D_{\mathrm{L}}$) in Mpc.
For systems with galaxy pairs we have assumed the same luminosity distance, calculated via the redshift $z$ taken from NED, by using the interface provided by \texttt{astropy.astroquery.ipac.ned} and assuming the cosmology stated in \cref{sec:paper_2_introduction}.
}
	\end{subtable}
\end{table}

\subsection{VLA Observations}
\rev{
We present VLA observations at 1.4, 6.0, and 33~GHz, comprising A-configuration (VLA-A) EVLA data (project code 23A-324) and are complemented by archival EVLA data. Our multiscale decomposition requires comparable angular resolutions at each spatial scale across the frequency range. At low angular resolution ($\sim 0.3''$--$1.0''$), we match high-frequency (33~GHz) observations from VLA-B/C configurations with lower-frequency ($1.4$ and $6.0$~GHz) VLA-A observations. At high angular resolution ($\sim 0.05''$--$0.2''$), we complement the \emph{e}-MERLIN observations at $1.4$ and $6.0$~GHz (\cref{sec:paper_2_eM_observations}) with VLA-A observations at $33$~GHz, which provide comparable spatial scales at the higher frequency. This approach allows us to analyse emission structures at consistent spatial scales across all three frequencies, at both resolution regimes, which is essential for our multiscale decomposition.

We used standard calibration strategies provided by the EVLA CASA pipeline \parencite{TheCASATeam_2022,CASAMcMullin2007}\footnote{See \href{https://science.nrao.edu/facilities/vla/data-processing/pipeline}{https://science.nrao.edu/facilities/vla/data-processing/pipeline}.}. For observations in which the EVLA CASA pipeline failed, primarily due to incompatibilities with early EVLA data formats from observations during and shortly after the VLA upgrade (between 2010 and 2012; project code AL746 in \cref{tab:data_vla}), we used our custom pipeline, \textsc{Synphly}\footnote{Documentation and examples available at \url{https://github.com/JBCA-Interferometry/Synphly/tree/gain_calibration}.}, which implements similar calibration strategies.
}

\subsection{\emph{e}-MERLIN Observations}
\label{sec:paper_2_eM_observations}
We present high-angular resolution, sensitive \emph{e}-MERLIN observations at $1.4$~GHz (L band) and $6.0$~GHz (C band). These data are part of the \emph{e}-MERLIN Legacy Project, a comprehensive campaign to acquire high-resolution and high-sensitivity data on U/LIRGs within the local Universe \parencite[PI's][]{Conway2008}. The observations achieve angular resolutions of $\sim 0.2''$ and $\sim 0.05''$, respectively, with average sensitivities of $\sim10-20 ~ \mu \text{Jybeam}^{-1}$ for an integration time of $\sim 10$ hours. These high-resolution observations below $10$~GHz are crucial for direct comparison with VLA-A detections at higher frequencies, allowing spectral analysis of emission at physical scales $\lesssim 250$~pc (Lucatelli et.al., in prep.). 

\rev{
To calibrate \emph{e}-MERLIN observations, we used the \emph{e}-MERLIN CASA Pipeline \parencite{eMCP2021} (v1.1.19) with CASA v.5.8.0, following the standard calibration strategy. The only additional step was applied to L-band observations, which required prior auto-flagging of raw data using \textsc{AOFlagger} (v.2.15) \parencite{AOFlagger_2010} due to significant radio frequency interference (RFI) before starting calibration within the pipeline. 
}

In \cref{fig:C_eM_native_1} we display all native high-angular resolution maps made with \emph{e}-MERLIN images at C-band. 
We note that in this study, we do not include \emph{e}-MERLIN L-band observations for UGC\,04881, VV\,250, and IRAS\,F17132+5313. 

\section{Image Processing}
\label{sec:paper_2_image_processing}
Based on our methodology in \citetalias{Lucatelli_2024}, we spatially decompose the radio emission of our sources at frequencies of $1.4$, $6.0$, and $33.0$~GHz. We analyse compact and diffuse structures at representative scales below and greater than $\sim~0.3''$, distinguishing between nuclear emission of compact cores (potentially from AGN activity), nuclear extended emission (nuclear SB characterised by CNDs) and diffuse emission at larger scales (likely from SF). This spatial decomposition allows us to investigate emission fractions in different regions, quantify the contributions of distinct physical processes, and examine their relationships across physical scales. 

\subsection{Multiscale and Multi-frequency Imaging Data}
\label{sec:paper_2_sed_strategy}
\label{sec:paper_2_high_res_data}
We have implemented an integrated approach to analyse radio spectra across various spatial scales, combining high- and low-resolution imaging on a component-by-component basis. Our methodology includes:
\begin{enumerate}
	\item \textbf{High-Resolution Imaging}: We used \emph{e}-MERLIN at 1.4 and 6~GHz, alongside VLA-A at 26-33~GHz (achieving 30-100~mas resolution) to characterise emission mechanisms within nuclear regions. These images enable the separation of compact cores and nuclear extended components (see \cref{sec:paper_2_image_decomposition}).
	\item \textbf{Low-Resolution Imaging}: We use VLA-A data at 1.4 and 6.0~GHz and VLA-B/C at 33~GHz to capture both nuclear regions and diffuse structures at scales $\gtrsim 250$~pc. 
	\item \textbf{Multiscale Decomposition}: With these data we decompose the radio emission at a multiscale level. In particular, compact-cores from nuclear extended components and off-nuclear emission with low-angular resolution imaging. 
	We are able to conduct a comprehensive analysis of the emission throughout the sources, in particular to characterise the total multiscale extended emission.
	The decomposition quantifies the fractions of diffuse versus compact emission across spatial scales and frequencies (see \cref{sec:multiscale_tracers,fig:radio_decomposition}).
\end{enumerate}

We use \emph{e}-MERLIN (1.4 and 6 GHz) with VLA-A (33 GHz) to achieve consistent overlap in the $uv$-plane along $\lambda$. The overlap is particularly optimal between \emph{e}-MERLIN at $\sim 6.0$~GHz and VLA-A at $33.0$~GHz. We use VLA observations, in A and B/C configurations, at reference frequencies of $1.4$, $6.0$ and $33.0$~GHz, to probe the diffuse emission. The overlap of the $uv$ coverage between VLA-A ($1.4$ and $6.0$~GHz) and VLA-C ($33.0$~GHz) provides similar angular resolutions and allows us to study both nuclear regions and diffuse emission. This data allows us to understand the morphological properties on larger scales, which can be compared with the properties of high-resolution data to determine whether they are linked among distinct spatial scales.

\subsection{Imaging of Interferometric Visibilities}
\label{sec:paper_2_imaging_visibilities}
For each source at a given frequency, we prepare two distinct visibilities for further processing. One is the native self-calibrated visibilities (see \cref{sec:paper_2_selfcalibration_strategy}), used to produce the final full resolution and sensitive radio maps (having the native $uv$ coverage). The other set of visibilities is a sub-product of the previous case, but computing the $uv$ coverage that is common between all multi-frequency visibilities for a given source at a specific angular resolution. These visibilities are going to be used to produce $uv$ and beam-matched multi-frequency images. To guarantee that these images have a similar restoring beam, a homogeneous circular beam is specified during the deconvolution of the visibilities. We use \textsc{WSClean} as the imager (v.~3.4) \parencite{offringa-wsclean-2014,offringa-wsclean-2017} to generate multi-frequency images from our observations. With \textsc{WSClean}, we run a Multi-Frequency Synthesis (MFS) deconvolution with three sub-bands, using the argument \code{-channels-out 3 -join-channels}. This means that the full bandwidth in each observing band is split into three parts.\footnote{The exact frequencies of each image depends on the data characteristics, but the default splits are evenly divided regarding the number of channels, and not spectral windows.} We keep this consistency between \emph{e}-MERLIN and VLA, unless said otherwise. In addition, one combined image for the full bandwidth is created, named MFS image.\footnote{Centred at the central frequency of the hole data.} 

\begin{figure*}
 	\centering
 	\includegraphics[width=1.0\linewidth]{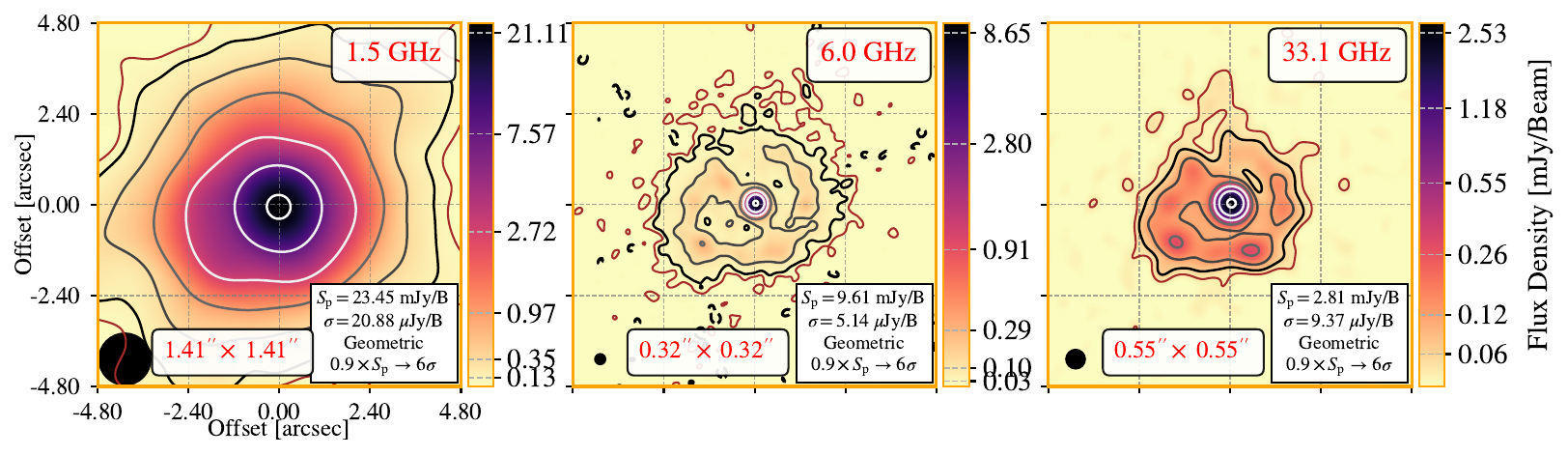}
 	\includegraphics[width=1.0\linewidth]{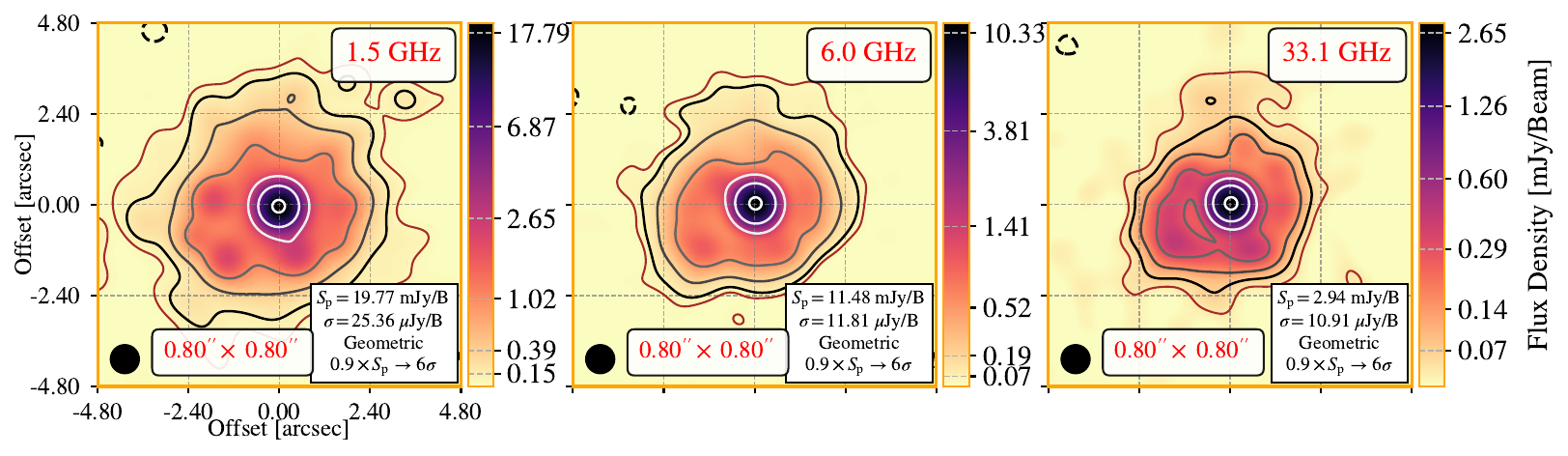}
 	\caption[Multi-frequency MFS radio maps of Mrk\,331, at $1.5$, $6.0$ and $33.0$~GHz at scales $\gtrsim 0.3''$.]{Multi-frequency MFS radio maps of Mrk\,331, at $1.5$, $6.0$ and $33.0$~GHz on scales $\gtrsim 0.3''$. The top row shows the native images, while the bottom row shows the $uv$ beam matched images with a common restoring beam of $0.8''$. 
 		\emph{Notes}: 
 		In the scale of these maps, $2.0''$ corresponds to $760~\mathrm{pc}$.
 		We show the contour levels in the radio maps and on the colour bar at the right side of each panel. We auto-generate these levels in log-space, using a geometric sequence with 6 contours, given by 
 		$c_i = 0.9 S_{\mathrm{p}}[({6\sigma})/({0.9S_{\mathrm{p}}})]^{({i-1})/({N-1})}$, with $i=1,2,3,4,5,6$, $N=6$, and $S_{\mathrm{p}}$ the peak intensity.
 		We add an extra contour in brown, marking the $3 \times \sigma$ boundary (not shown in the colour bar). Each panel displays: 
 		- the restoring beam size (lower-left); 
 		- the peak intensity $S_{\mathrm{p}}$ and the \emph{global} RMS noise level $\sigma_{\mathrm{rms}}$ (calculated in the residual image) (lower-right); 
 		- the central frequency of the image (upper-right).
 	}
 	\label{fig:Mrk331_multifreq_lowres}
 \end{figure*}

\begin{figure*}
 	\centering
 	\includegraphics[width=1.0\linewidth]{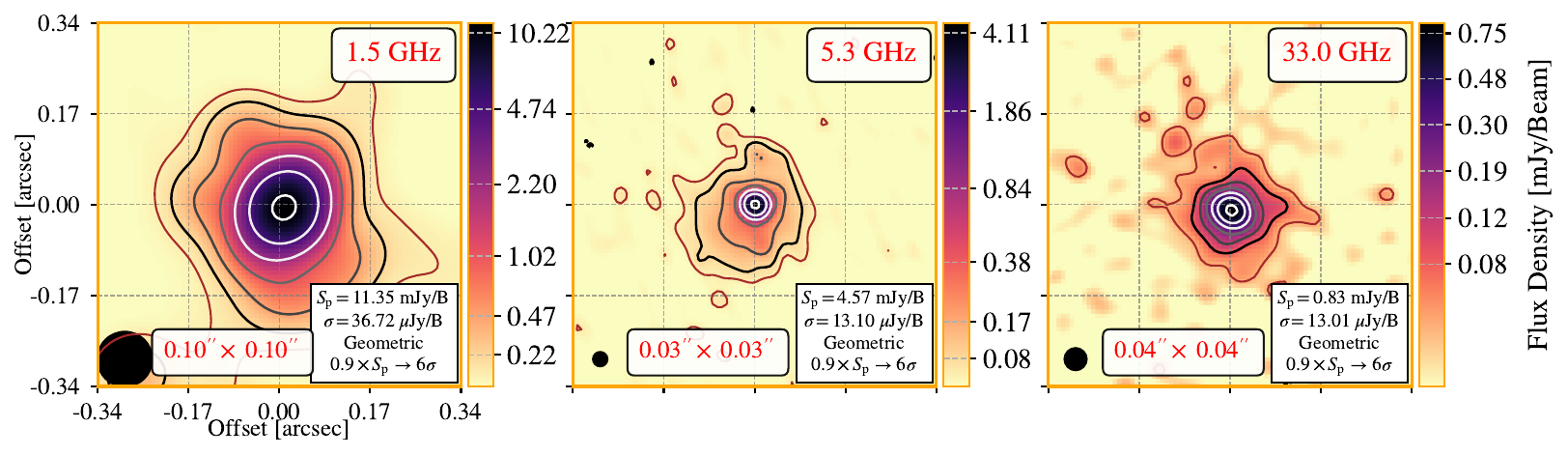}
 	\includegraphics[width=1.0\linewidth]{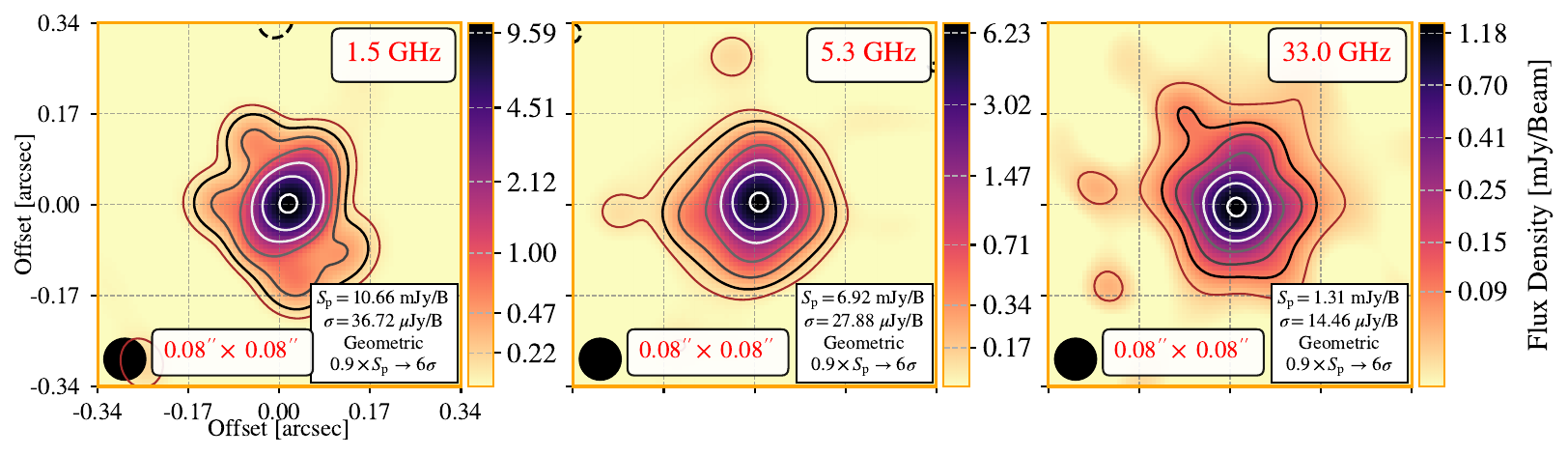}
 	\caption[Multi-frequency MFS radio maps of Mrk\,331, at 1.4, 6.0 and 33.0~GHz at scales $\lesssim 0.3''$.]{Multi-frequency MFS radio maps of Mrk\,331, at 1.4, 6.0 and 33.0~GHz at scales $\lesssim0.3''$. The top row shows the native images, while the bottom row shows the $uv$ beam matched images with a common restoring beam of $0.08''$.
 		\emph{Notes}: Same as \cref{fig:Mrk331_multifreq_lowres} (similarly, a scale of $0.2''$ corresponding to $76~\mathrm{pc}$).
 	}
 	\label{fig:Mrk331_multifreq_highres}
 \end{figure*}

Although we are using tree observing bands in this work, the $uv$ match was performed using all observing bands at the two respective angular scales with \emph{e}-MERLIN and VLA. In order to be consistent in producing the images, we use a common pixel scale at each angular scale. For example, in the low-angular resolution case (see \cref{tab:data_vla}), we produce all images with a pixel scale of $0.04''$. This is to accommodate the maximum resolution of the VLA-A X band. In the high-angular resolution case, we produce all images with a pixel scale of $0.008''$ to be suitable with \emph{e}-MERLIN-C. This allows us to easily process/analyse and compare the images directly, regardless of the frequency. We show in \cref{fig:Mrk331_multifreq_lowres} a set of images for Mrk\,331 at larger scales. The top row highlights native images (full angular resolution and sensitivity), while the bottom row shows the $uv$ and beam-matched images (with a restoring beam of $0.8"$). Similarly, in \cref{fig:Mrk331_multifreq_highres} we show the native and matched images for the nuclear region of Mrk\,331 using high-angular resolution data, probing structures on scales $\lesssim 0.3"$. We used a restoring beam with a size of $0.08''$ to match the images.

We performed self-calibration (phases and amplitudes) on all visibilities to significantly improve image quality and flux density measurements. We provide a brief summary of our strategy in \cref{sec:paper_2_selfcalibration_strategy}. In addition, for all systems composed of two nuclei with significant angular separation ($\gtrsim 20"$), {we applied primary beam correction because the flux density response varies across the field of view. This correction was necessary since the phase tracking centre was pointed at one of the two nuclei (or at a position between them), causing sources offset from the pointing centre to experience reduced sensitivity. For example, VV\,250 is constituted of two nuclei (E and WE), separated by $\sim 33"$. The observation at $33.0$~GHz with VLA-C pointed to the E galaxy, and the beam response at the position of VV\,250-WE is $\sim 50\%$ relative to the phase tracking centre.} However, for {the} homogeneity of data calibration and processing, we used the primary beam correction option in \textsc{WSClean} as a default. This was important during self-calibration in case outlier sources were used to improve gain solutions.

\subsection{Image Detection, Characterisation \emph{\&} Decomposition}
\label{sec:paper_2_image_decomposition}
Using continuum images generated by \textsc{WSClean}, we implement the image decomposition approach developed in \citetalias{Lucatelli_2024} to characterise compact/unresolved cores and extended structures, both at nuclear regions ($\lesssim 250 $~pc) and larger scales ($\gtrsim 0.3$~kpc). This method quantifies the frequency-dependent variation of flux density in different emitting regions. 
Complementary to the analysis presented here, in \cref{app:paper_2_morphometry} we provide some metrics to characterise the structure of our sources. The size calculations are based on \citetalias{Lucatelli_2024}, and for reference, \cref{fig:Mrk331_C_eM_CE_Lgrow_levels} demonstrates an updated example of the procedure compared to \citetalias{Lucatelli_2024}.

\subsubsection{Source Extraction}
Source extraction across multiple frequency bands presents unique challenges, as emission structures can vary significantly between images. The detection of components is complicated by both the intrinsic physics of the source and instrumental limitations (sensitivity and angular resolution), resulting in features that may be detectable in some images whilst falling below the detection threshold in others. This variation makes it particularly challenging to maintain a consistent component identification and labelling across the frequency range, especially for sources with complex morphologies. To optimise our source extraction and analysis, we selected images that balance angular resolution and sensitivity. In high-angular resolution images, we prioritise the use of \emph{e}-MERLIN observations at C-band to perform source extraction and deblending. In the low-angular resolution data, we use VLA-A C-band images. In either case, the MFS images (from \textsc{WSClean}) are used since they contain the highest signal-to-noise ratio in each band. As an example, in \cref{fig:example_source_multilabels_uvmatch} we highlight each nuclear component of {\color{black} NGC\,7674}, and in \cref{fig:source_detection_example} we show the regions obtained via source-extraction.

\subsubsection{Image Decomposition}
The detected regions guide our multi-frequency analysis by constraining the spatial distribution and properties of structures across all images. During image decomposition, we initialise each model component's parameters---including position, peak brightness, and size---based on these source-extracted regions (see \citetalias{Lucatelli_2024} for detailed methodology). As an example, consider the case of NGC\,7674, a classic Seyfert 2 AGN. In \cref{fig:example_source_multilabels_uvmatch} we show its radio map at $\sim 6.0$~GHz with \emph{e}-MERLIN. In \cref{fig:source_detection_example} we show the detection map from the same image. The identified regions of NGC\,7674 are labelled with an identification number (\texttt{ID}). We use these regions to specify the components to be modelled. From their prior properties (as mentioned previously), we employ the fitting to decompose the emission into individual components. We show in \cref{fig:grid_of_data_models} an example with the results from such procedure, highlighting the imaging data, models, and residuals, at distinct frequencies. We note that the model images provide a good description of the data, and thus we can access the flux densities and sizes for each identified region. 

The possible AGN (compact-core) component of this source is labelled as C1/AGN \texttt{ID2}). It is relatively faint ($\sim$ 1 mJy at 6 GHz) compared to the total source flux density. Our method allows us to accurately separate the flux density of this component in relation to the global emission. Although we do not discuss the properties of spectral indices in this paper, we report that C1/AGN has a flat integrated spectral index of $\alpha = -0.15\pm 0.03$ (in the frequency range of $6$--$33$~GHz). In \cref{tab:highres_native_decomposition} we present a preview of basic image properties {resulting} from the decomposition of native high-angular resolution images at 6.0~GHz with \emph{e}-MERLIN. Similarly, in \cref{tab:lowres_native_decomposition} we show the results of the low-angular resolution decomposition. For both cases, the complete tabular data are available online.

\section{Results \emph{\&} Discussion}\label{sec:paper_2_results}\label{sec:paper_2_discussion}
We now provide a systematic analysis of our sources, focusing on the discussion of basic properties, including flux distributions, radio luminosities, emission sizes, compactness, and surface densities. We then make comparisons with infrared luminosities, stellar masses, and merger stages to probe relationships between nuclear radio emission and galaxy-wide properties. We identify preliminary patterns and build an overview of the properties that needs further investigation to understand key interlinked and multiscale physical processes in U/LIRGs. We can extend this multiscale/wavelength investigation to a larger number of galaxies in upcoming surveys. 

\rev{
Throughout this section, we report the Kendall rank correlation coefficient $\tau_{\mathrm{K}}$ \parencite{Kendall_1938} on top of each diagram to quantify the statistical correspondence between the measured quantities. This non-parametric statistic is preferred for our sample size ($N \lesssim 30$), as it provides more robust estimates with smaller variance for small data sets \parencite[e.g.,][]{Croux_Dehon_2010}. The associated $p$-value indicates the probability of obtaining the observed correlation (or stronger) under the null hypothesis of no association between the two variables; values of $p < 0.05$ are considered statistically significant.
}

\subsection{Multiscale Tracers for the Radio Emission}\label{sec:multiscale_tracers}
The work by \textcite{Smith_1998} compares VLBI-scale sizes and integrated flux densities for a few sources that are presented in our current work. They find that there is no clear connection between the presence and/or strength of VLBI-size structures at the nuclear regions and other observed quantities at larger scales. They suggested that this happens because the processes at the VLBI and VLA scales are physically connected. We address some of these aspects in the discussion that follows.
\begin{figure}
	\centering
	\begin{tikzpicture}[scale=0.9]
		\node[draw, thick, ellipse, minimum width=5.0cm, minimum height=3.2cm, 
		fill=gray!20] (total) at (0,0) {};
		
		\node[draw, thick, circle, minimum size=1.2cm, 
		fill=gray!40] (nuclear) at (-0.6,0) {};
		
		\node[black, font=\large\bfseries] at (-2.2,1.5) {$S_{\nu}^{\mathrm{T}}$};
		\node[black, font=\large\bfseries] at (1.0,-0.9) {$S_{\nu}^{\mathrm{D}}$};
		\node[black, font=\large\bfseries] at (-0.6,0) {$S_{\nu}^{\mathrm{N}}$};
		
		\draw[line width=3pt, black] (-1.4,-2.0) -- (-0.2,-2.0);
		\node[below, black, font=\large] at (-0.8,-2.0) {1.0 kpc};
		
		\draw[-{Stealth[length=6pt]}, thick] (0.09,0) -- (3.45,0);
		
		\begin{scope}[shift={(5.0,0)}]
			\node[draw, thick, ellipse, minimum width=2.8cm, minimum height=1.8cm,
			fill=gray!10] (nuc_ext) at (0,0) {};
			
			\node[draw, thick, circle, minimum size=0.8cm,
			fill=gray!60] (comp_core) at (0,0) {};
			
			\node[black, font=\large\bfseries] at (-0.7,0.45) {$S_{\nu}^{\mathrm{ne}}$};
			\node[black, font=\large\bfseries] at (0,0) {$S_{\nu}^{\mathrm{cc}}$};
			
			\draw[line width=3pt, black] (-0.5,-1.4) -- (0.9,-1.4);
			\node[below, black, font=\large] at (0.2,-1.4) {100.0 pc};
			
			\node[above, black, font=\large\bfseries] at (0,1.3) {Nuclear Region};
			\node[above, black, font=\footnotesize] at (0,1.0) {\emph{e}-MERLIN};
		\end{scope}
	\end{tikzpicture}
	\caption[Scheme (not to scale) to represent the multiscale decomposition of radio emission in U/LIRGs.]{Scheme (not to scale) to represent the multiscale decomposition of radio emission in U/LIRGs, showing the hierarchical structure from large-scale diffuse emission $S_{\nu}^{\mathrm{D}}$ to nuclear components $S_{\nu}^{\mathrm{N}}$.
    The idealisation of a nuclear region contains both compact core $S_{\nu}^{\mathrm{cc}}$ and nuclear extended emission $S_{\nu}^{\mathrm{ne}}$. Note that, in principle, $S_{\nu}^{\mathrm{N}}$ can be recovered by both low- and high-angular resolution imaging, since $S_{\nu}^{\mathrm{N}} \simeq S_{\nu}^{\mathrm{cc}} + S_{\nu}^{\mathrm{ne}}$. Note that we intepret the total emission attributed to SF as $S_{\nu}^{\mathrm{SF}} \approx S_{\nu}^{\mathrm{D}} + S_{\nu}^{\mathrm{ne}} \approx S_{\nu}^{\mathrm{T}} - S_{\nu}^{\mathrm{cc}}$.
    This highlights the different physical processes across spatial scales of $\sim$10--250\,pc (nuclear) to $\sim$0.5--3.0\,kpc (total, diffuse).}
	\label{fig:radio_decomposition}
\end{figure}
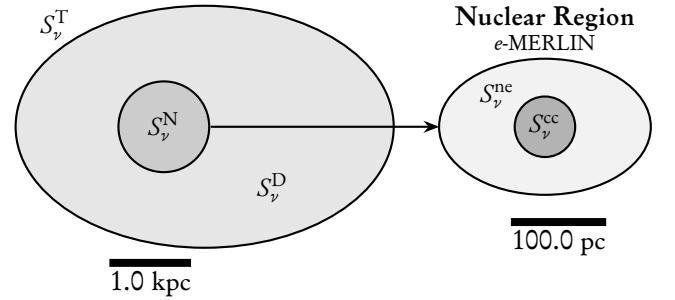

To understand the role of nuclear radio emission, we investigate the relative flux contributions from distinct spatial scales. In the following, we define some quantities that characterise the radio emission at distinct scales. To guide these definitions, we refer the reader to \cref{fig:radio_decomposition}. We label the total radio emission on VLA scales as $S_{\nu}^{\mathrm{T}}$, {which represents the integrated flux density of all emission detected by the VLA. The nuclear region is denoted by $S_{\nu}^{\mathrm{N}}$ whilst the diffuse structures on VLA scales is $S_{\nu}^{\mathrm{D}}$.} When we zoom into the nuclear region, we can resolve the compact core $S_{\nu}^{\mathrm{cc}}$ and the nuclear extended components $S_{\nu}^{\mathrm{ne}}$. The diffuse component $S_{\nu}^{\mathrm{D}}$ is simply evaluated by
\begin{align}
	S_{\nu}^{\mathrm{D}}  \approx S_{\nu}^{\mathrm{T}} - S_{\nu}^{\mathrm{N}}.
\end{align}
The nuclear extended emission is evaluated via
\begin{align}
	S_{\nu}^{\mathrm{ne}}  \approx S_{\nu}^{\mathrm{N}} - S_{\nu}^{\mathrm{cc}}.
\end{align}
where $S_{\nu}^{\mathrm{N}}$ is measured from high-angular resolution images.
We define the multiscale tracer for the total extended emission of a galaxy as
\begin{align}
	S_{\nu}^{\mathrm{SF}} \approx S_{\nu}^{\mathrm{T}} - S_{\nu}^{\mathrm{cc}},
\end{align}
where $S_{\nu}^{\mathrm{SF}}$ is the integrated flux density that represents the total extended emission related to SF (the sum $S_{\nu}^{\mathrm{ne}}+S_{\nu}^{\mathrm{D}}$). We note, however, that this nuclear extended emission is a representation of processes that are not associated with an AGN. Thus, it can include compact components related to supernova remnants and core-collapse supernovae. In this work, we do not attempt to disentangle these components, but we acknowledge that they can contribute to the nuclear extended emission. Thus, components that are off-centre of the main core {region are} still {included} in the nuclear extended emission, though {we do not draw} conclusions {about} their nature. Sources that contain such characteristics are III\,Zw\,35, UGC\,04881, II\,Zw096 and IRAS\,F17132+5313, see \cref{fig:C_eM_native_1}.

To analyse the relative contributions of different emission components, we introduce two groups of dimensionless parameters, following a systematic notation. We use the letter $\xi$ to denote emission fractions where only SF-related quantities (no AGN contribution) appear in the numerator, while the letter $\Upsilon$ represents fractions where AGN-associated quantities are included in the numerator. Specifically, we define the total extended, compact core, and nuclear emission fractions relative to the total radio emission as
\begin{align}
    \label{eq:xi_SF_u_cc_u_N}
	\xi^{\mathrm{SF}}_{\mathrm{T}} \equiv \dfrac{S_{\nu}^{\mathrm{SF}}}{S_{\nu}^{\mathrm{T}}}, 
	\qquad 
	\Upsilon^{\mathrm{cc}}_{\mathrm{T}} \equiv \dfrac{S_{\nu}^{\mathrm{cc}}}{S_{\nu}^{\mathrm{T}}},
	\qquad 
	\Upsilon^{\mathrm{N}}_{\mathrm{T}} \equiv \dfrac{S_{\nu}^{\mathrm{N}}}{S_{\nu}^{\mathrm{T}}}.
\end{align}
These ratios provide quantitative measures of how radio emission is distributed across different spatial scales, providing insight into the dominant physical processes operating at each scale. We also define the fractions of nuclear extended emission relative to the total nuclear emission and over the total radio emission, respectively.
\begin{align}
	\xi^{\mathrm{ne}}_{\mathrm{N}} \equiv \dfrac{S_{\nu}^{\mathrm{ne}}}{S_{\nu}^{\mathrm{N}}}, \qquad 
	\xi^{\mathrm{ne}}_{\mathrm{T}} \equiv \dfrac{S_{\nu}^{\mathrm{ne}}}{S_{\nu}^{\mathrm{T}}}.
\end{align}
In \cref{tab:fractional_emission} we provide the median values of some of the previous quantities, across all sources, for the three frequencies of study. We generate these metrics by linking each low-resolution emission from \cref{tab:lowres_native_decomposition} with the corresponding high-resolution measurements in \cref{tab:highres_native_decomposition} where \emph{e}-MERLIN detections are available. In these cases, we do this for each pair $S_{\nu}^{\mathrm{N}} \leftrightarrow S_{\nu}^{\mathrm{T}}$, $S_{\nu}^{\mathrm{cc}} \leftrightarrow S_{\nu}^{\mathrm{T}}$ and $S_{\nu}^{\mathrm{ne}} \leftrightarrow S_{\nu}^{\mathrm{T}}$.

\rev{
In later sections/analysis, we also discuss the sizes of these components at a multiscale level (as shown in \cref{fig:radio_decomposition}). For instance, we define the total source size as $R_{95}^{\mathrm{T}}$ measured in low-angular resolution maps,  and $R_{95}^{\mathrm{N}}$ as the total size for the nuclear region, measured in high-angular resolution maps.
}
\begin{table}
	\centering
	\caption[Median values of fractional multiscale emission tracers.]{Median values of fractional multiscale emission tracers, and physical properties of our sources. We distinguish the measurements at 1.4, 6.0 and 33.0~GHz. For the physical sizes, we only show values at 6.0~GHz.}
	\label{tab:fractional_emission}
	\begin{subtable}[h]{0.99\linewidth}
		\centering
        \scalebox{0.9}{%
		\begin{tabular}{l|cc|cc|cc}
			\hline
			& \multicolumn{2}{c|}{1.4~GHz} & \multicolumn{2}{c|}{6.0~GHz} & \multicolumn{2}{c}{33.0~GHz} \\
			Quantity 											 & Med. & $2\sigma$ & Med. & $2\sigma$ & Med. & $2\sigma$ \\[0.5ex] 
			\hline 
			$\langle \Upsilon_{\mathrm{T}}^{\mathrm{N}}\rangle$         & $0.38$ & $0.14$--$0.60$ & $0.56$ & $0.33$--$0.63$ & $0.52$  & $0.38$--$0.70$   \\[0.5ex]
			$\langle \Upsilon_{\mathrm{T}}^{\mathrm{cc}}\rangle$        & $0.08$ & $0.03$--$0.21$ & $0.11$ & $0.05$--$0.20$ & $0.15$  & $0.08$--$0.25$   \\[0.5ex]
			$\langle \Upsilon_{\mathrm{D}}^{\mathrm{N}}\rangle$         & $0.52$ & $0.18$--$1.42$ & $0.80$ & $0.56$--$1.52$ & $0.87$  & $0.53$--$1.83$   \\[0.5ex]
			$\langle \Upsilon_{\mathrm{SF}}^{\mathrm{cc}}\rangle$       & $0.09$ & $0.03$--$0.27$ & $0.13$ & $0.06$--$0.25$ & $0.18$  & $0.09$--$0.33$   \\[0.5ex]
			$\langle \xi_{\mathrm{T}}^{\mathrm{ne}}\rangle$             & $0.21$ & $0.08$--$0.37$ & $0.26$ & $0.17$--$0.37$ & $0.21$  & $0.18$--$0.39$   \\[0.5ex]
			$\langle \xi_{\mathrm{T}}^{\mathrm{D}}\rangle$              & $0.56$ & $0.33$--$0.67$ & $0.53$ & $0.35$--$0.58$ & $0.49$  & $0.35$--$0.60$   \\[0.5ex]
			$\langle \xi_{\mathrm{D}}^{\mathrm{ne}}\rangle$             & $0.43$ & $0.11$--$0.96$ & $0.54$ & $0.34$--$0.86$ & $0.71$  & $0.33$--$1.00$   \\[0.5ex] \hline
			$\langle L_{\mathrm{R}}^{\mathrm{N}}\rangle$                & $0.26$ & $0.10$--$0.64$ & $1.02$ & $0.38$--$1.73$ & $0.55$  & $0.36$--$0.93$   \\[0.5ex] 
			$\langle \Sigma_{\mathrm{L}}^{\mathrm{N}}\rangle$
																	    & $0.84$ & $0.17$--$2.43$   & $3.61$ & $0.60$--$9.72$  & $2.59$  & $0.45$--$4.44$\\[0.5ex] \hline
			$\langle R_{95}^{\mathrm{N}}\rangle$                        & \multicolumn{2}{c|}{---}  & $96.4$ & $83.6$--$145.5$ & \multicolumn{2}{c}{---} \\[0.3ex]
			$\langle R_{95}^{\mathrm{T}}\rangle$                        & \multicolumn{2}{c|}{---}  & $0.79$ & $0.53$--$1.03$  & \multicolumn{2}{c}{---} \\[0.3ex]
			\hline
		\end{tabular}
        }
	\caption*{\emph{Notes}: The luminosities are given in units of $[10^5 \times \Lsun]$, and the surface densities in units of [$10^{6} \times \Lsun~\mathrm{kpc^{-2}}$].
	The median of the total nuclear radius $\langle R_{95}^{\mathrm{N}}\rangle$ (the radius enclosing 95\% of the total nuclear flux density) is given in units of [pc] while the median of the total source radius $\langle R_{95}^{\mathrm{T}}\rangle$ (the radius enclosing 95\% of the total source flux density in low-angular resolution images) is given in units of [kpc]. See \cref{app:paper_2_morphometry} for details on the calculation of the radii. 
	}
	\end{subtable}
\end{table}

\subsection{Radio Luminosities}
The spectral luminosity $L_{\nu}$ refers to the luminosity of a source per unit of frequency. At a specific frequency, we may denote that the luminosity per unit of frequency at frequency $\nu$ is ${L_{\nu}}(\nu)$. The relation between $L_{\nu}(\nu)$ and the flux density is \parencite[e.g.,][]{Condon2018}
\begin{align}
	\label{eq:L_nu_chap4}
	L_{\nu}(\nu) =  \dfrac{4 \pi D_{\mathrm{L}}^2}{(1+z)^{\alpha+1}} S_{\nu}(\nu)
\end{align}
where $D_{\mathrm{L}}$ is the luminosity distance (see \cref{tab:source_info}). The integrated luminosity in the radio band is obtained by integrating \cref{eq:L_nu_chap4} over the frequency range of interest. 
{\color{black} 
	For a composite spectrum, $\alpha$ changes with frequency, hence the term $(1+z)^{-(\alpha+1)}$ needs to be integrated with $S_\nu(\nu)$. However, we can assume that the radio spectrum is a composition of different emission processes, given by power-laws (PL) of the form $S_{\nu}^{i}(\nu) \propto \nu^{\alpha_i}$, and hence constant spectral indices $\alpha_i$. This gives us 
	\begin{align}
		\label{eq:L_R_chap4}
		L_{\mathrm{R}}^{i} = \int\limits_{\nu_1}^{\nu_2} L_{\nu}^{i}(\nu) \dd \nu
		=
		\dfrac{4 \pi D_{\mathrm{L}}^2}{(1+z)^{\alpha_i+1}} \int\limits_{\nu_1}^{\nu_2} S_{\nu}^{i}(\nu) \dd \nu.
	\end{align}
}

{\color{black}
	The following discussion considers the radio luminosities over the intervals $1.0$--$2.0$, $4.0$--$8.0$ and $28.0$--$35.0$~GHz. We solve \cref{eq:L_R_chap4} using numerical integration through \code{scipy.integrate.quad}. The spectral index ($\alpha$) and thermal fraction parameters used in these calculations are derived from comprehensive SED fitting of our multi-frequency radio observations. This approach accounts for the varying emission mechanisms across our sample and provides more accurate luminosity estimates. 
	However, we note that adopting typical fixed values ($\alpha = -0.85$, thermal fraction $= 0.1$ at $1.4$~GHz ~\textcite{Murphy2012}) yields luminosities with a standard deviation of $\sim 8\%$ from those derived with the SED parameters.}
\begin{figure}
	\centering
	\includegraphics[width=0.8\linewidth]{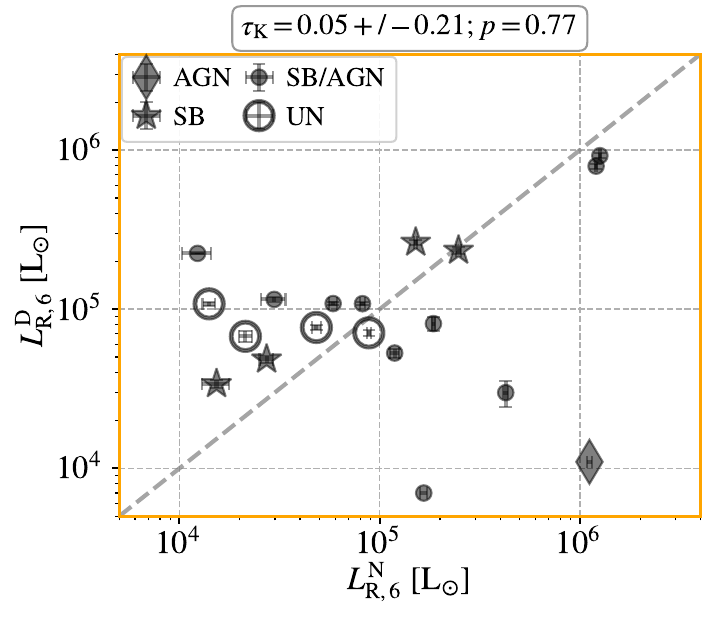}
	\includegraphics[width=0.8\linewidth]{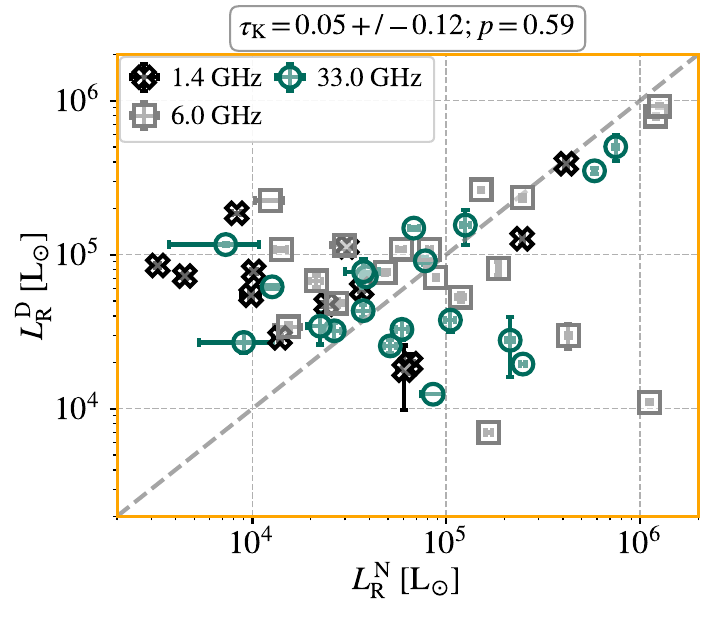}
	\caption[Relationship between the radio luminosities of nuclear and diffuse structures.]{Relationship between the radio luminosities of nuclear regions, $L_{\mathrm{R}}^{\mathrm{N}}$, at linear physical scales of $\lesssim 250$~pc with the luminosities of the diffuse emission $L_{\mathrm{R}}^{\mathrm{D}}$, at linear physical scales of $\gtrsim 250$~pc. \emph{Upper}: We show the $6.0$~GHz radio luminosities ($L_{\mathrm{R,6}}$), distinguishing the points by the radio class taken from the LIRGI sample. \emph{Lower}: Same as the left panel, but we differentiate the luminosities calculated at $1.4$, $6.0$ and $33.0$~GHz, symbolised with crosses, open squares, and open circles, respectively. 
		\emph{Notes}: The dashed diagonal line represents the one-to-one relation.
	}
	\label{fig:LR_nuc_vs_LR_tot}
\end{figure}

\subsection{Nuclear Emission (AGN+SB)}
We find in \cref{tab:fractional_emission} that the median contribution of nuclear emission between sources with pairs of high- and low-angular resolution detections is 
$\langle \Upsilon_{\mathrm{T}}^{\mathrm{N}}(\mathrm{6~GHz}) \rangle = 0.56~(0.33,~0.63)$. 
The median value of the emission sizes is 
$\langle R_{50}^{\mathrm{N}} \rangle = 48.0(36.9,70.3)$~pc and 
$\langle R_{95}^{\mathrm{N}} \rangle = 96.4(83.6,147.4)$~pc, 
with radio luminosities of 
$\langle L_{\mathrm{R}}^{\mathrm{N}}(\mathrm{6~GHz})\rangle = 1.02(0.70,1.73) \times 10^5~\Lsun$, 
and luminosity surface densities of 
$\langle \Sigma_{\mathrm{L}}^{\mathrm{N}}\rangle = 3.61(0.60,9.72) \times 10^6~\Lsun~\mathrm{kpc^{-2}}$. 
The most extreme sources are UGC\,05101 and Mrk\,331, as seen in \cref{fig:C_eM_native_1}. The radio sizes of nuclear regions ($R_{95}^{\mathrm{N}}$) are compatible with previous studies of nuclear SB properties in individual galaxies \parencite[e.g.,][]{Klockner2004,Winkel_2022,Satoh_2022}, for studies of sub-sample of galaxies \parencite[e.g.,][]{Davies_2007,Watabe_2008}, and for theoretical results \parencite[e.g.,][]{Kawakatu_2008}. 

In \cref{fig:LR_nuc_vs_LR_tot} we compare the radio luminosities of nuclear regions ($L_{\mathrm{R}}^{\mathrm{N}}$) with the radio luminosities of diffuse emission ($L_{\mathrm{R}}^{\mathrm{D}}$) from low-angular resolution VLA maps. In the {left} panel, the symbols/colours represent distinct radio classes. We do not see a clear trend that links the classes to a connection between nuclear and diffuse luminosities. In the {right} panel of \cref{fig:LR_nuc_vs_LR_tot}, we compare the same quantities, but highlighting the luminosities at 1.4, 6.0 and 33.0~GHz. We see that in the regime of $L_{\mathrm{R}}^{\mathrm{N}}\lesssim 5\times 10^{4}\Lsun$ the nuclear luminosity contributes minimally {to} $L_{\mathrm{R}}^{\mathrm{D}}$, {indicating} that nuclear SBs/AGNs have little correlation with it. However, there is some hint that brighter nuclear regions are linked to more significant diffuse emission, although the scatter is large throughout the range. This suggests that $L_{\mathrm{R}}^{\mathrm{D}}$ at scales $\gtrsim 250$~pc is not clearly influenced by $L_{\mathrm{R}}^{\mathrm{N}}$, and further investigation is needed to see whether SB/AGN cases have significantly higher luminosities at larger scales, triggered by nuclear activity compared to the SB-only (including UN) classes \parencite[see also][]{Farrah2003}. We also compared the same scatter plot, colour coded by merger stage, but no clear correlation was found.
{\color{black} For comparison, in \textcite{Farrah2003} the authors highlight a significant correlation between the total and SB infrared luminosities for a sample of SFGs and U/LIRGs. In addition, \textcite{Asmus_2015} demonstrate a strong correlation between the nuclear infrared luminosity (at $12~\mu$m) with the total observed X-ray luminosity (in the range $2$--$10$~keV). 
}

\subsubsection{Nuclear Extended Emission (nuclear SB)}
\begin{figure}
	\centering
	\includegraphics[width=0.80\linewidth]{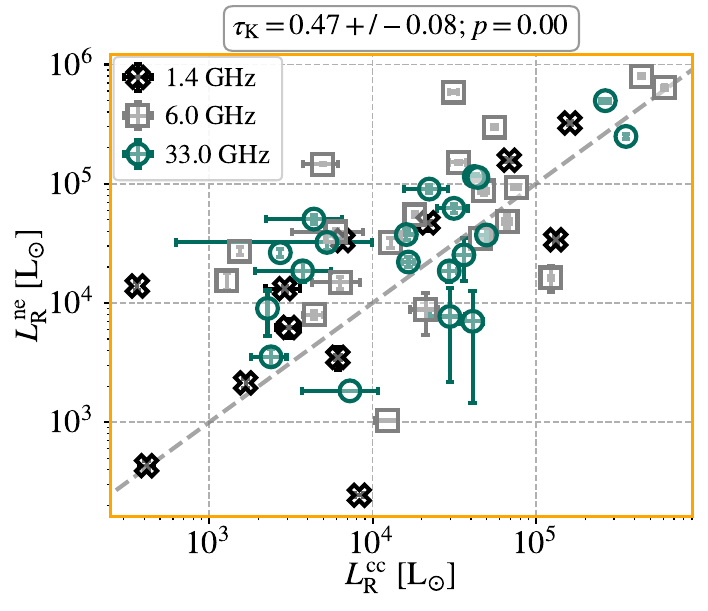}
	\caption[Radio luminosities of compact cores and nuclear extended regions.]{Relations between the radio luminosities of compact cores, $L_{\mathrm{R}}^{\mathrm{cc}}$, and nuclear extended regions, $L_{\mathrm{R}}^{\mathrm{ne}}$, at linear physical scales of $\lesssim 250$~pc. We use symbolised crosses, open squares, and open circles to indicate the luminosities calculated at $1.4$, $6.0$ and $33.0$~GHz, respectively. Given the close one-to-one correlation (as indicated by the dashed line), we see the interplay between the emission between AGN cores and nuclear SBs.}
	\label{fig:LR_ne_vs_LR_cc_color_freq}
\end{figure}

In the context of galaxy evolution during merger interactions, accretion and feedback can change the energy balance between an AGN and a nuclear SB (probed via $S_{\nu}^{\mathrm{cc}}$ and $S_{\nu}^{\mathrm{ne}}$, respectively) depending on the accretion efficiency. Their contribution to the nuclear activity may change during evolutionary phases and also with merger stage. In \cref{fig:LR_ne_vs_LR_cc_color_freq} we demonstrate that $L_{\mathrm{R}}^{\mathrm{ne}}$ and $L_{\mathrm{R}}^{\mathrm{cc}}$ are related to each other close to a 1:1 relation despite the scatter. In
\textcite{Kawakatu_2008} the authors found that this correlation becomes more bounded for brighter AGNs, and that for low-luminosity AGNs, $L_{\mathrm{R}}^{\mathrm{ne}}$ and $L_{\mathrm{R}}^{\mathrm{cc}}$ are almost independent.

\begin{figure}
	\centering
	\includegraphics[width=0.80\linewidth]{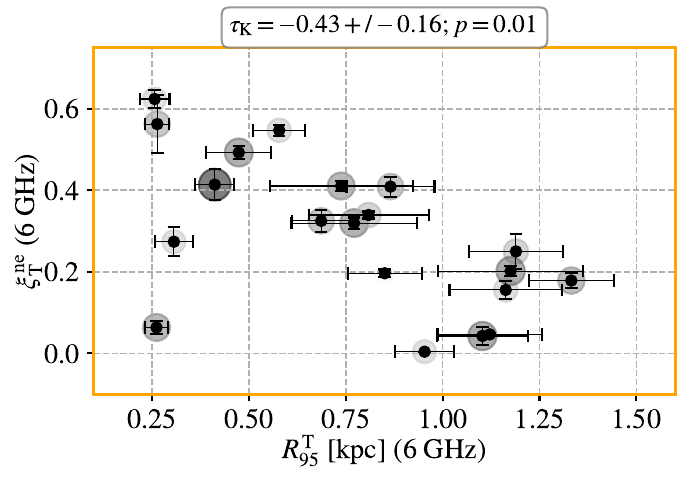}
	\caption[Relationship between the nuclear extended emission fraction and the total large-scale sizes.]{Connection between the nuclear extended emission fraction $\xi_{\mathrm{T}}^{\mathrm{ne}}$ and the large-scale sizes $R_{\mathrm{R}}^{\mathrm{T}}$, both at 6.0~GHz. 
		This result indicates that an increased flux density contribution from nuclear extended emission (nuclear SB --- $S_{\nu}^{\mathrm{ne}}$) links to smaller global sizes of the sources. 
		\emph{Notes}: Size and opacity of the data points increases with the luminosity distance $D_{\mathrm{L}}$ to the systems (see \cref{tab:source_info}).}
	\label{fig:xi_ne_tot_vs_LR_tot}
\end{figure}

We investigate in \cref{fig:xi_ne_tot_vs_LR_tot} the fraction $\xi_{\mathrm{T}}^{\mathrm{ne}}$ with the total sizes of the sources $R_{95}^{\mathrm{T}}$. We see a trend that sources that have significant $\xi_{\mathrm{T}}^{\mathrm{ne}}$ are related to smaller $R_{95}^{\mathrm{T}}$. We have also investigated the relationship between $\xi_{\mathrm{N}}^{\mathrm{ne}}$ and $L_{\mathrm{R}}^{\mathrm{T}}$, with the aim of understanding whether the balance between the SB and the AGN is related to the total power, $L_{\mathrm{R}}^{\mathrm{T}}$. However, we could not find a clear correlation. If the compact core fraction $\Upsilon_{\mathrm{N}}^{\mathrm{cc}}$ also does not show such a pattern, it could mean that the nuclear emission as a whole is connected with the total radio luminosity (a connection between AGN and total SF activity). {\color{black} However, the alternative interpretation is that nuclear activity is not associated with large-scale SF at all evolutionary stages. For instance, if SF activity at larger scales is significant compared to nuclear regions, it could mean that the infall of gas has not effectively moved into the nuclear regions (e.g. larger source sizes at larger scales), which only in later states could be triggered by the infall. Hence, this is a question of the order in which the two components are linked and at which stages. }

\begin{figure}
	\centering
	\includegraphics[width=0.85\linewidth]{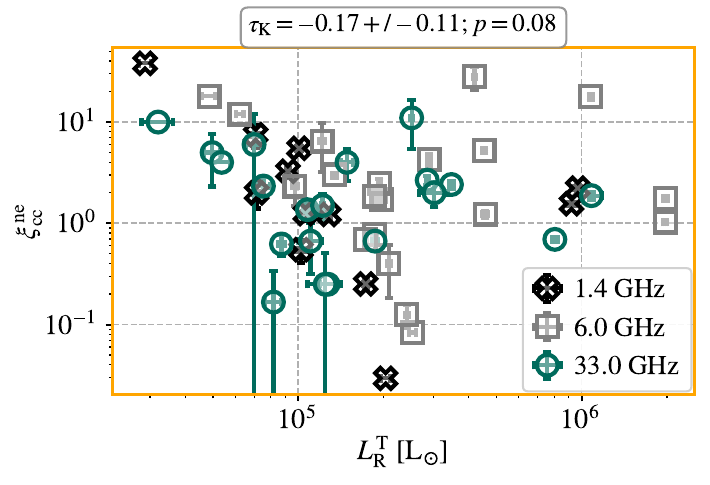}
	\caption[Fraction of nuclear extended emission to compact-core emission.]{Fraction of nuclear extended emission to compact-core emission ($\xi_{\mathrm{cc}}^{\mathrm{ne}} = S_{\nu}^{\mathrm{ne}} / S_{\nu}^{\mathrm{cc}}$), at linear scales of $\lesssim 250$~pc, compared with the total radio luminosity $L_{\mathrm{R}}^{\mathrm{T}}$ at linear scales of $\gtrsim 250$~pc, at the three central frequencies of $1.4$, $6.0$ and $33.0$~GHz. 
		We observe a lack of correlation for the energy balance of SBs and AGNs and its impact with the total radio power. Although, a tentative interpretation suggests that $\xi_{\mathrm{cc}}^{\mathrm{ne}}$ decreases along the most luminous systems.
	}
	\label{fig:xi_ne_cc_vs_LR_color_freq}
\end{figure}

In addition to what was stated above, in \cref{fig:xi_ne_cc_vs_LR_color_freq} we show the nuclear extended to compact core fraction $\xi_{\mathrm{cc}}^{\mathrm{ne}}$ against the total radio luminosity $L_{\mathrm{R}}^{\mathrm{T}}$. We note that there is no clear correlation, but a tentative conclusion is that $\xi_{\mathrm{cc}}^{\mathrm{ne}}$ slightly decreases with $L_{\mathrm{R}}^{\mathrm{T}}$, suggesting that higher luminosities are induced by bright emission of AGN cores in relation to SB luminosity. In contrast, we have also inspected the relationship between the individual components $L_{\mathrm{R}}^{\mathrm{ne}}$ and $L_{\mathrm{R}}^{\mathrm{cc}}$ with the total nuclear luminosity $L_{\mathrm{R}}^{\mathrm{N}}$, and observed that $L_{\mathrm{R}}^{\mathrm{N}}$ is more closely bounded to $L_{\mathrm{R}}^{\mathrm{ne}}$ than $L_{\mathrm{R}}^{\mathrm{cc}}$, highlighting a scenario where nuclear SBs become increasingly more relevant than AGNs. In conclusion, the complexity of these systems tells us that AGN, SB, and large-scale diffuse emission act altogether to define the total radio luminosities in U/LIRGs. This suggests that the relative contribution of the components $S_{\nu}^{\mathrm{ne}}$ (and also $S_{\nu}^{\mathrm{cc}}$) is not clearly reflected in the total radio luminosity. Hence, suggesting that the contribution fraction of nuclear activity as a whole may be fundamentally connected \emph{or not} with the global properties of the galaxies.

\subsubsection{Compact-Cores (AGN emission)}
\label{sec:paper_2_compact_cores_AGN}
We consider compact core components with exceptionally high brightness temperatures\footnote{However, compact SNe and SNRs may also show relatively high $T_{\mathrm{b}}$ \parencite[see][]{torres2009}.} (e.g., $T_{\mathrm{b}}^{6~\mathrm{GHz}} \gtrsim 10^6$~K) as signatures of AGN activity. However, in general, we consider that the primary compact compact core components in the \emph{e}-MERLIN images are associated with AGNs. This allows us to study their physical properties. These high temperatures originate either from direct emission from the AGN core itself or from unresolved components associated with sub-parsec scale jets. However, this is not general since AGN cores can be radio-quiet \parencite[e.g.,][]{Kellermann_1989A,Wilson_1995} and completely enshrouded by dust \parencite[e.g.,][]{Hickox_2018,Falstad_2021}, which can be related to distinct evolutionary phases \parencite[e.g.,][]{Sanders1988,Hickox_2009}. In these cases, the apparent AGN contribution to the radio emission may not be measured directly, resulting in a null (or small) AGN fraction. In addition, the contribution of AGN can change with frequency resulting in different fractions \parencite[e.g.,][]{Vega_2008,Ciesla_2015,Martinez_2024}, and also change over time \parencite[e.g.,][]{Dietrich2018}.
\begin{figure}
	\centering
	\includegraphics[width=0.80\linewidth]{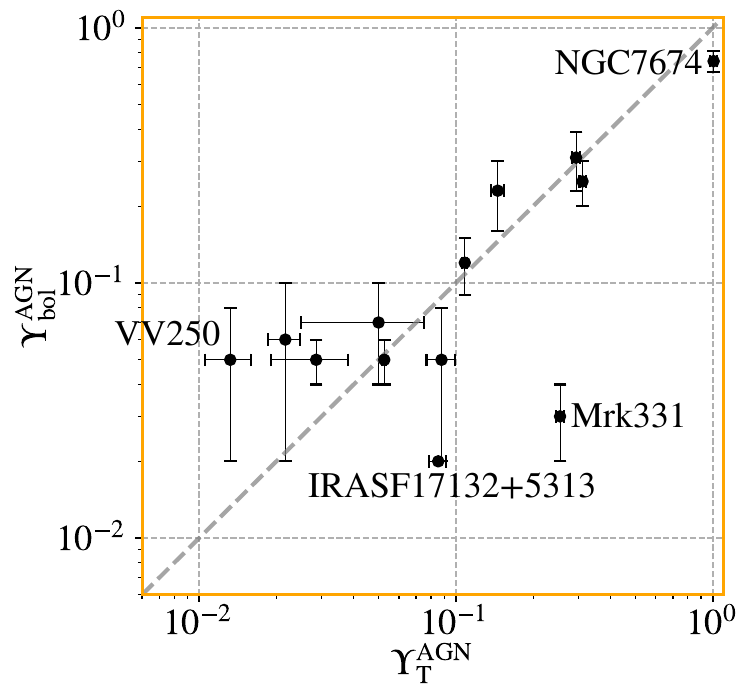}
	\caption[Comparisons of AGN fractions with literature.]{Comparison between our estimates for the AGN fraction $\Upsilon_{\mathrm{T}}^{\mathrm{AGN}}$ against the bolometric AGN fraction from \protect\textcite{DiazSantos_2017}. Note that in this comparison we are interested in the total AGN contribution, which includes both core and jet emission. This occurs for NGC\,7674, thus we have considered that both core and jets contributes to $\Upsilon_{\mathrm{T}}^{\mathrm{AGN}}$, which gives $\sim 1.0$.}
	\label{fig:Upsilon_comparison}
\end{figure}

We assume that the flux densities from detectable compact regions are a proxy for AGN activity, and thus estimate its relative contribution, or AGN/core fraction, as $\Upsilon_{\mathrm{T}}^{\mathrm{cc}}$. We have used the decomposed fluxes to calculate $\Upsilon_{\mathrm{T}}^{\mathrm{cc}}$, and also we have applied spectral corrections to each corresponding frequency ($1.4$, $6.0$ and $33.0$~GHz). The intrinsic spectral index of each component was used to make this correction. \cref{tab:fractional_emission} lists the average value for the AGN/core fraction among our sources, which at $6.0$~GHz is $\langle \Upsilon_{\mathrm{T}}^{\mathrm{cc}} \rangle(\mathrm{6~GHz}) = 0.11$ with $2\sigma$ variance of $(0.05, 0.20)$. However, these values {are} lower limits, as stated above, {due to} the different phases of AGN evolution and obscuration. Moreover, we find that our tracer of {the} AGN/core fraction is in good agreement with the work by \textcite{DiazSantos_2017}, using their bolometric AGN fractions, as shown in \cref{fig:Upsilon_comparison}. {Note that we distinguish between the compact-core fraction $\Upsilon_{\mathrm{T}}^{\mathrm{cc}}$ (which includes only the unresolved core emission) and the AGN fraction $\Upsilon_{\mathrm{T}}^{\mathrm{AGN}}$ (which encompasses all AGN-related emission, including the compact core and any extended jet structures). Both quantities represent the contribution of AGN activity relative to the total radio emission.}
\begin{figure}
	\centering
	\includegraphics[width=0.85\linewidth]{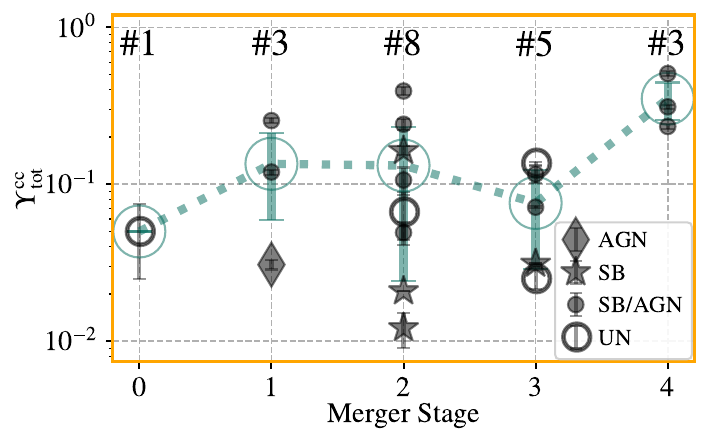}
	\caption[Comparison of the compact-core fraction with merger stages.]{Compact-core fraction ($\Upsilon_{\mathrm{T}}^{\mathrm{cc}}$) at $6.0$~GHz, compared with the merger stages taken from \textcite{Stierwalt2013} (compiled in \cref{tab:source_classes} under ``MS''). We observe a small increase of the core fraction at the merger stage 4, but no clear variation in the other stages. We note however that composite systems (AGN/SB) have relatively larger fractions $\Upsilon_{\mathrm{T}}^{\mathrm{cc}}$.
		\emph{Notes}: The dotted line (with large circles) represents the median value of $\Upsilon_{\mathrm{T}}^{\mathrm{cc}}$ per merger stage bin. The top part of the figure shows the number of sources per bin.
	}
	\label{fig:Upsilon_cc_nuc_vs_ms}
\end{figure}

\begin{figure}
	\centering
	\includegraphics[width=0.85\linewidth]{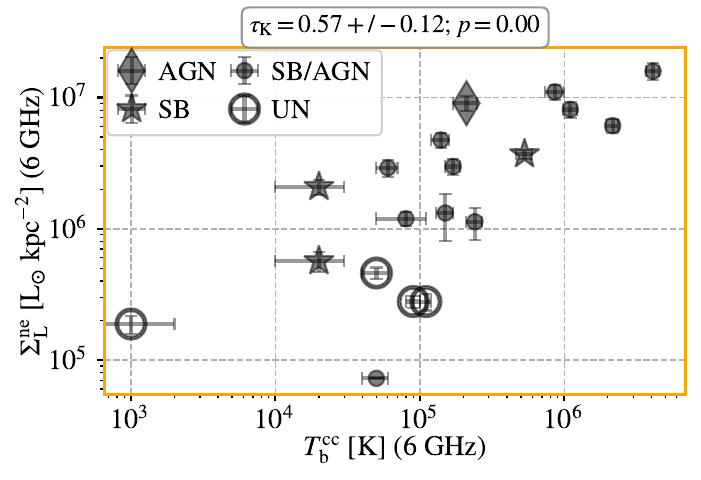}
	\caption[Relation between the SB luminosity surface density and the brightness temperature of the compact-core emission.]{
		Relation between the SB luminosity surface density $\Sigma_{\mathrm{L}}^{\mathrm{ne}}$ and the brightness temperature of the compact-core emission $T_{\mathrm{b}}^{\mathrm{cc}}$ at $6.0$~GHz. Similarly as in \cref{fig:LR_nuc_vs_LR_tot}, we differentiate the source classes by distinct symbols, indicated in the figure label (see \cref{tab:source_classes}).
	}
	\label{fig:sLR_ne_vs_Tb_cc}
\end{figure}
For additional comparison, our estimates of the AGN fraction are substantially lower than those determined by \citet{Dietrich2018}. The most notable discrepancy is observed in UGC\,05101, where they attributed an AGN fraction of $0.76\pm 0.04$, approximately three times higher than our determination. The \emph{e}-MERLIN radio map for UGC\,5101 shown in \cref{fig:C_eM_native_1} clearly contains characteristics of a nuclear SB, where the compact core is well distinguished from the surrounding emission related to the SB. Our multiscale analysis enables better separation of compact nuclear SB and large-scale diffuse emission from AGN activity. In particular, compact SBs can appear similar to AGN signatures in observations with insufficient spatial resolution, leading to potential AGN fraction overestimation.

As a contextualisation, we also note that the AGN luminosity fraction in the radio band, derived by \textcite{Morabito_2022}, is on average $\sim 0.5$, among sources with distinct classes. One of the probable reasons for these larger AGN fractions could be limitations in the {observations with lower} angular resolution ($\sim 0.3''$) used by \textcite{Morabito_2022}. Considering that VLA observations have a similar angular resolution at $6.0$~GHz, we calculated the mean values of the nuclear-to-total fraction $\Upsilon_{\mathrm{T}}^{\mathrm{N}}$. As shown in \cref{fig:radio_decomposition}, we can use $S_{\nu}^{\mathrm{N}}$ from either \emph{e}-MERLIN and VLA. Using the VLA fluxes for $S_{\nu}^{\mathrm{N}}$ in \cref{eq:xi_SF_u_cc_u_N} we obtained $\langle \Upsilon_{\mathrm{T}}^{\mathrm{N}} \rangle(\mathrm{6.0~GHz}) = 0.45~(0.38, ~0.62)$. Using the \emph{e}-MERLIN emission, we obtained $\langle \Upsilon_{\mathrm{T}}^{\mathrm{N}} \rangle(\mathrm{6.0~GHz}) = 0.56~(0.34, ~0.63)$. This details a scenario in which the total nuclear emission at $\sim 0.3''$ angular resolution, identified as bright and unresolved cores, is attributed solely to AGN activity. However, high-resolution imaging from \emph{e}-MERLIN and the VLA indicates that approximately half could be from SF (SB) emission, as \cref{fig:LR_ne_vs_LR_cc_color_freq} suggests. In contrast, the work by \textcite{Nardini_2009} and \textcite{Nardini_2010} found that the global fraction of AGN/SB in their sources is $\sim 0.33$, but it is unclear which spatial scales of the SBs are probed, i.e., whether the SB emission considers scales $\lesssim 250$~pc (nuclear) or $\lesssim 1$~kpc (total/diffuse components).

In \cref{fig:Upsilon_cc_nuc_vs_ms} we compare the merger stages obtained from \textcite{Stierwalt2013} with $\Upsilon_{\mathrm{T}}^{\mathrm{cc}}$ at $6.0$~GHz. We see a small increase in the fraction $\Upsilon_{\mathrm{T}}^{\mathrm{cc}}$ with advanced merger stages. This trend aligns with the findings of \textcite{haan2011,Petric_2011,Ricci2017}, who reported a small increase in compact-core emission as mergers progress. However, other studies \parencite{Stierwalt2013,Paspaliaris2021,Castillo_2024} found less definitive correlations. Our multiscale approach enables us to detect subtle variations in AGN contribution across the merger sequence. This approach is particularly valuable for distinguishing between compact SBs and AGN activity \parencite[e.g.,][]{Herrero_Illana_2017}, compared to single-scale observations, which often cannot distinguish between the two.

To further access the physical conditions of nuclear SBs, we can investigate the luminosity surface density (see \cref{app:paper_2_morphometry}) of individual regions of the radio emission. In particular, in \cref{fig:sLR_ne_vs_Tb_cc} we compare the SB luminosity surface density $\Sigma_{\mathrm{L,6}}^{\mathrm{ne}} = L_{\mathrm{R,6}}^{\mathrm{ne}} / A_{95}^{\mathrm{ne}}$ (see \cref{tab:highres_native_decomposition}) with the brightness temperature of the compact-core emission $T_{\mathrm{b}}^{\mathrm{cc}}$ at $6.0$~GHz. The quantity $T_{\mathrm{b}}^{\mathrm{cc}}$ was derived from $S_{6}^{\mathrm{cc}}$ (see \cref{fig:radio_decomposition}), using the deconvolved half-light sizes obtained from the image decomposition (column $R_{50,\rm d}^{6.0}$ in \cref{tab:highres_native_decomposition}). To properly measure $\Sigma_{\mathrm{L,6}}^{\mathrm{ne}}$, we have removed the compact-core component from the images to isolate the nuclear extended emission, which allowed us to better account for its luminosity surface density, with minimal contamination from the AGN/core. \cref{fig:sLR_ne_vs_Tb_cc} shows a correlation between $T_{\mathrm{b}}^{\mathrm{cc}}$ and $\Sigma_{\mathrm{L,6}}^{\mathrm{ne}}$, which implies two interpretations, that stronger SB activity increases compact-core brightness temperature, or that compactness of compact-cores increases SB activity. This aligns with expectations, as both are coupled to the total nuclear activity, although they originate from different emission mechanisms. This direct evidence has not been reported so far in the literature.

\rev{
\subsubsection{Brightness temperature as a complementary AGN diagnostic}
As a complement to the literature-based morphological classification in \cref{tab:source_classes}, we examine whether the brightness temperature of the compact-core emission, $T_{\mathrm{b}}^{\mathrm{cc}}$, can serve as an independent diagnostic of AGN activity. High brightness temperatures ($T_{\mathrm{b}} > 10^5$~K) in compact radio cores at mas scales have been demonstrated to cleanly trace AGN activity \parencite[e.g.,][]{Condon1991,Morabito_2022}, as thermal processes from star formation alone cannot sustain such temperatures on these spatial scales.

Using the $T_{\mathrm{b}}^{\mathrm{cc}}$ values derived from the deconvolved sizes and flux densities of the compact-core components at $6.0$~GHz (\cref{tab:highres_native_decomposition}), we classify sources into high-$T_{\mathrm{b}}$ ($T_{\mathrm{b}}^{\mathrm{cc}} > 10^5$~K) and low-$T_{\mathrm{b}}$ ($T_{\mathrm{b}}^{\mathrm{cc}} < 10^5$~K) categories. This classification is broadly consistent with the literature-based radio morphological classes. All sources with $T_{\mathrm{b}}^{\mathrm{cc}} > 10^6$~K (UGC\,05101, Mrk\,331, and UGC\,08696 NWE) are classified as SB/AGN composites, while most SB-only and unclassified (UN) sources have $T_{\mathrm{b}}^{\mathrm{cc}} < 10^5$~K. However, a few cases reveal discrepancies. VV\,705~N, classified as SB in the literature, exhibits $T_{\mathrm{b}}^{\mathrm{cc}} = 5.3 \times 10^5$~K, suggesting a possible embedded AGN. Conversely, NGC\,5256 (both components) and IIZw096~NE are classified as SB/AGN composites but have $T_{\mathrm{b}}^{\mathrm{cc}} < 10^5$~K, consistent with a radio-quiet or heavily obscured AGN phase, as discussed above.

Regarding the relationship between $\xi_{\mathrm{cc}}^{\mathrm{ne}}$ and $L_{\mathrm{R}}^{\mathrm{T}}$ (\cref{fig:xi_ne_cc_vs_LR_color_freq}), the $T_{\mathrm{b}}$-based classification reveals that the sources with the most extreme values of $\xi_{\mathrm{cc}}^{\mathrm{ne}}$ (i.e., nuclear extended emission strongly dominating over the compact core) are exclusively low-$T_{\mathrm{b}}$ systems. These SF-dominated sources predominantly occupy the lower luminosity regime ($L_{\mathrm{R}}^{\mathrm{T}} \lesssim 5 \times 10^5$~$\mathrm{L_{\odot}}$), whilst high-$T_{\mathrm{b}}$ sources span the full luminosity range with more moderate $\xi_{\mathrm{cc}}^{\mathrm{ne}}$ values. This supports the interpretation that the tentative decrease of $\xi_{\mathrm{cc}}^{\mathrm{ne}}$ with $L_{\mathrm{R}}^{\mathrm{T}}$ is partly driven by the increasing dominance of AGN cores in more luminous systems. 

The strong correlation between $T_{\mathrm{b}}^{\mathrm{cc}}$ and the nuclear SB luminosity surface density $\Sigma_{\mathrm{L}}^{\mathrm{ne}}$ (\cref{fig:sLR_ne_vs_Tb_cc}; $\tau_{\mathrm{K}} = 0.57$, $p < 0.01$) further validates the use of $T_{\mathrm{b}}^{\mathrm{cc}}$ as a physically meaningful diagnostic. This correlation indicates that more intense nuclear starbursts are associated with brighter compact cores, strengthening the co-evolution between AGN and nuclear SF activity at scales $\lesssim 250$~pc. Overall, the $T_{\mathrm{b}}$-based classification complements the literature-based morphological classes, and provides an additional, observationally derived diagnostic that can be applied independently of multi-wavelength ancillary data.
}

\subsection{Link between nuclear and large-scale emission}
\begin{figure}
	\centering
	\includegraphics[width=0.49\linewidth]{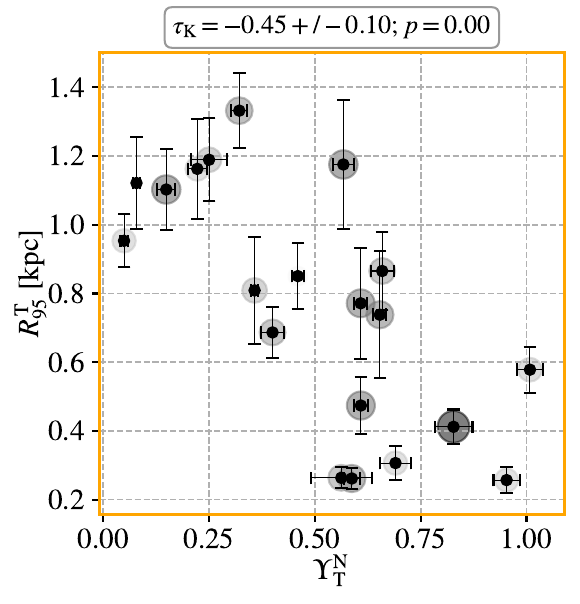}    
	\includegraphics[width=0.49\linewidth]{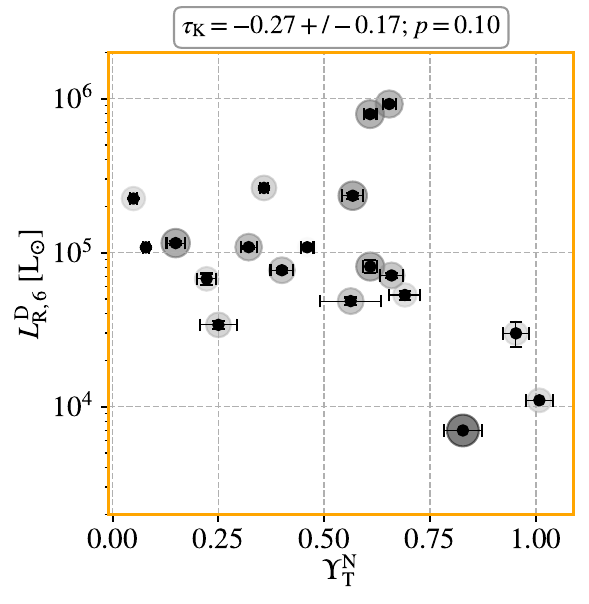}
	\caption[Large-scale diffuse emission compared with the nuclear emission fraction.]{
		Fraction of nuclear flux density $\Upsilon_{\mathrm{T}}^{\mathrm{N}}$ compared with the {total source sizes $R_{95}^{\mathrm{T}}$ and with the} $6.0$~GHz radio luminosity of diffuse structures $L_{\mathrm{R}}^{\mathrm{D}}$. 
		{\emph{Left:} Relationship between $\Upsilon_{\mathrm{T}}^{\mathrm{N}}$ and $R_{95}^{\mathrm{T}}$ highlighing a tentative correlation between the two quantities, indicating that a higher fraction of nuclear emission relates to smaller the total source sizes.}
		{\emph{Right:}} We note that for our sources, $L_{\mathrm{R}}^{\mathrm{D}}$ is quite independent on the nuclear activity probed by $\Upsilon_{\mathrm{T}}^{\mathrm{N}}$, suggesting that distinct physical processes operates at nuclear and large-scale regions.
		\emph{Notes}: Size and opacity of the data points increases with the luminosity distance $D_{\mathrm{L}}$ to the systems (see \cref{tab:source_info}).
	}
	\label{fig:Upsilon_nuc_tot}
\end{figure}
To highlight the relevance of nuclear radio emission, in \cref{fig:Upsilon_nuc_tot} we compare $\Upsilon_{\mathrm{T}}^{\mathrm{N}}$ against the sizes $R_{95}^{\mathrm{T}}$ and the diffuse luminosity $L_{\mathrm{R}}^{\mathrm{D}}$. In the left panel, we see a behaviour in which a higher fraction of nuclear emission is related to smaller wide-scale source sizes, $R_{95}^{\mathrm{T}}$. 
In particular, no correlation is found when we compare $\Upsilon_{\mathrm{T}}^{\mathrm{N}}$ with $L_{\mathrm{R}}^{\mathrm{D}}$ (right panel). These results suggest that multiple processes affect the radio luminosity at scales $\gtrsim 250$~pc, and not only the nuclear one. As stated previously, one possible interpretation is that the contribution of diffuse structures at larger scales, likely related to SF, can be relevant to define the total source luminosities.

\begin{figure}
	\centering
	\includegraphics[width=0.49\linewidth]{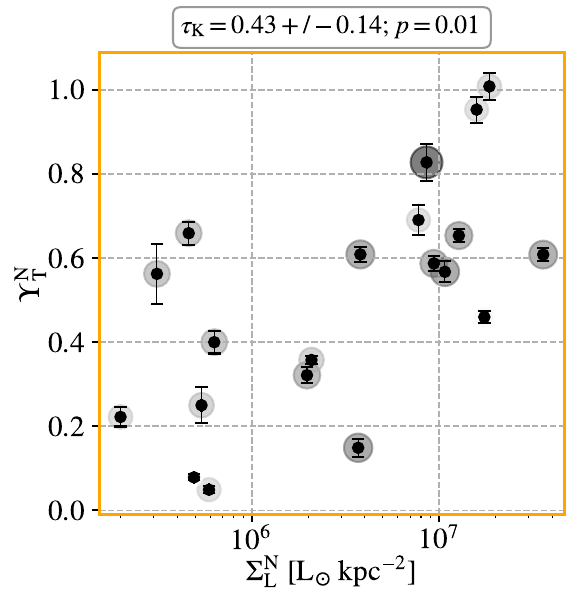}
	\includegraphics[width=0.49\linewidth]{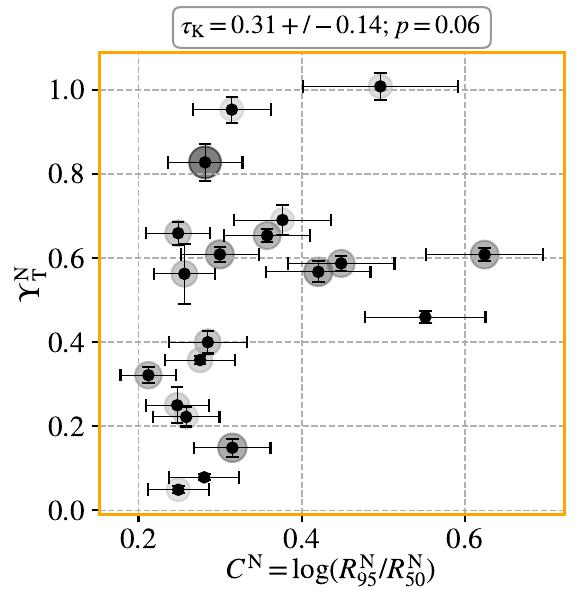}
	\caption[Nuclear emission fraction compared with nuclear luminosity surface density and concentration.]{
			Fraction of nuclear flux density $\Upsilon_{\mathrm{T}}^{\mathrm{N}}$ compared with the nuclear luminosity surface density $\Sigma_{\mathrm{L}}^{\mathrm{N}}$ (left panel) and nuclear concentration $C^{\mathrm{N}}$ (right panel), both at $6.0$~GHz.
			\emph{Left:} A positive correlation between $\Upsilon_{\mathrm{T}}^{\mathrm{N}}$ and $\Sigma_{\mathrm{L}}^{\mathrm{N}}$ indicates that systems with more energetically intense nuclear regions contribute larger fractions to the total radio emission.
			\emph{Right:} No clear correlation between $\Upsilon_{\mathrm{T}}^{\mathrm{N}}$ and $C^{\mathrm{N}}$ suggests that the spatial compactness of nuclear regions does not directly determine their contribution to the total emission.
			\emph{Notes}: Size and opacity of the data points increases with the luminosity distance $D_{\mathrm{L}}$ to the systems (see \cref{tab:source_info}).
	}
	\label{fig:Upsilon_nuc_tot_sLR_CN}
\end{figure}
To further investigate the physical conditions that govern nuclear activity, in \cref{fig:Upsilon_nuc_tot_sLR_CN} we compare the nuclear-to-total fraction $\Upsilon_{\mathrm{T}}^{\mathrm{N}}$ with the nuclear luminosity surface density $\Sigma_{\mathrm{L}}^{\mathrm{N}}$ (left panel) and the nuclear concentration $C^{\mathrm{N}}$ (right panel), both measured at $6.0$~GHz. The left panel demonstrates a positive correlation between $\Upsilon_{\mathrm{T}}^{\mathrm{N}}$ and $\Sigma_{\mathrm{L}}^{\mathrm{N}}$, indicating that systems with higher nuclear luminosity surface densities tend to have larger nuclear contributions to the total radio emission. This suggests that when nuclear regions are more energetically intense per unit area, they become increasingly dominant relative to the extended emission. Sources with the highest surface densities ($\Sigma_{\mathrm{L}}^{\mathrm{N}} \gtrsim 10^7$ L$_\odot$ kpc$^{-2}$) exhibit nuclear fractions approaching or exceeding $\Upsilon_{\mathrm{T}}^{\mathrm{N}} \sim 0.6$--$1.0$, whilst sources with lower surface densities show more diverse nuclear contributions.

In contrast, the right panel reveals no clear correlation between $\Upsilon_{\mathrm{T}}^{\mathrm{N}}$ and the nuclear concentration $C^{\mathrm{N}}$. This indicates that the spatial compactness of nuclear regions alone does not determine their contribution to the total radio emission. Sources can exhibit high nuclear concentrations (compact nuclear SBs) without necessarily dominating the total emission, and vice versa. This distinction highlights that the intensity of nuclear activity (as measured by $\Sigma_{\mathrm{L}}^{\mathrm{N}}$) is more relevant than its spatial distribution (as measured by $C^{\mathrm{N}}$) in determining the nuclear contribution to the total radio luminosity. These results support a picture in which local physical conditions at nuclear scales, such as gas density and SF efficiency, govern the nuclear contribution more strongly than morphological properties.

\begin{figure}
	\centering
	\includegraphics[width=0.85\linewidth,trim=0 0 0 0]{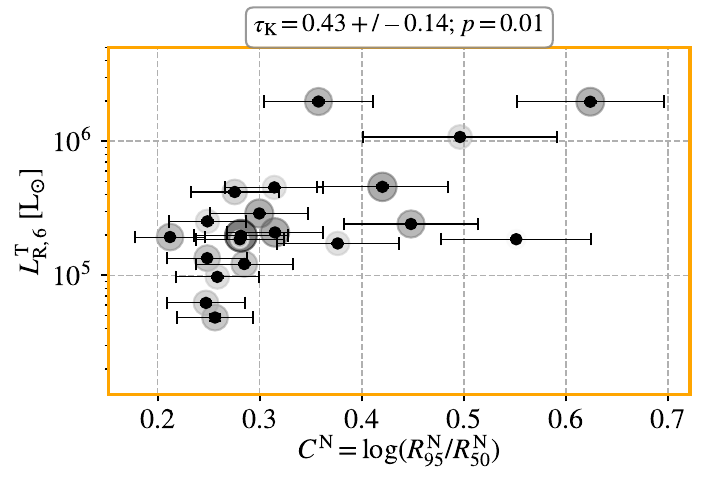}
	\caption[Relation between nuclear concentration and total radio luminosity.]{
		Relation between the nuclear concentration $C^{\mathrm{N}} = \log(R_{95}^{\mathrm{N}} / R_{50}^{\mathrm{N}})$ and the total radio luminosity at $6.0$~GHz, $L_{\mathrm{R}}^{\mathrm{T}}$. This plot shows tentative evidence that higher compactness of nuclear regions may be linked with brighter sources with respect to their radio luminosities.
	}
	\label{fig:Cnuc_vs_R95_tot}
\end{figure}

In \cref{fig:Cnuc_vs_R95_tot} we compare the nuclear concentration $C^{\mathrm{N}}$ with the total size of the sources $R_{95}^{\mathrm{T}}$ and the total radio luminosities $L_{\mathrm{R}}^{\mathrm{T}}$. We note insufficient statistics to conclude that both quantities are correlated. However, the test for $C^{\mathrm{N}}$ and $L_{\mathrm{R}}^{\mathrm{T}}$ is significant in inferring that these two quantities are likely related to each other. What this highlights is that the spatial concentration of nuclear regions is not directly linked to the wide-scale sizes, but it is linked to the total radio luminosity, {with more concentrated nuclear regions corresponding to higher radio luminosities.}

The diagnostics we have presented so far indicate that nuclear activity (whether AGN or SB) does not always dominate the total radio emission. This suggests that both nuclear and distributed SF throughout the galaxy contribute significantly to $L_{\mathrm{R}}^{\mathrm{T}}$. The scattered influence of nuclear activity on extended emission could suggest that feedback may be less efficient than assumed in some models.


\subsection{Stellar masses and multiscale extended emission}
From the correlation between SF with IR and stellar masses $M_{*}$ we expect that the extended emission at larger scales (through the radio and infrared luminosities) is also related to the stellar masses. Taking the total stellar masses measured by \textcite{Shangguan_2019} and our disentangled radio luminosities, $L_{\mathrm{R}}^{\mathrm{SF}}$
and $L_{\mathrm{R}}^{\mathrm{D}}$, we could not find a clear correlation between these quantities, suggesting that the extended emission at larger scales is not directly linked to the stellar mass of the host galaxies. Since our analysis comprises only 15 sources, we also compared the total radio luminosities with the stellar masses for all 42 LIRGI sources, and no clear relation between the $1.4$~GHz total radio luminosity $L_{\mathrm{R, 1.4}}^{\mathrm{T}}$ and $M_{*}$ was found.
In a future work we will investigate the relationship between SF and the total extended emission.

\subsection{Which regions are mostly linked to IR luminosities?}
A quantitative assessment of the radio-infrared correlation is through the parameter $q_{\mathrm{IR}}$ \parencite{Helou_1985,Yun_2001,Bell_2003}, 
\begin{align}
	q_{\mathrm{IR}} = \log_{10}
	\parenthesis{\dfrac{L_{\mathrm{IR}}~\mathrm{[W]}}{3.75 \times 10^{12}~\mathrm{[Hz]}}}
	-
	\log_{10}(L_{1.4~\mathrm{GHz}}~\mathrm{[W~Hz^{-1}]}).
\end{align}
The correlation reflects a remarkably tight relationship between infrared and radio continuum luminosities in galaxies with active SF. This relationship spans over five orders of magnitude in luminosity with typical scatter of only 0.2--0.3 dex \parencite{Condon1992,Yun_2001,Bell_2003}. Physically, this correlation arises because both infrared and radio emission trace recent SF, the infrared part through dust heated by young stars and radio through synchrotron emission from cosmic rays accelerated in supernova remnants \parencite{Murphy_2009,Schober_2017}, and then diffuse in the galaxy, where they interact with the gas and produce the observed emission \parencite[e.g.,][]{Kornecki_2022}.

Different source classes exhibit characteristic $q_{\mathrm{IR}}$ values. Star-forming galaxies typically show $q_{\mathrm{IR}} \approx 2.3$--$3.0$ \parencite{Molnar_2021}, while radio-loud AGNs are readily identified by their significant radio excess ($q_{\mathrm{IR}} < 2.3$) \parencite{Del_Moro_2013}. Radio-quiet AGN generally maintain values similar to star-forming galaxies, though often with slightly depressed $q_{\mathrm{IR}}$ \parencite{Padovani_2011,Delvecchio_2017}. Several studies have investigated correlations between $q_{\mathrm{IR}}$ and other galaxy properties, including redshift evolution \parencite[e.g.,][]{Magnelli_2015,Delhaize_2017}, stellar masses and SF rates \parencite{Gurkan_2018,Read_2018,Molnar_2018}, and black-hole accretion rates \parencite{Wong_2016}. A systematic decrease in $q_{\mathrm{IR}}$ with increasing redshift ($\Delta q_{\mathrm{IR}} \propto (1+z)^{-0.19}$) has been observed, potentially reflecting evolving magnetic field strengths or cosmic-ray electron cooling mechanisms. This parameter thus is essential as a diagnostic tool for identifying AGN contamination and studying radio emission mechanisms across diverse galaxy populations.

\begin{figure}
	\centering
	\includegraphics[width=0.85\linewidth]{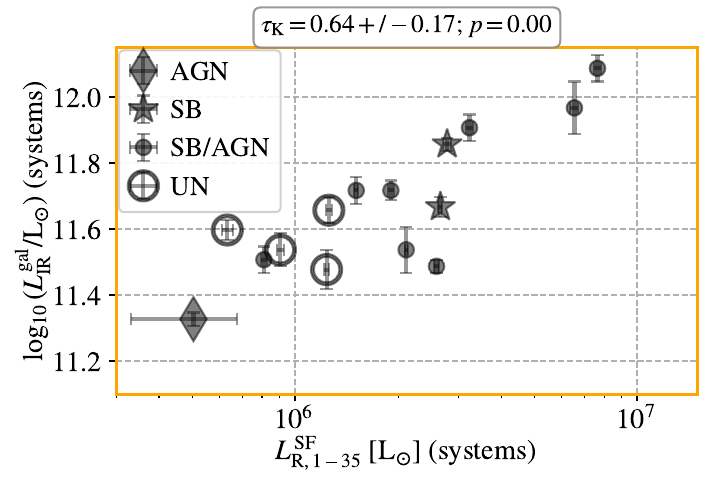}
	\caption[Infrared-radio correlation for the total IR luminosities and total extended radio luminosities.]{Infrared-radio correlation, between the total radio luminosity in the range $1.0$--$35.0$ and the total host-galaxy infrared luminosity $L_{\mathrm{IR}}^{\mathrm{gal}}$ taken from \protect\textcite{Shangguan_2019}. Note that both quantities are global-integrated, summed over for systems with two or more galaxies.}
	\label{fig:LR_nuc_vs_LIR_tot_all}
\end{figure}
The radio-infrared correlation empirically connects infrared luminosities to radio luminosities if the energy produced is solely due to SF activity. By comparing the radio emission from different scales with $L_{\mathrm{IR}}$, we can understand which one correlates the most with it. We have compared $L_{\mathrm{IR}}$ with the radio luminosity from all the scales and components studied here. Among the comparisons of $L_{\mathrm{IR}}$ with $L_{\mathrm{R}}^{\mathrm{T}}$, $L_{\mathrm{R}}^{\mathrm{N}}$, $L_{\mathrm{R}}^{\mathrm{D}}$, $L_{\mathrm{R}}^{\mathrm{ne}}$ and $L_{\mathrm{R}}^{\mathrm{SF}}$, we report here that the highest correlation corresponds to the radio luminosity attributed to the total extended emission, e.g. $L_{\mathrm{R}}^{\mathrm{SF}}$. We highlight this result in \cref{fig:LR_nuc_vs_LIR_tot_all}, where we show  the integrated radio luminosities (in the range $1.0$--$35.0$~GHz) against the total host-galaxy infrared luminosities $L_{\mathrm{IR}}^{\mathrm{gal}}$, taken from \textcite{Shangguan_2019}. 

In a similar study, \textcite{U_2019} discuss the relationship between the nuclear SF surface density $\Sigma_{\mathrm{SFR}}^{\mathrm{N}}$ against the total infrared luminosity, merger stages, and nuclear separation between the merging galaxies (their Fig.~8). They conclude that there is little correlation between these quantities and $\Sigma_{\mathrm{SFR}}^{\mathrm{N}}$, but there is some correlation with nuclear SF rates. In an analogous comparison, we did not see a clear relation between $\Sigma_{\mathrm{L}}^{\mathrm{N}}$ and $L_{\mathrm{IR}}^{\mathrm{gal}}$. However, we note that the total IR luminosities are calculated from all sources for each system. Hence, the individual characteristics of each galaxy could not reflect in the global properties (infrared and radio luminosities). Therefore, we would need to determine the infrared properties for each source individually in a future study.

\begin{figure}
	\centering
	\includegraphics[width=0.85\linewidth]{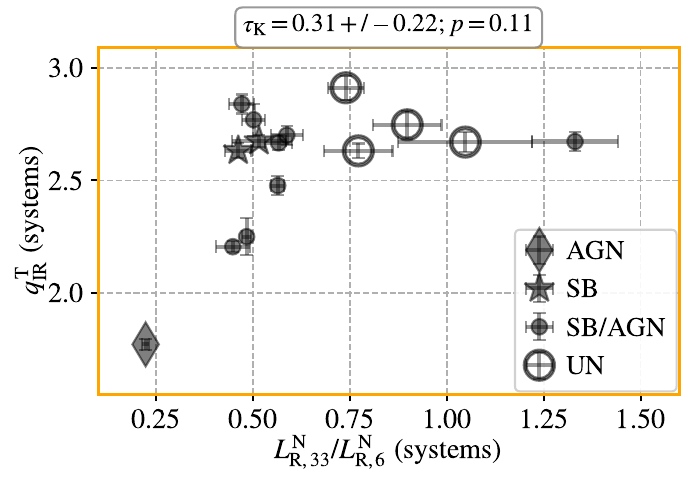}
	\includegraphics[width=0.85\linewidth]{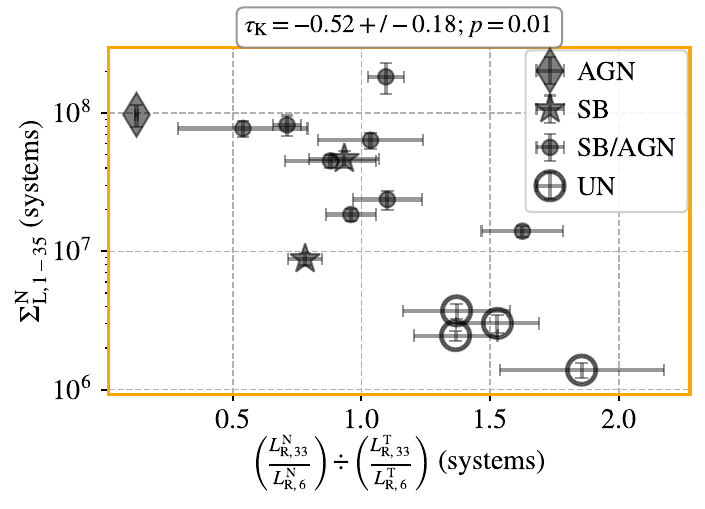}
	\caption[Diagnostics of radio spectral properties and their connection to the radio-infrared balance.]{
    \rev{
    Diagnostics of the radio spectral properties and their connection to the radio-infrared balance.
		\emph{Upper}: Radio-infrared parameter $q_{\mathrm{IR}}$ compared with the nuclear luminosity ratio $L_{\mathrm{R,33}}^{\mathrm{N}}/L_{\mathrm{R,6}}^{\mathrm{N}}$, showing that sources with lower spectral ratios tend to exhibit radio excess (lower $q_{\mathrm{IR}}$).
		\emph{Lower}: Nuclear luminosity surface density $\Sigma_{\mathrm{L}}^{\mathrm{N}}$ compared with the ratio of $L_{\mathrm{R,33}}/L_{\mathrm{R,6}}$ at nuclear and total scales, probing whether the spectral behaviour identified in the upper panel arises from nuclear-scale processes. Sources where both ratios are similar (values near 1) indicate uniform spectral properties across scales, whilst large deviations suggest distinct emission mechanisms operating at nuclear and large-scale regions.
    }
	}
	\label{fig:qIR_T_vs_fracLR_33_LR_6_N_all}
\end{figure}

\rev{
The top panel of \cref{fig:qIR_T_vs_fracLR_33_LR_6_N_all} demonstrates that the nuclear luminosity ratio $L_{\mathrm{R,33}}^{\mathrm{N}}/L_{\mathrm{R,6}}^{\mathrm{N}}$ traces deviations in $q_{\mathrm{IR}}$, linking the spectral properties of nuclear emission to the radio-infrared balance. Since the 33~GHz radio luminosity is known to directly probe SF activity \parencite{Murphy2011,Murphy2012}, in particular recent SF, it can be used to investigate these deviations. We see some signs that both quantities decrease for sources containing AGN or a mix of AGN/SB. Smaller fractions of $L_{\mathrm{R,33}}/L_{\mathrm{R,6}}$ suggest compact and AGN-related emission (radio excess, given also the lower $q_{\mathrm{IR}}$), as SF activity results in increased $33.0$~GHz luminosities, leading to higher ratios. A natural extension is to ask whether this spectral behaviour at nuclear scales differs from the global one, and how this relates to the physical conditions of the nuclear regions. In the bottom panel of \cref{fig:qIR_T_vs_fracLR_33_LR_6_N_all}, we address this by comparing the ratio of $L_{\mathrm{R,33}}/L_{\mathrm{R,6}}$ between nuclear and total scales against the nuclear luminosity surface density $\Sigma_{\mathrm{L}}^{\mathrm{N}}$ integrated in the range $1$--$35$~GHz. This comparison serves as a diagnostic to probe systems with extreme conditions and intense SF activity at the nuclear regions, and to understand whether the spectral trends identified in the upper panel are driven by nuclear-scale processes.

We see that sources with the highest nuclear luminosity surface densities ($\Sigma_{\mathrm{L}}^{\mathrm{N}} \sim 10^8$ L$_\odot$) tend to have more uniform spectral properties between nuclear and total scales (ratio values near 1), whilst sources with lower surface densities show significantly greater spectral differences between scales. This inverse correlation suggests that in the most extreme nuclear environments, the emission mechanisms at nuclear and galactic scales become more similar, potentially because of the dominance of a single emission process. AGN-classified sources consistently show higher $\Sigma_{\mathrm{L}}^{\mathrm{N}}$ ($\gtrsim 5 \times 10^7$ L$_\odot$), while unclassified sources (UN), believed to be SBs or sources with large-scale SF, populate the lower luminosity and higher spectral ratio difference region of the plot. This indicates less extreme nuclear conditions with distinct emission mechanisms compared to their host global properties. Together, both panels of \cref{fig:qIR_T_vs_fracLR_33_LR_6_N_all} connect the radio-infrared balance (via $q_{\mathrm{IR}}$) to the spectral structure of the emission across spatial scales. The upper panel identifies the nuclear spectral ratio as a tracer of $q_{\mathrm{IR}}$ deviations, while the lower panel shows that the degree to which nuclear and total spectral ratios diverge is governed by the intensity of the nuclear activity itself.
}

\subsection{The complexity of U/LIRGs}
By comparing the properties of the emission at nuclear regions and larger scales,
\cref{fig:Upsilon_nuc_tot,fig:Upsilon_nuc_tot_sLR_CN,fig:Cnuc_vs_R95_tot}, we can comment on some preliminary conclusions. We observe that the fraction $\Upsilon_{\mathrm{T}}^{\mathrm{N}}$ has some correlation with $\Sigma_{\mathrm{L}}^{\mathrm{N}}$ but is not clearly correlated with $C^{\mathrm{N}}$. Similarly, we do not have enough data to infer a correlation between of $\Upsilon_{\mathrm{T}}^{\mathrm{N}}$ with $L_{\mathrm{R}}^{\mathrm{D}}$. 
These patterns collectively confirm that U/LIRGs represent complex systems where nuclear and extended processes can operate semi-independently, with nuclear contribution depending more on local physical conditions than on global galaxy properties. The results may reflect different merger stages, where early-stage mergers show distributed activity with reasonable nuclear concentration, whilst later stages develop intense, spatially concentrated nuclear regions that can dominate total emission regardless of the properties of extended components.
The lack of clear correlation between $\Upsilon_{\mathrm{T}}^{\mathrm{N}}$ and $L_{\mathrm{R}}^{\mathrm{D}}$ particularly suggests that AGN triggering and SF enhancement can follow multiple pathways in these complex systems, with the balance between nuclear and diffuse emission determined by local gas dynamics, magnetic field configurations, and feedback efficiency rather than simply the overall energy budget of the system. In the next step, we have to extend this systematic study to a larger sample in order to constrain these arguments in a more quantitative way. The ideal case should be to compare the nuclear properties with the diffuse emission only, at radio and at infrared wavelengths. We propose this question to be investigated in future multiscale studies with a larger number of galaxies, employing the same methodology in infrared data.

\subsection{Limitations}
With the VLA, the observations used in this study are sensitive to structures with scales $\lesssim 3$~kpc. Therefore, we may have resolved out the emission at scales larger than that. It is worth mentioning that the scales probed with the VLA L band are larger than in bands C and Ka. Some sources did not show significant differences in the morphology of diffuse structures between the images at these bands. However, for other sources, such as IRAS20351+2521, the differences are significant. For example, the low-angular resolution central region of this source, with a radius of $R_{95} \approx 0.7$~kpc, is embedded in a much larger disk of diffuse emission, with a diameter of $\approx 19$~kpc. This structure is recovered only at the L-band and not in the other bands --- even with attempts to use a taper function during deconvolution. We have conducted a simple check to estimate the predicted flux density of the diffuse structure at $6.0$~GHz. Using the integrated flux density at $1.4$~GHz (and also the median values), a spectral index of $\alpha = -1.0$, and considering the difference in spatial sensitivity, the expected emission falls below the noise level at 6~GHz with the VLA. For cases like this, we studied only the emission at similar scales. 

We also observed that in sources such as NGC\,7674, within the errors, the total integrated flux density in low-angular resolution maps is similar as in the high-angular resolution maps. In these scenarios, the integrated flux density of diffuse structures is relatively small. Hence, we have proceeded with the assumption that the flux density present in the residual map (obtained through the image decomposition) is representative of $S_\nu^{\mathrm{D}}$. That can also be checked via the difference between VLA and \emph{e}-MERLIN maps, $S_\nu^{\mathrm{D}} \sim S_{\nu}^{\mathrm{T}} - S_{\nu}^{\mathrm{N}}$.

The image-based decomposition method gives standard errors for the fitted parameters. For a given modelled structure with multiple components, we would need to propagate these parameters' uncertainties of each component to the total flux densities of each component. This can become a complex process that requires the use of optimised MCMC sampling algorithms for multiple components, leading to many free parameters (e.g. $\gtrsim 40$ for NGC\,7674, shown in \cref{fig:grid_of_data_models}). Consequently, we have opted not to adopt this approach. We argue that the use of prior information from the data through source extraction to constrain models and their parameters represents a good trade-off to obtain physical models to decompose the radio emission.
Thus, the error associated with each model component was determined from their individual images in the same way as when computing the flux densities of the original images (see \cref{fig:Mrk331_C_eM_CE_Lgrow_levels}). 


\section{Conclusions}
\label{sec:paper_2_conclusion}
In this work we used a comprehensive synthesis of methods and \rev{recent} observational data to conduct a detailed study of the structure of local U/LIRGs along three frequencies and distinct spatial scales. Using simultaneous high- and low-angular resolution imaging, we have decomposed the radio emission into two spatial scales: nuclear emission (compact cores and nuclear extended emission) and total and diffuse regions. We have used source characterisation to constrain model components to decompose the images, allowing us to model sources even with complex morphologies. We have presented a systematic way to properly match the $uv$ plane of multiscale and multi-frequency visibilities, and to generate beam-matched images. Our self-calibration strategy was imperative to derive optimal, multi-frequency images that will be used to derive broadband SEDs for our future work, and support upcoming studies.

\subsection{Summary}
In the following, we summarise our main conclusions.
\begin{enumerate}
    \item As part of the e-MERLIN Legacy Project LIRGI, we presented new \emph{e}-MERLIN C-band images of local U/LIRGs, which simultaneously offers the highest angular resolution and sensitive observations available to date. 
    \item With our decomposition strategy in the deconvolved image plane we have recovered  the physical sizes of the emitting regions down to a factor of $2\sim10 $ relative to the angular resolution of the images (given by the restoring beam). The image characterisation also allowed us to robustly measure physical sizes for multiple components and angular scales. In addition, we were able to decompose/fit the radio emission for a wide range of signal-to-noise ratios, and found that our method is able to recover well the total flux densities for structures down to $\sim$ $0.1$~mJy.
    \item For the first time, using a synthesis of observations and methods, we were able to systematically decompose and characterise nuclear extended emission on scales smaller than $\sim 250$~pc. This sets a new paradigm for understanding the interplay between SB and AGN activity of U/LIRGs, particularly in nuclear regions.
    \item The fractions of emission at distinct scales/components does not correlate well the total power of the sources, but it does with the sizes. The lack of correlation suggests that physical processes at \emph{all} scales regulates the total power of the sources. Also, we observe some evidence that the shape (sizes) and power (luminosity) of the large-scale structures are simultaneously dependent on the compact (AGN) and nuclear extended emission (SB) processes, and not in one in particular. 
    \item In general, we note a lack of clear correlation between nuclear emission and large-scale properties. The total radio luminosity $L_{\mathrm{R}}^{\mathrm{T}}$ correlates (weakly) with $\Sigma_{\mathrm{L}}^{\mathrm{N}}$, an indication that the energy density at nuclear regions has partial impact in defining the total energy budget. The correlation has a less clear trend with $L_{\mathrm{R}}^{\mathrm{D}}$, suggesting that processes at their corresponding scales may be independent. Thus, additional multiscale and multi-frequency observations are required to investigate this property further. In contrast, $\Upsilon_{\mathrm{T}}^{\mathrm{N}}$ surprisingly does not show clear signs to be correlated with $L_{\mathrm{R}}^{\mathrm{D}}$, suggesting that emission processes at \emph{all} scales are relevant to define both $L_{\mathrm{R}}^{\mathrm{D}}$ and $L_{\mathrm{R}}^{\mathrm{T}}$.
    \item The nuclear concentration $C^{\mathrm{N}}$ does not show a clear correlation with the large-scale source sizes $R_{95}^{\mathrm{T}}$, an indication that the emission activity at scales $\gtrsim 250$~pc has unique characteristics, and is probably regulated by diffuse SF. A point of investigation in this matter is to study systems with (and without) clear evidence of outflows and inflows, and understand how the compactness of nuclear regions relates to the large-scale emission properties. The nuclear concentration $C^{\mathrm{N}}$, however, has some links with the total radio luminosity, which is expected, meaning that more compact nuclear regions correlate with higher total radio luminosities.
    \item  The fraction of SB emission (probed by $\xi_{\mathrm{T}}^{\mathrm{ne}}$) showed an anti-correlation with the large-scale sizes of the sources, $R_{95}^{\mathrm{T}}$. This suggests that the emission activity at scales $\gtrsim 250$~pc is suppressed, and the nuclear activity becomes dominant. This could be a sign of different evolution stages for gas transportation from galactic-scales to nuclear regions.
    \item Measurement of AGN fractions (or compact-core fractions) without comparing high- and low-angular resolution imaging may lead to overestimates of the AGN contribution, and may not offer clues of how it changes with the merger stage,  $L_{\mathrm{IR}}$ or other properties. In this study, we have properly separated the AGN/cc contribution from the total emission using multiscale radio observations, observing that its fractional contribution slightly increases with the merger stage. 
    \item As traced by $\xi_{\mathrm{T}}^{\mathrm{SF}}$, the total multiscale extended emission $S_{\nu}^{\mathrm{SF}}$ (SF related) is the main driver of the flux density/luminosity of our sources, with the exception of NGC\,7674. 
    \item This multiscale study revealed nuclear regions with complex structures. Some sources absent of compact cores (e.g. IRAS\,23436+5257 S), while others with no detection of nuclear-extended emission (e.g. NGC\,5256 SWE), but simultaneously showing complex large-scale diffuse emission.
\end{enumerate}

\subsection{Follow-up Work}
Our subsequent work will involve a comprehensive multiscale and spectral energy distribution analysis of the sources studied here, using the complete wideband coverage from $1.0$ to $35.0$~GHz. The synthesis presented here offers clear applications for future facilities such as the Square Kilometre Array and the Next Generation Very Large Array, where multiscale and multi-band observations will be routinely available. The homogeneity capability of these instruments will enable systematic application of the strategy developed here, in order to broaden the statistics of derived quantities, advancing our understanding of the physical processes governing galaxy evolution in extreme environments.

\section*{Acknowledgements}
We would like to thank the reviewer for their insightful comments in the original manuscript. We acknowledge financial support from Grant LINKB20064 (Spanish National Research Council Program of Scientific Cooperation for Development i-LINK+2020).
G.L. and K.W acknowledges financial support for a PhD studentship from STFC.
J.M., A.A. and M.P.T. acknowledges financial support from the Severo Ochoa grant
CEX2021-001131-S and from the Spanish grant PID2023-147883NB-C21, funded by MCIU/AEI/ 10.13039/501100011033, as well as support through the ``ERDF A way of making Europe'' and by the ``European Union''. 
S.d.P. acknowledges support from ERC Advanced Grant 789410.
J.M. and G.L. acknowledges the Spanish Prototype of an SRC (SPSRC) service and support funded by the Spanish Ministry of Science, Innovation and Universities, by the Regional Government of Andalusia, by the European Regional Development Funds and by the European Union NextGenerationEU/PRTR. The SPSRC acknowledges financial support from the State Agency for Research of the Spanish MCIU through the ``Center of Excellence Severo Ochoa'' award to the Instituto de Astrofísica de Andalucía (SEV-2017-0709) and from the grant CEX2021-001131-S funded by MCIN/AEI/ 10.13039/501100011033.
This project has received funding from the European Union's Horizon 2020 research and innovation programme under grant agreement No 101004719.
We thank Sotirios Sanidas and Anthony Holloway for helping with computer infrastructure.
We thank David Williams-Baldwin for providing support with the \emph{e}-MERLIN data. 
We would like to acknowledge the support of the \emph{e}-MERLIN Legacy project ``LIRGI'', upon which this study is based. \emph{e}-MERLIN and, formerly, MERLIN, is a National Facility operated by the University of Manchester at Jodrell Bank Observatory on behalf of the STFC.
The National Radio Astronomy Observatory is a facility of the National Science Foundation operated under cooperative agreement by Associated Universities, Inc.
This research has made use of the NASA/IPAC Extragalactic Database (NED), which is operated by the Jet Propulsion Laboratory, California Institute of Technology, under contract with the National Aeronautics and Space Administration. 

\section*{Data Availability}
Data used in this work is available under request to the corresponding author, in multiple formats, {such as phase-calibrated and self-calibrated visibilities (single and combined) and images.} Code documentation, development notes and Jupyter Notebooks to conduct the analysis are publicly available at the following GitHub repositories: \url{https://github.com/lucatelli/morphen} and \url{https://github.com/lucatelli/ph4ser}.


\bibliographystyle{mnras}
\bibliography{Bibliography.bib}

\clearpage
\appendix
\section{Multi-frequency observations}\label{app:paper_2_data_projects}
In \cref{tab:data_vla} we list the EVLA project codes for all observations used in this study.
\begin{table}
\centering
\caption[EVLA project code observations with the VLA.]{Available data with the VLA, using A and B/C configurations.}
\label{tab:data_vla}
\begin{subtable}{0.99\linewidth}
    \centering
    \scalebox{0.90}{%
    \begin{tabular}{l|ccc|c}
    \hline 
                          & \multicolumn{3}{c|}{\textbf{Low Res ($\gtrsim $ 0.3")}}    & \multicolumn{1}{c}{\textbf{High Res ($\lesssim $ 0.3")}} \\[0.3ex]
        Source Name       &  1.4 GHz & 6.0 GHz & 33 GHz       & 33 GHz     \\[0.3ex] \hline
        UGC\,05101        & 16B-063  & 19A-076 & AL746        & AL746$^{*}$\\[0.3ex]  
        IRAS23436+5257    & 23A-324  & 23A-324 & 22B-267      & 23A-375    \\[0.3ex]  
        MCG+12-02-001     & 23A-324  & 23A-324 & 22B-267      & 23A-375    \\[0.3ex]  
        NGC\,6670         & 23A-324  & 23A-324 & multi        & 22A-314    \\[0.3ex]  
        Mrk\,331          & 23A-324  & 23A-324 & 22B-267      & 23A-375    \\[0.3ex]  
        NGC\,5256         & 23A-324  & multi   & 22B-267      & 23A-375    \\[0.3ex]  
        UGC\,08696        & 22A-426  & multi   & AL746        & AL746$^{*}$\\[0.3ex]  
        UGC\,04881        & 23A-324  & AL746   & 22B-267      & 23A-375    \\[0.3ex]  
        NGC\,7674         & 23A-324  & 23A-324 & 14A-471      & 14A-471    \\[0.3ex]  
        VV\,250           & 23A-324  & AL746   & multi        & 23A-375    \\[0.3ex]  
        VV\,705           & 23A-324  & multi   & AL746        & AL746      \\[0.3ex]  
        III\,Zw\,035      & 23A-324  & AL746   & 14A-471      & 14A-471    \\[0.3ex]  
        IRAS\,F17132+5313 & 23A-324  & AL746   & multi        & multi      \\[0.3ex]  
        IRAS\,20351+2521  & 23A-324  & 23A-324 & 22B-267      & 23A-375    \\[0.3ex]  
        II\, Zw\, 096     & 23A-324  & AL746   & 14A-471      & multi      \\[0.3ex] 
        \hline
        \hline
    \end{tabular}
    }
    \caption*{
    $^{*}$We used the VLA-B configuration at 33.0~GHz for UGC\,08696 and UGC\,05101.
    }
\end{subtable}
\end{table}

\section{Image Processing}\label{app:paper_2_image_processing}
In this section we summarise relevant aspects taken to process the images used in this work. We also provide details on the self-calibration strategy and the source extraction strategies that supports the characterisation and decomposition of radio emitting regions.
\subsection{Self-Calibration Strategy}
\label{sec:paper_2_selfcalibration_strategy}
\subsubsection{Automated Self-Calibration}
Standard phase-referencing calibration often leaves residual phases and amplitudes that do not produce images with optimal quality. To address this, we employ self-calibration, a technique that uses the science visibility data to correct itself. This approach is described in detail by \textcite{selfcal_memo2022}, with broader discussions in \textcite{Pearson_1984} and \textcite{Taylor_1999}. Although the cited examples of imaging improvements from \textcite{Andrews_2018} and \textcite{Isella_2019} relate to protoplanetary disks, they offer a comprehensive picture of how self-calibration is essential to achieve optimal image quality.

Self-calibration employs an iterative approach to improve image quality. The process begins by deconvolving the visibilities to generate a model that represents the observed emission structure. This initial model is used to correct visibility phases and subsequent iterations to refine both the model and the phase corrections. Once sufficient signal-to-noise ratio is achieved, amplitude self-calibration may also be performed to further enhance data quality and image fidelity.

To enhance reproducibility of scientific results from interferometric observations, we developed an automated self-calibration pipeline designed for multi-instrument and multi-frequency interferometric data at single pointings. We have successfully tested this pipeline with \emph{e}-MERLIN at $1.4$ and $6.0$~GHz, and with all VLA configurations from $1.4$ to $45.0$~GHz, for sources with total integrated flux densities ranging from $\sim 1~\mathrm{mJy}$ to $\sim 1.0~\mathrm{Jy}$. 
Full documentation, examples and development notes are available in our open-source package \textsc{ph4ser}\footnote{See at \href{https://github.com/lucatelli/ph4ser}{https://github.com/lucatelli/ph4ser}.}.

\subsubsection{Joint Self-calibration of VLA and \emph{e}-MERLIN}
\label{sec:paper_2_joint_selfcalibration}
\begin{figure*}
\centering
\includegraphics[width=1.0\linewidth]{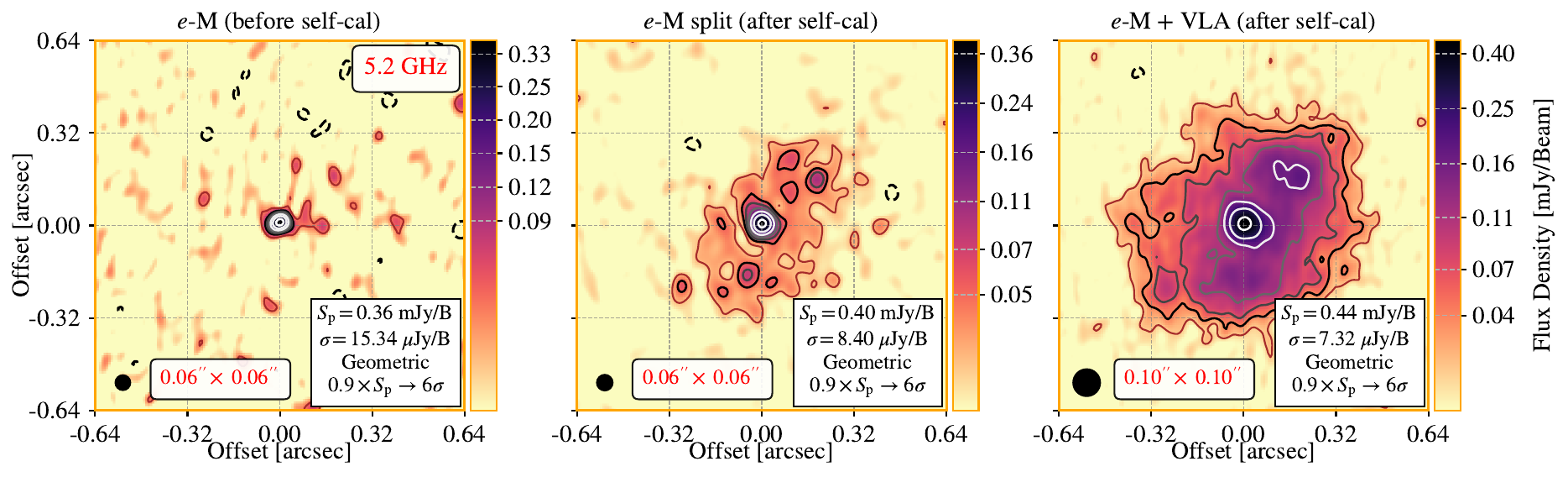}
\caption{Improving image fidelity with joint self-calibration (phase + amplitude) of \emph{e}-MERLIN and VLA observations. We use VLA to fill the short spacings present in \emph{e}-MERLIN.
The left panel shows the phase-referenced calibrated image of IRAS\,23436+5257 (N). The middle panel shows a pure \emph{e}-MERLIN image, resulting from the joint self-calibration with the VLA. The combined image between \emph{e}-MERLIN and VLA is displayed in the right panel. 
\emph{Notes}: The contour levels are show in the colour bar at the right side of each panel, auto-generated in log-space, using a geometric sequence with 6 contours, given by 
$c_i = 0.9 S_{\mathrm{p}}[({6\sigma})/({0.9S_{\mathrm{p}}})]^{({i-1})/({N-1})}$, 
with $i=1,2,3,4,5,6$, $N=6$, and $S_{\mathrm{p}}$ the peak intensity.
    We add an extra contour in brown, marking the $3 \times \sigma$ boundary. The values of $S_{\mathrm{p}}$ and $\sigma$ are given in each panel.}
\label{fig:example_selfcal_combined_em_VLA_Cband}
\end{figure*}
Joint self-calibration involves simultaneously self-calibrating combined interferometric visibilities from different instruments (e.g., VLA and \emph{e}-MERLIN) or multiple VLA configurations. This technique offers significant advantages: it minimises potential phase offsets between different datasets by generating a combined model during deconvolution and self-calibration loops, and it can correct for small amplitude differences between arrays, configurations, and observational epochs. A critical prerequisite for this approach is to verify that the source is non-variable across datasets, as variations would compromise self-calibration.

Self-calibration of \emph{e}-MERLIN observations typically yields good results for sources with integrated flux densities above $10$~mJy when all antennas are available. While self-calibration remains possible for fainter sources, it often results in excessive data flagging, prone to lose antennas, which is prohibitive. An alternative approach is to begin self-calibration with a high-quality model derived from other observations, although this requires non-variability.

In this work, we employ VLA data to fill the short spacings of \emph{e}-MERLIN measurements and perform joint self-calibration. VLA observations serve to constrain the model at larger scales, whilst also helping constraining in-beam positions of clean components and total flux densities of unresolved structures. The key idea of joint self-calibration lies in using deconvolution weights that incorporate structures at scales sensitive to both arrays.

For concatenated \emph{e}-MERLIN and VLA visibilities with homogeneous weights (see \citetalias{Lucatelli_2024}), the weighting parameter \code{robust} in WSClean determines which array's characteristics dominate the resulting image. Setting \code{robust}$\lesssim$\code{-1.0} produces an \emph{e}-MERLIN-alike image, whilst \code{robust}$\gtrsim$\code{0.0} yields a more VLA-alike image. The interval from \code{-1.0} to \code{0.0} represents a regime in which imaging produces a combined image between the two arrays. We take advantage of this behaviour by initiating phase-only self-calibration with \code{robust=-0.25} (or \code{robust=0.0} depending on data quality and integrated flux density), then changing to \code{robust=0.0-0.5} for subsequent iterations to account for more diffuse emission.

We show in \cref{fig:example_selfcal_combined_em_VLA_Cband} the substantial improvements achieved through joint self-calibration using two iterations (\code{p} + \code{ap}). The left panel shows an image generated after standard phase-referencing calibration (no self-calibration) with the \emph{e}-MERLIN CASA Pipeline (eMCP). The middle panel displays the final self-calibrated \emph{e}-MERLIN image produced through joint self-calibration with the VLA, and then obtained by splitting the \emph{e}-MERLIN observation from the final self-calibrated combined visibility. The right panel presents the final combined and self-calibrated image. The improved \emph{e}-MERLIN image not only reveals significantly more structure than the original, but also contains reduced RMS levels by a factor of two.

We noted that the extended emission present in the pure joint self-calibrated \emph{e}-MERLIN image at the middle panel of \cref{fig:example_selfcal_combined_em_VLA_Cband} can be recovered without self-calibration. However, that is only achieved by applying a taper during deconvolution (usually a Gaussian sky taper of $\sim$0.05" or higher), but resulting in a much worse RMS, $\sigma_{\mathrm{rms}} \sim 50$--$100$ $\mu\mathrm{Jy~beam}^{-1}$,  and with significant side-lobe effects. This demonstrates the significant advantages of joint self-calibration.

\subsection{Characterisation of emitting regions}
\label{app:paper_2_morphometry}
\label{app:paper_2_source_detection}
We have implemented a modular way to perform source extraction. The source extraction routines can use either \textsc{photutils} or \textsc{sep}, but other methods can be implemented. The general idea is to perform source detection in a reference radio image (\cref{fig:example_source_multilabels_uvmatch}), where the detected components, as shown in \cref{fig:source_detection_example}, are used to define the number of structures to be fitted in the multi-frequency images.

\begin{figure}
\centering
\includegraphics[width=0.99\linewidth]{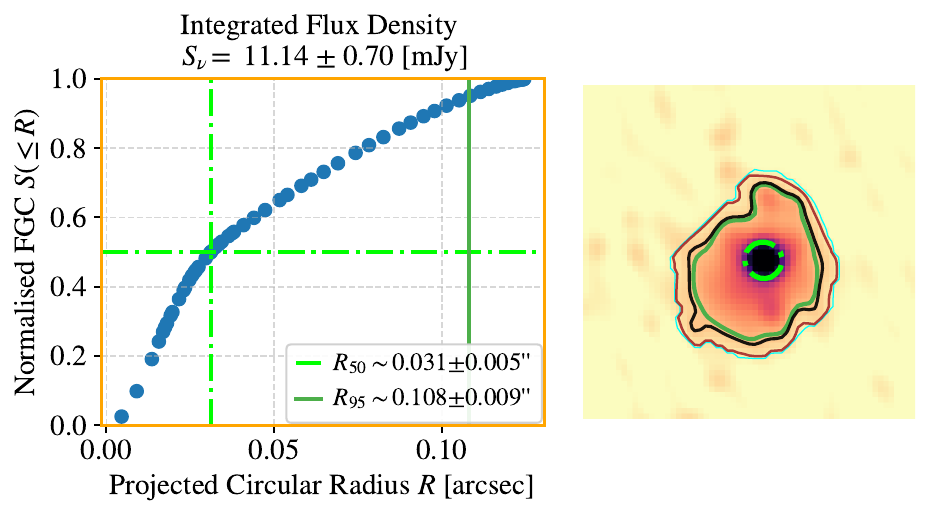}
    \caption{
    Determination of the sizes of emitting regions, through the flux growth curve (FGC), derived from the brightness distribution of emission levels. 
    \emph{Left}: Normalised FGC $S(\leq R)$ as a function of projected radius $R$. The vertical dot-dashed line represents $R_{50}$ and the solid one represents $R_{95}$. \emph{Right}: Map with contours, highlighting the $50\%$ and $95\%$ contours. We also show extra contours as black -- $6\sigma_{\mathrm{mad}}$, brown -- $3\sigma_{\mathrm{mad}}$ and cyan -- the mask contour. The mask is generated by dilating the $6\sigma_{\mathrm{mad}}$ region (see \protect\citetalias{Lucatelli_2024}).
    }
\label{fig:Mrk331_C_eM_CE_Lgrow_levels}
\end{figure}

\begin{figure}
\centering
\includegraphics[width=0.80\linewidth]{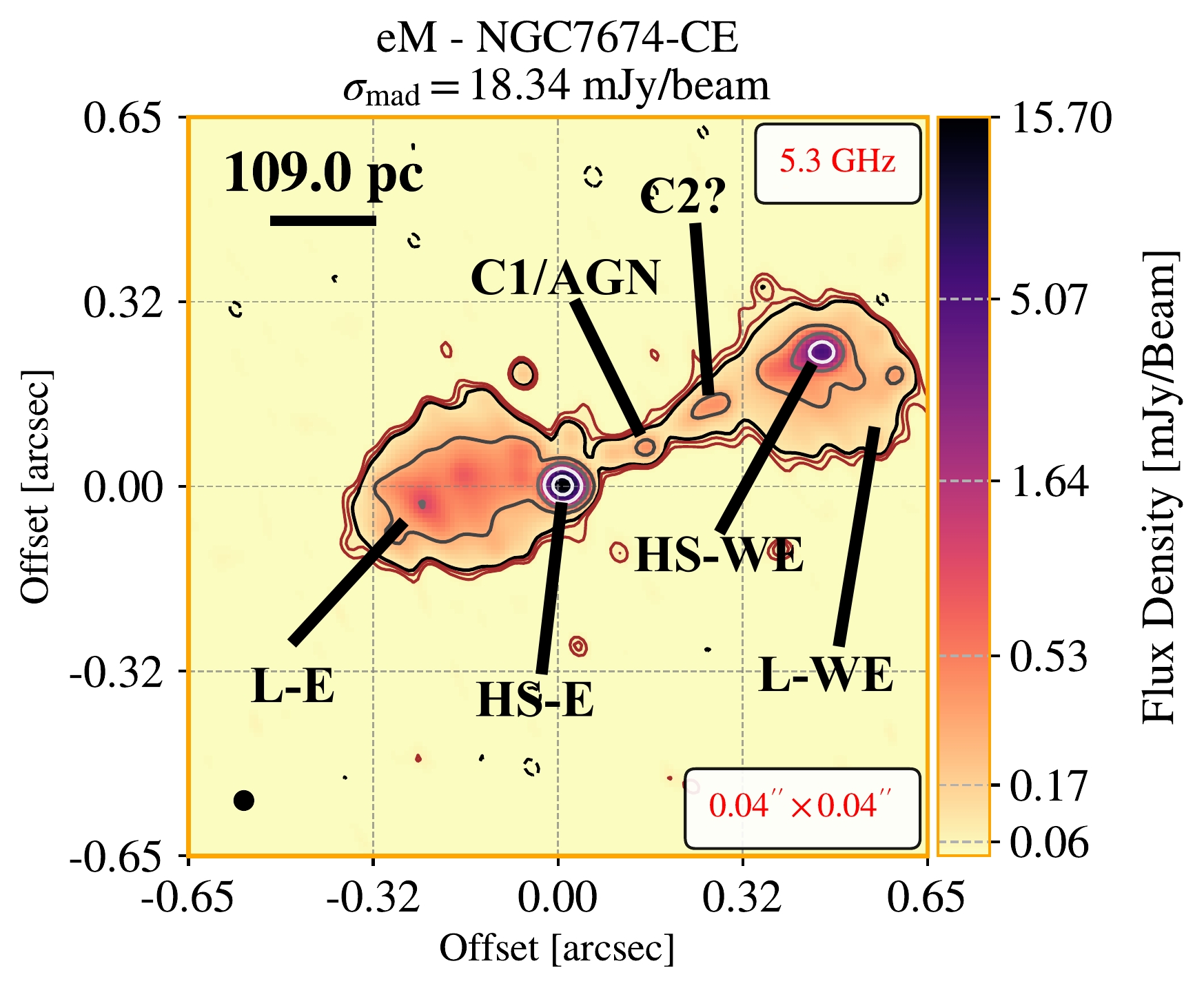}
\caption[Radio image of NGC\,7674 with \emph{e}-MERLIN at C-band.]{Radio image of NGC\,7674 with \emph{e}-MERLIN at C-band, highlighting its individual components: the AGN core (C1), the hot-spots HS-WE and HS-E, and the lobes L-E and L-WE. The component C2 appear to be part of the WE  jet. We use this source as a good example to highlight our image decomposition method. The results are shown in \cref{fig:source_detection_example} and \cref{fig:grid_of_data_models}.}
\label{fig:example_source_multilabels_uvmatch}
\end{figure}

\begin{figure}
    \centering
    \includegraphics[width=0.475\linewidth]{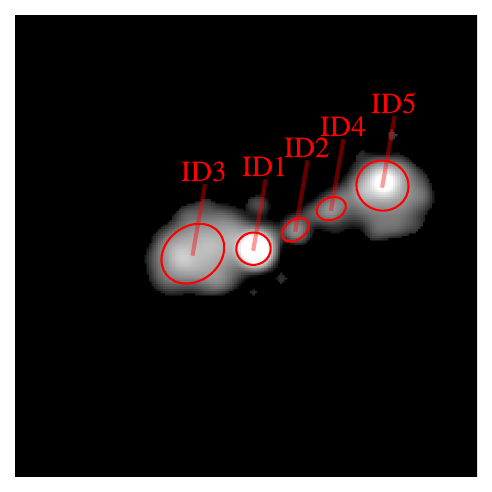}
    \includegraphics[width=0.475\linewidth]{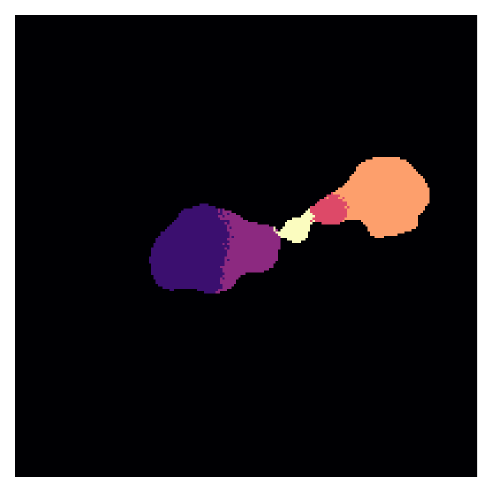}
    \caption{Example of source extraction and deblending (for NGC\,7674). Each region is assigned an ID (left panel), which are then populated into the fitting strategy, alongside with their respective masks (right side). We use these constraints to facilitate complex image decomposition problems. The results for this example are shown in \cref{fig:grid_of_data_models}. For more details, see \protect\citetalias{Lucatelli_2024}.}
    \label{fig:source_detection_example}
\end{figure}

We determine the sizes of emitting regions by using the pixel count of generated masks and converting them to areas, and thus to circular aperture radii. From the radio emission within those masks, we generate the flux density growth curve (FGC) so that the fractional circular apertures can be determined, providing, for example, the half-flux radius $R_{50}$ and the total flux radius $R_{95}$. An example is given in \cref{fig:Mrk331_C_eM_CE_Lgrow_levels}. As explained in \citetalias{Lucatelli_2024}, we take the radius enclosing $95\%$ of the flux as the largest spatial extent of any given emitting region (any image or component/region of our sources). For example, by applying this method to deconvolved images (fitted image model components), we can extract their respective deconvolved sizes. This is imperative for determining the deconvolved sizes of compact-core components, required to measure brightness temperatures, for example. The derived sizes are given in \cref{tab:lowres_native_decomposition} and \cref{tab:highres_native_decomposition}, for low- and high-angular resolution imaging, respectively. 

To complement our analysis, we adopt some metrics for image characterisation. The concentration index is a good indicator of the compactness of a given region/source. We define the nuclear concentration $C^{\mathrm{N}} = \log(R_{95}^{\mathrm{N}} / R_{50}^{\mathrm{N}})$ as a measure of the spatial compactness of nuclear regions. We also define the source concentration $C^{\mathrm{S}} = \log (R_{95}^{\mathrm{T}} / R_{95}^{\mathrm{N}})$. However, $C^{\mathrm{S}}$ should be taken as a measure of the ``decoupling'' between emission on larger scales and emission on smaller scales. 

To assess the emission compactness in terms of luminosity or flux density 
(discussed in \cref{sec:paper_2_discussion}), 
we calculate luminosity surface densities by dividing the luminosity by the projected enclosed area. For example, $\Sigma_{\mathrm{L}}^{\mathrm{N}} = L_{\mathrm{R}}^{\mathrm{N}}/{{A}}_{95}^{\mathrm{N}}$ and $\Sigma_{\mathrm{L}}^{\mathrm{T}} = L_{\mathrm{R}}^{\mathrm{T}}/{{A}}_{95}^{\mathrm{T}}$, where ${{A}}_{95}^{\mathrm{N}}$ and ${{A}}_{95}^{\mathrm{T}}$ represent the projected areas enclosing $95\%$ of nuclear and total emission, respectively. These surface density measurements provide insights into the compactness of radio emission processes operating at different spatial scales. 

\subsection{Multi-frequency Image Decomposition}
The image decomposition used in this study uses a strategy similar to that used in other algorithms previously used in optical image decomposition \parencite[e.g.,][]{Vika_2013,Haussler_2013}. The idea is to use the minimisation parameters of an image with good signal-to-noise ratio to provide initial conditions and constrain the minimisation of other images where the signal-to-noise may be limited.
\begin{figure*}
\centering
    \centering
    \includegraphics[width=1.0\textwidth]{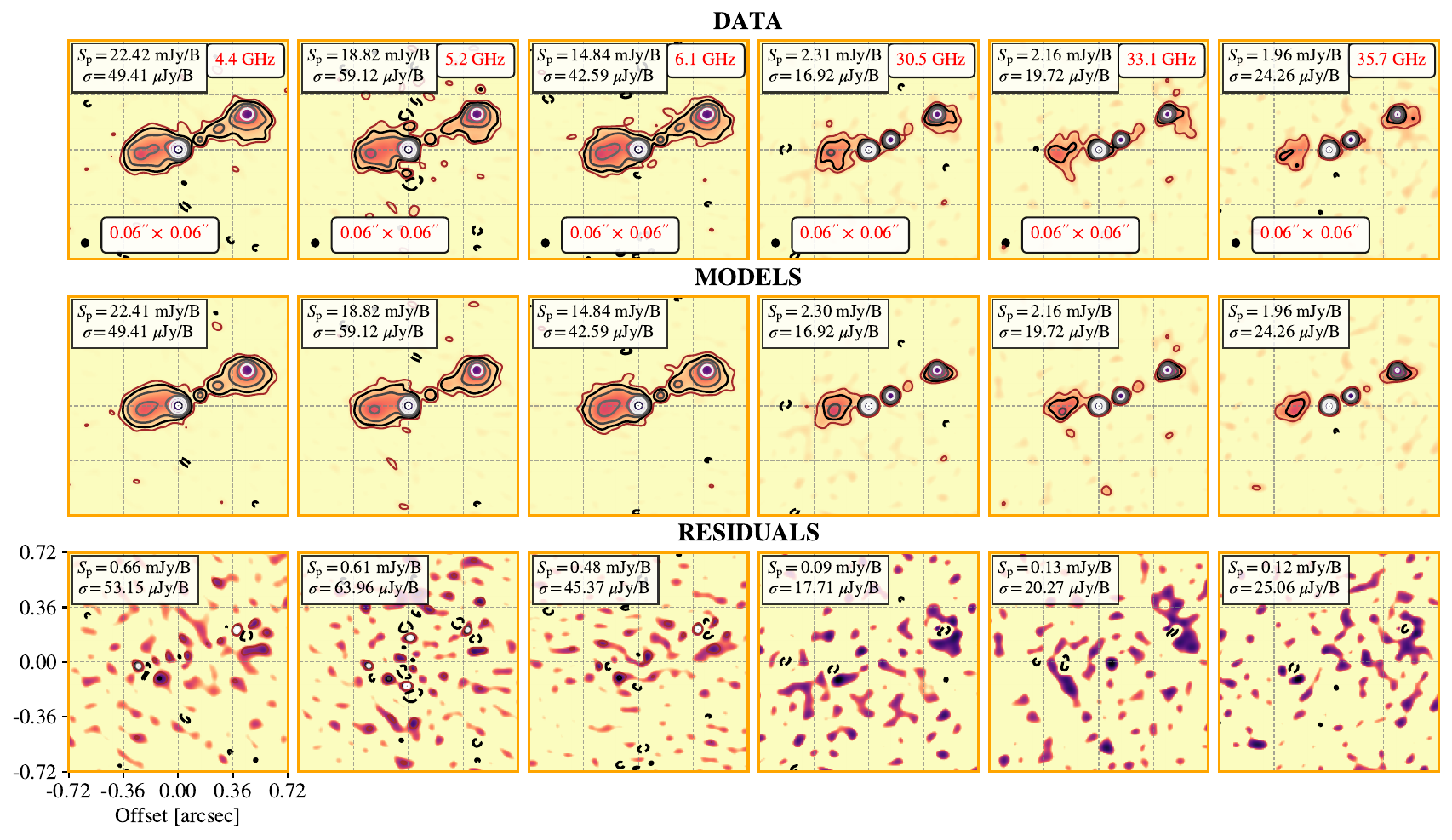}
    \caption{
        Multi-frequency image decomposition of NGC\,7674 as a result of providing constrained information during fitting. The model images also appear more realistic to the data since, during fitting, we perform the background modelling using the clean residual maps provided by \textsc{WSClean}. Hence, we make sure that the appropriate level of noise is accounted into the models, thus giving more realistic flux densities (in beam and integrated) for each model component. Note that the peak intensity differences are often below $1\%$. \emph{Notes}: Contour levels procedure is the same as discussed in \cref{fig:Mrk331_multifreq_lowres}. }
    \label{fig:grid_of_data_models}
\end{figure*}

Using the \textsc{lmf{}it} library, it facilitates us to construct minimisation objects in Python. Once minimisation is completed for a reference image, the fitting results contained in the object can be passed to the minimisation function that iterates over the rest of multi-frequency images. 
In \cref{fig:grid_of_data_models} we provide a clear example of this procedure. 
A source with complex structure can be well modelled with a constraint fit along various frequencies.

To fit the unresolved cores in VLA images, we have adopted a strategy to fix the maximum value of the S\'ersic effective radius to $R_n=1$. This is equivalent (but not equal) to a delta-function. This approach proved essential to minimise potential overestimations of the integrated flux density of unresolved nuclear regions.  For cases where the total emission size was close to the restoring beam of the image, this approach allowed us to fit a two-component model ($S_{\nu}^{\mathrm{N}} + S_{\nu}^{\mathrm{D}}$) in order to recover the diffuse component.

\section{Images and Derived Tabular Data}\label{app:paper_2_images_and_tabular_data}
In \autoref{tab:highres_native_decomposition} and \autoref{tab:lowres_native_decomposition} we present a summary version of the tabulated properties of the image characterisation analysis. The complete machine-readable versions are available online. High-angular resolution images are displayed in \autoref{fig:C_eM_native_1}. We will make available all visibilities and images once our next paper is completed.
\begin{figure*}
\centering
\includegraphics[width=0.33\linewidth,trim=0 0 0 0]{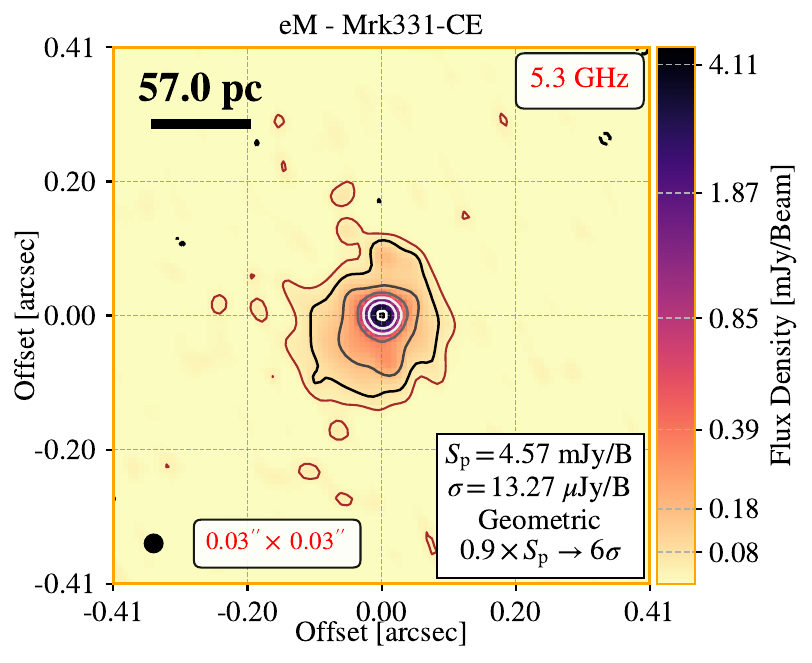}
\includegraphics[width=0.33\linewidth,trim=0 0 0 0]{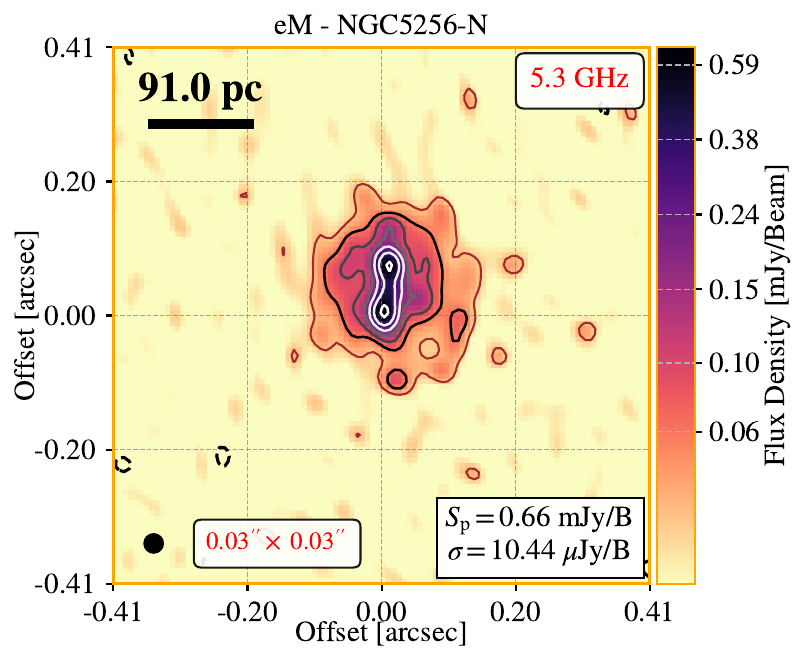}
\includegraphics[width=0.33\linewidth,trim=0 0 0 0]{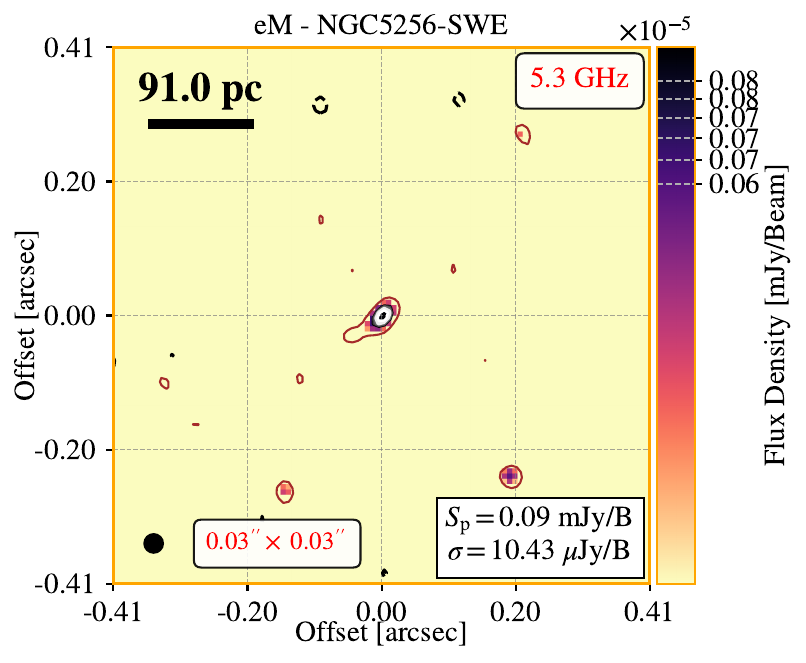}
\includegraphics[width=0.33\linewidth,trim=0 0 0 0]{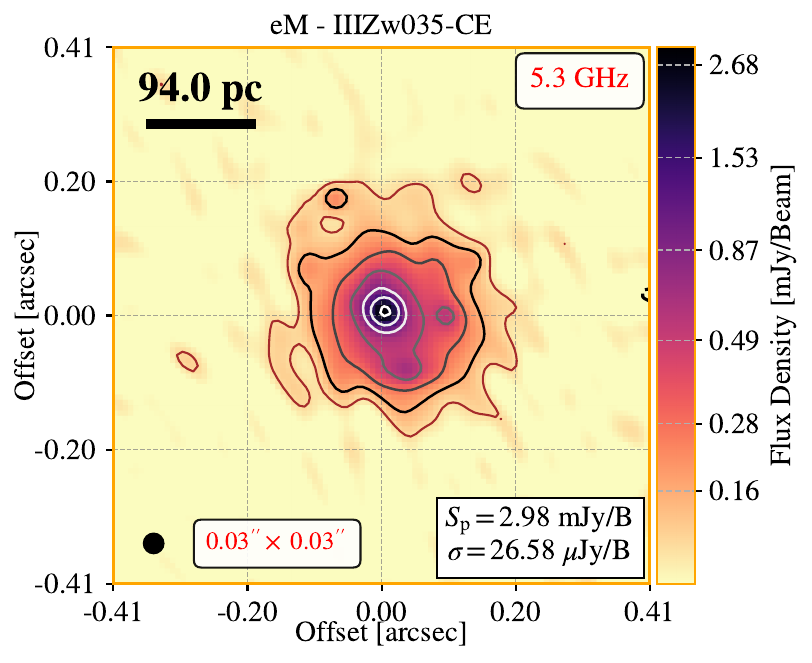}
\includegraphics[width=0.33\linewidth,trim=0 0 0 0]{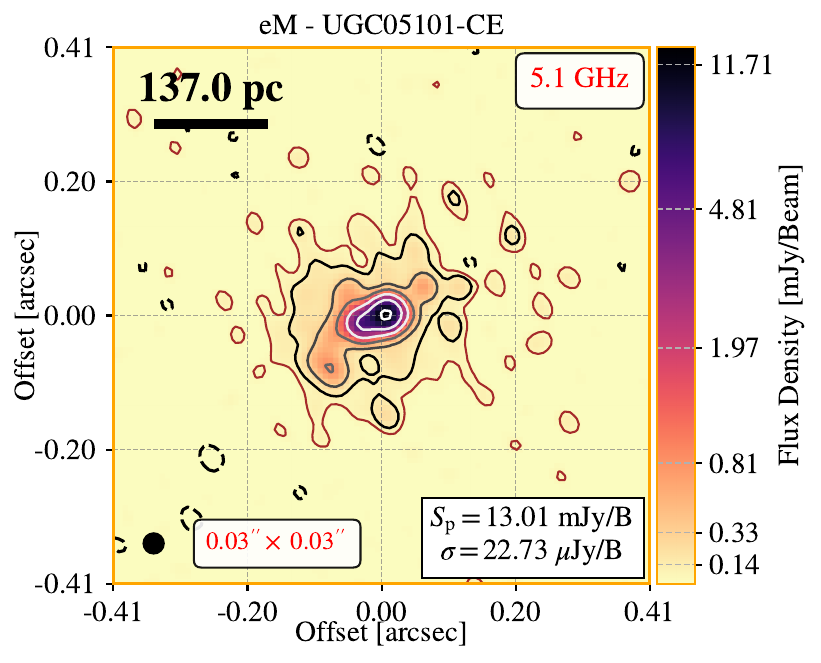}
\includegraphics[width=0.33\linewidth,trim=0 0 0 0]{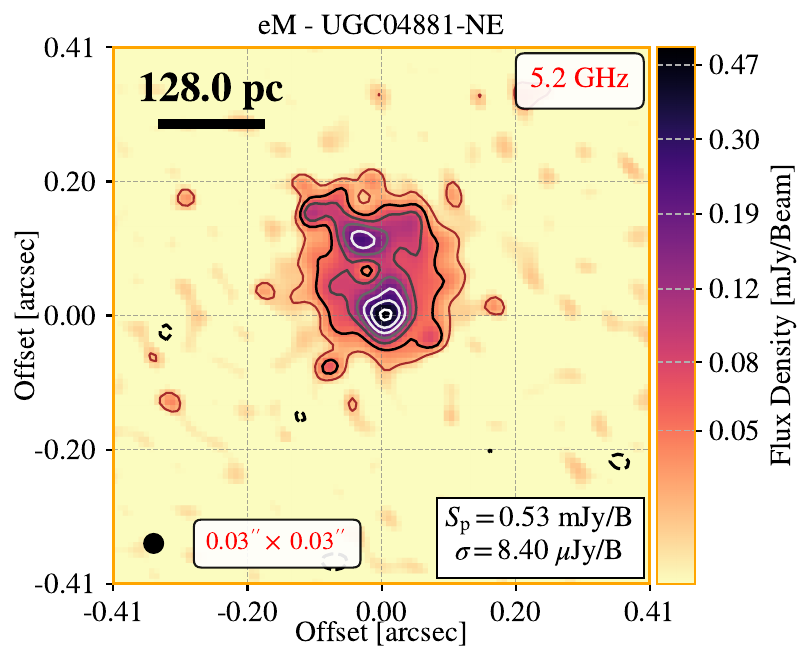}
\includegraphics[width=0.33\linewidth,trim=0 0 0 0]{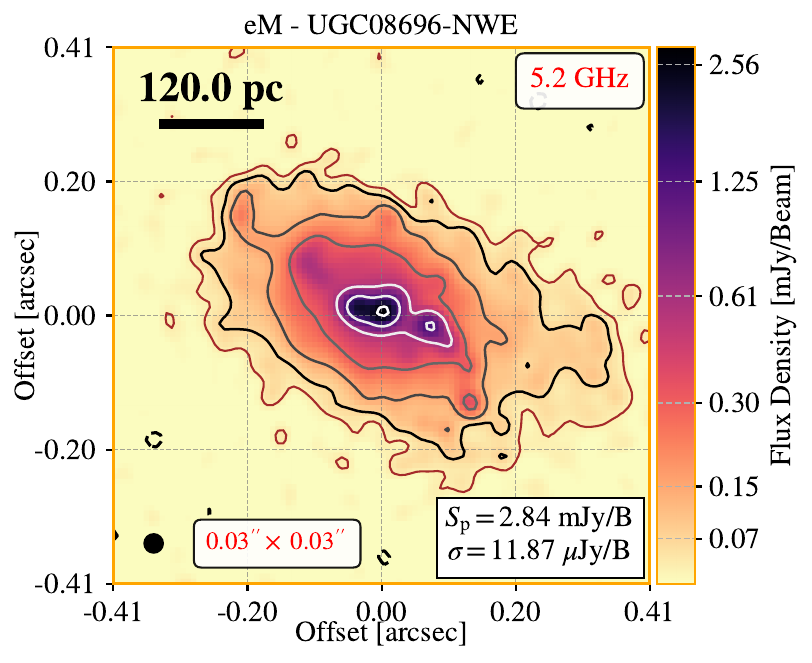}
\includegraphics[width=0.33\linewidth,trim=0 0 0 0]{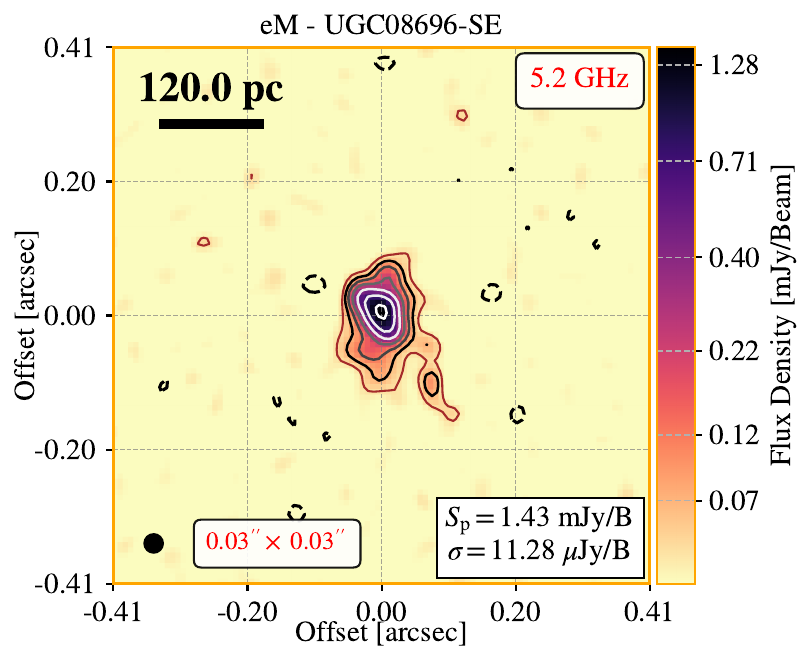}
\includegraphics[width=0.33\linewidth,trim=0 0 0 0]{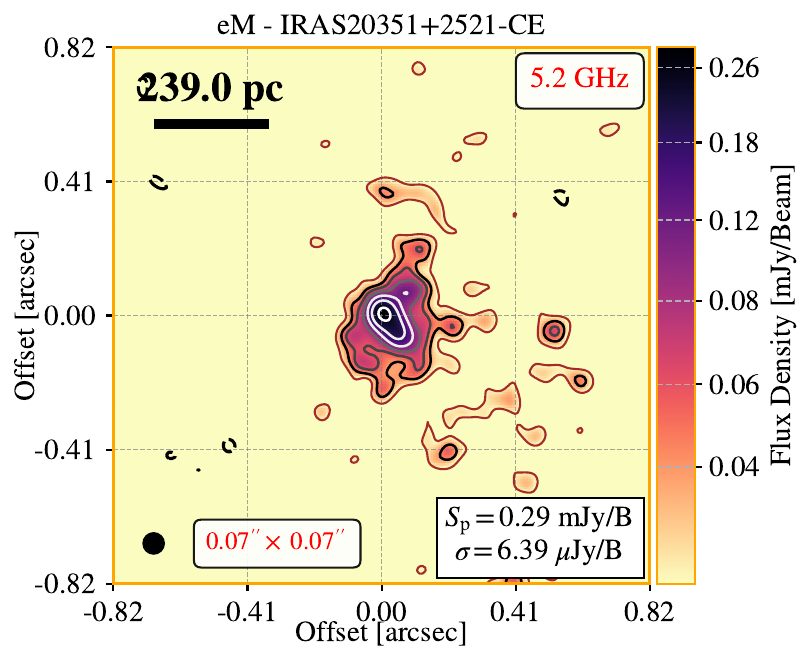}
\includegraphics[width=0.33\linewidth,trim=0 0 0 0]{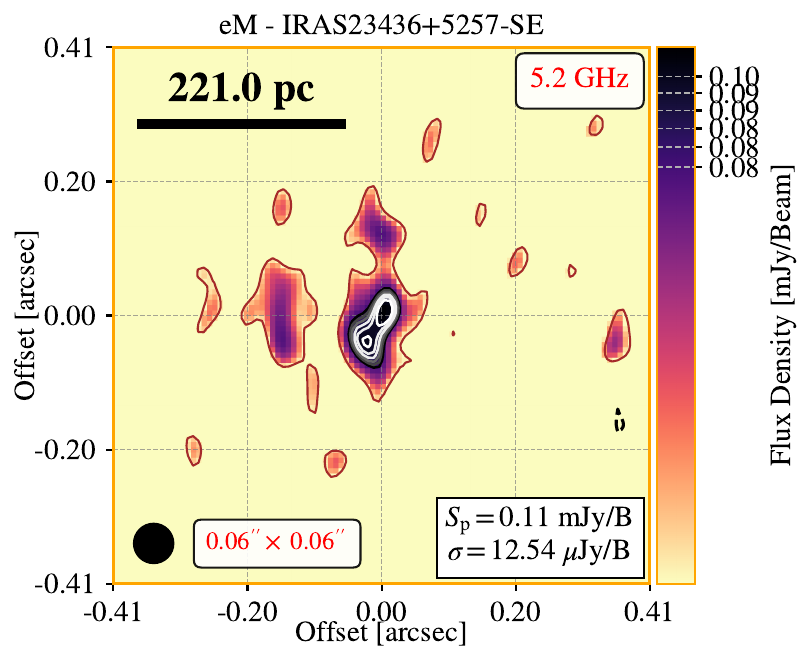}
\includegraphics[width=0.33\linewidth,trim=0 0 0 0]{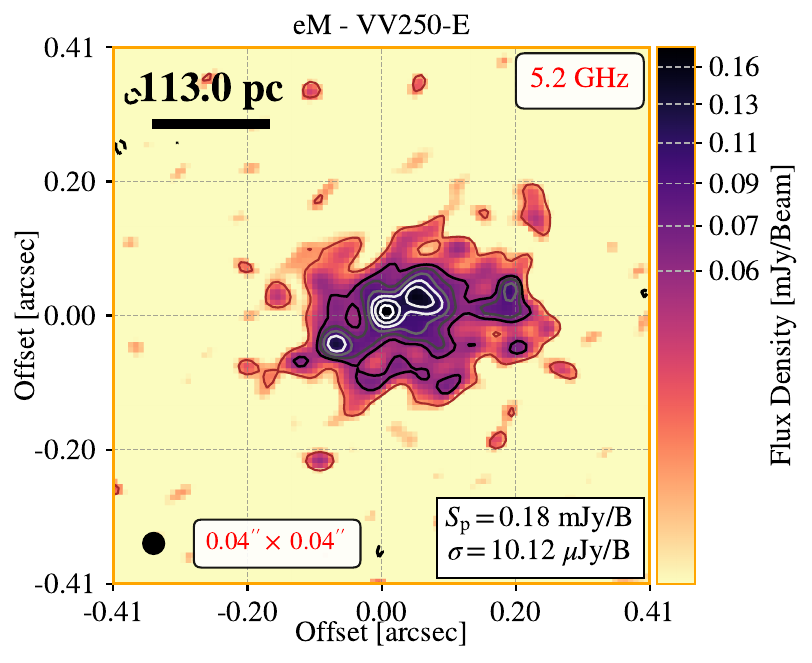}
\includegraphics[width=0.33\linewidth,trim=0 0 0 0]{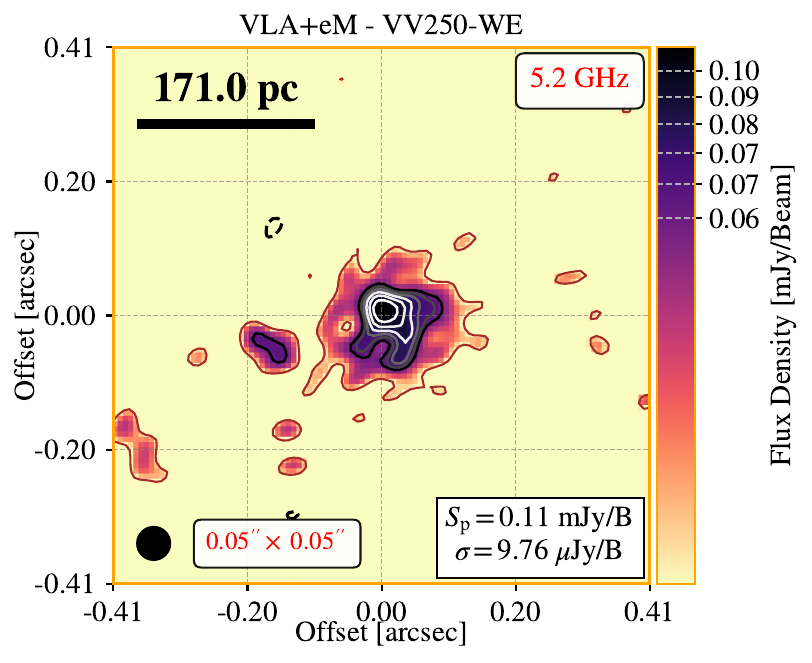}
\includegraphics[width=0.33\linewidth,trim=0 0 0 0]{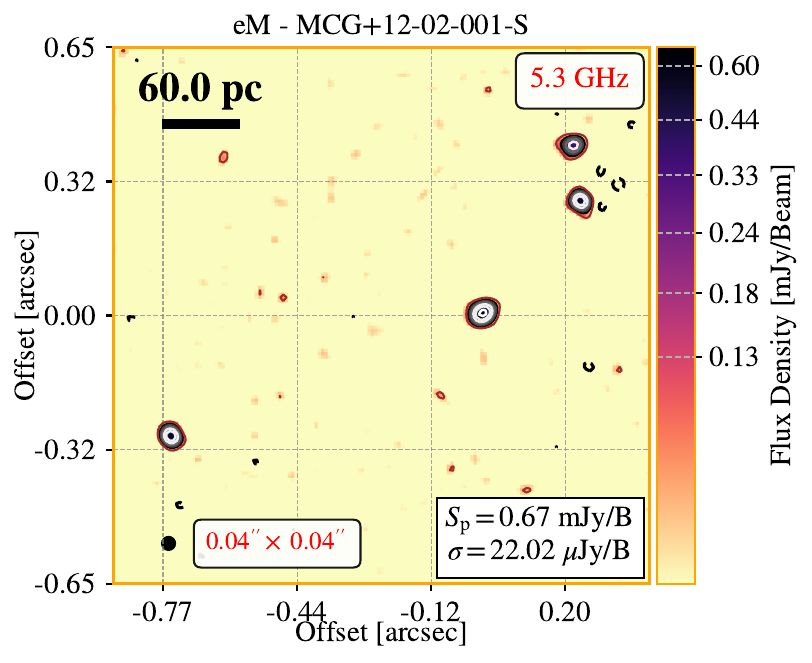}
\includegraphics[width=0.33\linewidth,trim=0 0 0 0]{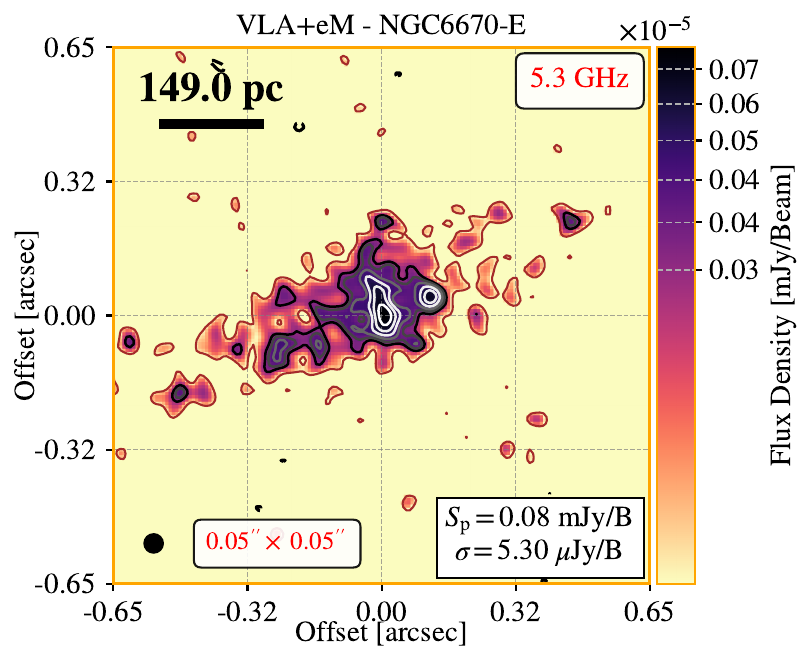}
\includegraphics[width=0.33\linewidth,trim=0 0 0 0]{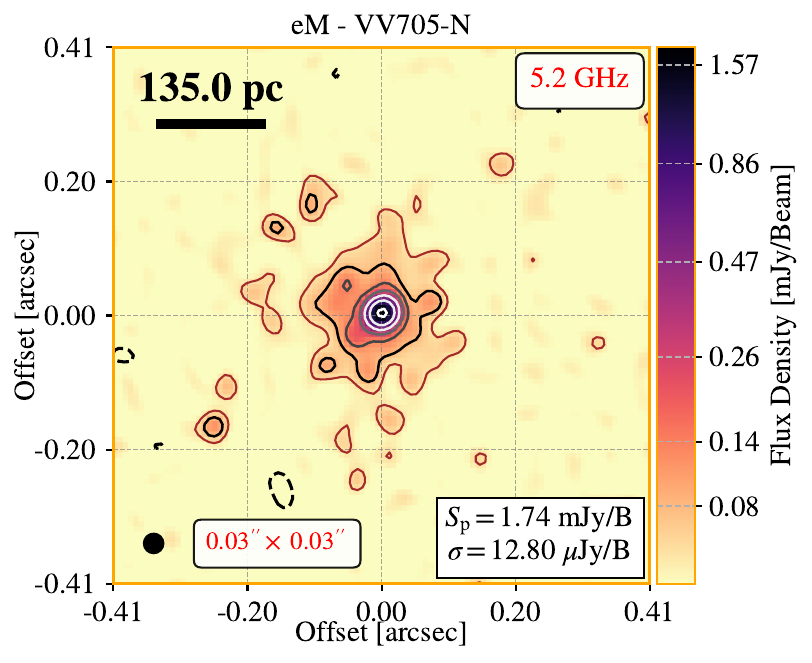}
\end{figure*}
\begin{figure*}
\centering
\includegraphics[width=0.33\linewidth,trim=0 0 0 0]{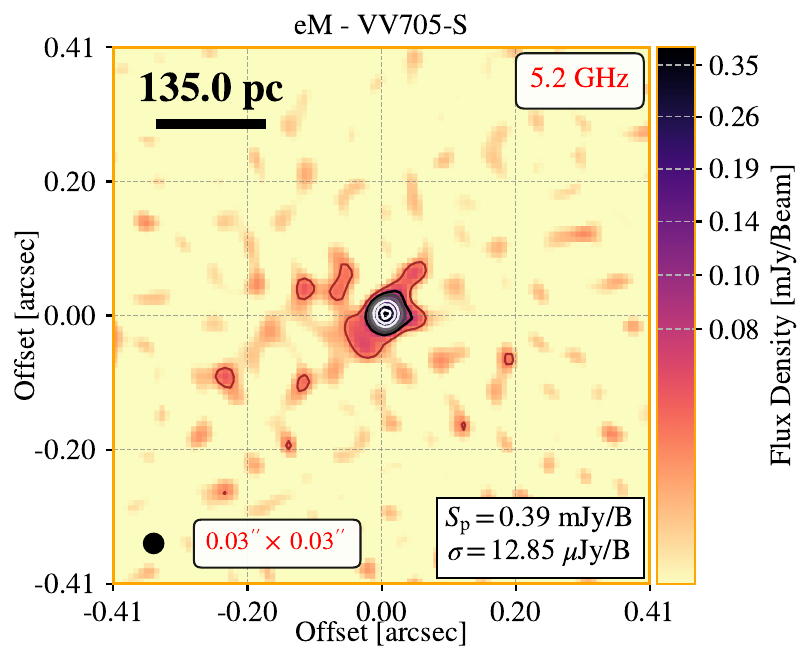}
\includegraphics[width=0.33\linewidth,trim=0 0 0 0]{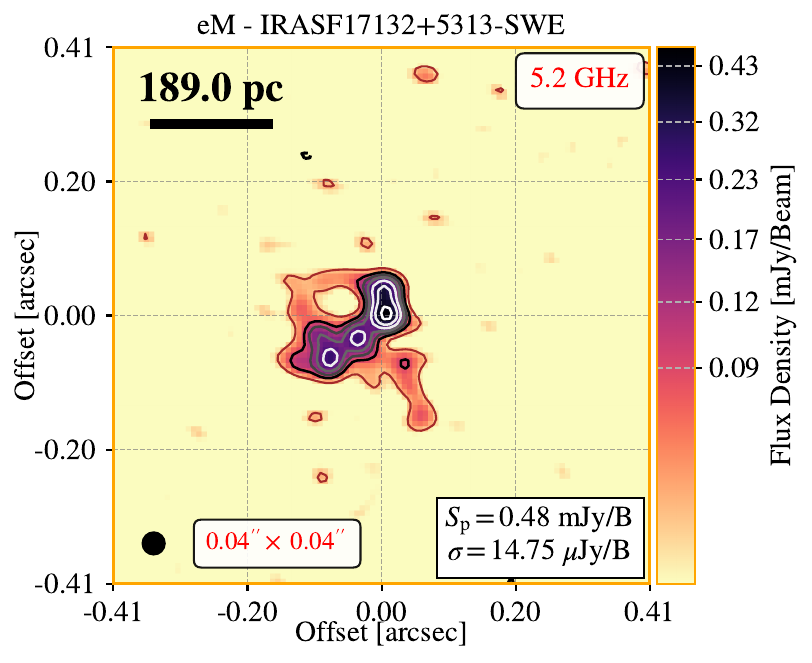}
\includegraphics[width=0.33\linewidth,trim=0 0 0 0]{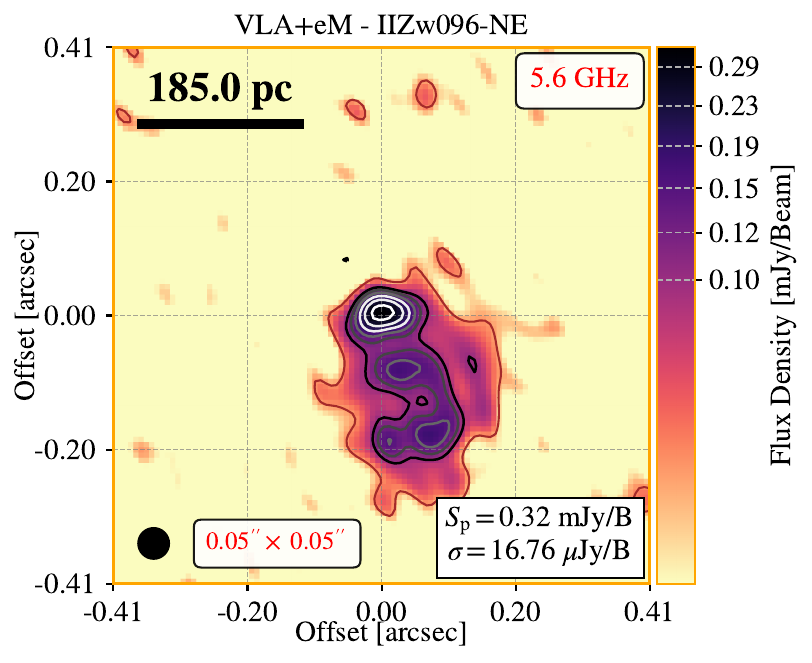}
\caption{Full resolution MFS radio maps with \emph{e}-MERLIN at C-band (centred at $\sim ~5.3$~GHz). IRAS\,23436+5257 (N) is shown in \autoref{fig:example_selfcal_combined_em_VLA_Cband} and NGC\,7674 is shown in \autoref{fig:example_source_multilabels_uvmatch}. For VV\,250 (WE), NGC\,6670 (E) and IIZw\,096 (NE) we display combined images (i.e. \emph{e}-MERLIN + VLA). We generated all images with a circular beam, using the option \texttt{-circular-beam} which fits a circular Gaussian beam to the point spread function, instead of an elliptical Gaussian beam.
\emph{Notes}: 
Each panel displays: 
- Source name and position (CE, S, N, SWE, SE, NE, NWE, WE) (title);
- the restoring beam size (lower-left); 
- the peak intensity $S_{\mathrm{p}}$ and the \emph{global} RMS noise level $\sigma_{\mathrm{rms}}$ (calculated in the residual image) (lower-right); 
- the central frequency of the image (upper-right); 
- the scale bar (upper-left), with a length equal to $5\times$ the size of the restoring beam. 
Contour evaluation is the same as in \autoref{fig:Mrk331_multifreq_lowres}.
}
\label{fig:C_eM_native_1}
\end{figure*}

\begin{table*}
\centering
\caption{Decomposition of the radio emission in the low-resolution data, using images at 6~GHz.}
\label{tab:lowres_native_decomposition}
\begin{subtable}[h]{0.99\textwidth}
    \centering
    \scalebox{0.99}{%
    \begin{tabular}{l|cccccccccccc}
    \hline 
    Source/Component			   & $S_{6}$           & $R_{50}^{6.0}$       & $R_{95}^{6.0}$         & $A_{50}^{6.0}$  			& $A_{95}^{6.0}$            \\[0.3ex]        \rule{0pt}{1.0em}
    (1)						       &  (2) [mJy]        & (3) [pc]             & (4) [pc]               & (5) [${\rm kpc^{2}}$]      & (6) [${\rm kpc^{2}}$]     \\[0.3ex] \hline \rule{0pt}{1.0em}
    Mrk331 (CE-tot)                & $22.8\pm 1.2$     & $185.5\pm 62.9$      & $850.5\pm 95.1$        & $0.110\pm 0.080$			& $2.270\pm 0.500$          \\[0.3ex]        \rule{0pt}{1.0em}
    Mrk331 (CE-uc)                 & $10.5\pm 0.5$     & $20.0\pm 1.5$        & $41.9\pm 4.8$          & $0.001\pm 0.000$		    & $0.005\pm 0.001$          \\[0.3ex]        \rule{0pt}{1.0em}
    Mrk331 (CE-ext)                & $12.1\pm 0.7$     & $484.1\pm 34.8$      & $904.8\pm 65.9$        & $0.736\pm 0.105$			& $2.571\pm 0.368$          \\[0.3ex]        \rule{0pt}{1.0em}
    \qquad$\vdots$                 & $\vdots$          & $\vdots$      	      & $\vdots$               & $\vdots$     		        & $\vdots$                  \\[0.3ex] \hline
    \end{tabular}
    }
    \caption*{\textbf{Notes:} 
    The sizes and areas of all unresolved-core components (with the label ``uc'') are deconvolved quantities. For extended and total components (``ext'' and ``tot'', respectively), the sizes and areas are convolved quantities. The complete table is available online in digital readable format.}
\end{subtable}
\end{table*}
\begin{table*}
\centering
\caption{Decomposition of the radio emission in the native high-resolution data with \emph{e}-MERLIN at $\sim 6$~GHz.}
\label{tab:highres_native_decomposition}
\begin{subtable}[h]{0.99\textwidth}
    \centering
    \resizebox{\textwidth}{!}{%
    \begin{tabular}{l|cccccccccccc}
    \hline 
    Source/Component                & $S_{6}$       & $R_{50,\rm d}^{6.0}$  & $R_{50}^{6.0}$          & $R_{95,\rm d}^{6.0}$ & $R_{95}^{6.0}$          & $T_{\rm b}^{6.0}$     & $A_{50}^{6.0}$                         & $A_{95}^{6.0}$                         \\[0.3ex] 
    (1)                             & (2) [mJy]     &  (3) [pc]             & (4) [pc]                & (5) [pc]     	     & (6) [pc]                & (7) [$\times 10^5$ K] & (8) [$\times 10^{-3}\ {\rm kpc^{2}}$]  & (9) [$\times 10^{-3}\ {\rm kpc^{2}}$]  \\[0.3ex] \hline
    Mrk331 (CE-nuc-tot)             & $10.5\pm 0.7$ & $10.9\pm 1.6$	        & $10.9\pm 1.6$           & $38.8\pm 3.3$		 & $38.8\pm 3.3$           & ---                   & $0.37\pm 0.11$	                        & $4.72\pm 0.79$                         \\[0.3ex]  
    Mrk331 (CE-nuc-cc)              & $5.3\pm 0.3$  & $2.4\pm 0.1$ 	        & $6.1\pm 0.4$            & $4.5\pm 0.8$		 & $12.9\pm 1.6$           & $21.6\pm 2.1$         & $0.02\pm 0.00$	                        & $0.06\pm 0.02$                         \\[0.3ex]  
    Mrk331 (CE-nuc-ne)              & $5.0\pm 0.6$  & $24.1\pm 1.6$	        & $23.4\pm 1.7$           & $41.5\pm 2.0$	     & $42.6\pm 2.4$           & $0.2\pm 0.0$          & $1.72\pm 0.24$ 	                    & $5.70\pm 0.63$                         \\[0.3ex]
    \qquad$\vdots$                  & $\vdots$      & $\vdots$      	    & $\vdots$                & $\vdots$     		 & $\vdots$                & $\vdots$              & $\vdots$   	                        & $\vdots$                               \\[0.3ex] \hline
    \end{tabular}
    }
    \caption*{\textbf{Notes:} 
    The sizes and areas of all compact-core components (with the label ``cc'') are deconvolved quantities. For nuclear extended and total components (``ne'' and ``tot'', respectively), the sizes and areas are convolved quantities. The complete table is available online in digital readable format.}
\end{subtable}
\end{table*}


\bsp	
\label{lastpage}
\end{document}